\documentclass[twocolumn]{aastex7}
\usepackage{float}
\usepackage{graphicx}
\usepackage{amsmath}
\usepackage{natbib}
\usepackage{color}
\usepackage[dvipsnames]{xcolor}
\usepackage{verbatim}
\usepackage[utf8]{inputenc}
\usepackage{mathtools}
\usepackage{amssymb}
\usepackage{textcomp}
\usepackage{enumitem}
\usepackage{braket}
\usepackage{caption}
\usepackage{booktabs}
\usepackage{multirow}
\usepackage{totcount}

\newcommand{\magsqarc}{mag\,arcsec\,$^{-2}$}

\renewcommand{\thetable}{\arabic{table}}

\shortauthors{Smercina \textit{et al.}}

\newtotcounter{citnum} 
\def\oldbibitem{} \let\oldbibitem=\bibitem
\def\bibitem{\stepcounter{citnum}\oldbibitem}

\begin{document}
\title{The Star Formation History and Evolution of the Ultra-Diffuse M81 Satellite F8D1}

\author[0000-0003-2599-7524]{Adam Smercina}
\thanks{Hubble Fellow}
\email{adam.smercina@tufts.edu}
\affiliation{Space Telescope Science Institute, 3700 San Martin Dr., Baltimore, MD 21218, USA}
\affiliation{Department of Physics and Astronomy, Tufts University, 574 Boston Avenue, Medford, MA 02155, USA}

\author[0000-0002-5564-9873]{Eric F. Bell}
\email{}
\affiliation{Department of Astronomy, University of 
Michigan, 323 West Hall, 1085 S. University Ave., Ann Arbor, MI, 48105-1107, USA} 

\author[0000-0002-7502-0597]{Benjamin F. Williams}
\email{}
\affiliation{Department of Astronomy, University of Washington, Box 351580, Seattle, WA 98195-1580, USA}

\author[0009-0001-1147-6851]{Benjamin N. Velguth}
\email{benjamin.n.velguth.gr@dartmouth.edu}
\affiliation{Department of Physics and Astronomy, Dartmouth College, Hanover, NH 03755, USA}

\author[0000-0003-0256-5446]{Sarah Pearson}
\email{}
\affiliation{DARK, Niels Bohr Institute, University of Copenhagen, Jagtvej 155A, 2200 Copenhagen,  Denmark}

\author[0000-0001-6380-010X]{Jeremy Bailin}
\email{}
\affiliation{Department of Physics and Astronomy, University of Alabama, Box 870324, Tuscaloosa, AL 35487-0324, USA}

\author[0000-0003-2544-054X]{Tsang Keung Chan}
\email{}
\affiliation{Department of Physics, The Chinese University of Hong Kong, Shatin, Hong Kong, China}

\author[0000-0002-1264-2006]{Julianne J.\ Dalcanton}
\email{}
\affiliation{Center for Computational Astrophysics, Flatiron Institute, 162 Fifth Ave, New York, NY 10010, USA}
\affiliation{Department of Astronomy, University of Washington, Box 351580, Seattle, WA 98195-1580, USA}

\author[0000-0001-6982-4081]{Roelof S. de Jong}
\email{}
\affiliation{Leibniz-Institut f\"{u}r Astrophysik Potsdam (AIP), An der Sternwarte 16, D-14482 Potsdam, Germany}

\author[0000-0001-9269-8167]{Richard D'Souza}
\email{}
\affiliation{Vatican Observatory, Specola Vaticana, V-00120, Vatican City State}

\author[0000-0001-8416-4093]{Andrew Dolphin}
\email{}
\affiliation{Raytheon Technologies, 1151 E. Hermans Road, Tucson, AZ 85756, USA}
\affiliation{Steward Observatory, University of Arizona, Tucson, AZ 85726, USA}

\author[0000-0001-8867-4234]{Puragra Guhathakurta}
\email{}
\affiliation{UCO/Lick Observatory, Department of Astronomy \& Astrophysics, University of California Santa Cruz, 1156 High Street, Santa Cruz, California 95064, USA}

\author[0000-0001-5538-2614]{Kristen B.W. McQuinn}
\email{}
\affiliation{Space Telescope Science Institute, 3700 San Martin Dr., Baltimore, MD 21218, USA}
\affiliation{Department of Physics and Astronomy, Rutgers, The State University of New Jersey, 136 Frelinghuysen Rd, Piscataway, NJ 08854, USA}

\author[0000-0003-2325-9616]{Antonela Monachesi}
\email{}
\affiliation{Departamento de Astronom\'ia, Universidad de La Serena, Av. Juan Cisternas 1200 N, La Serena, Chile}

\author[0000-0002-0558-0521]{Colin T. Slater}
\email{}
\affiliation{Department of Astronomy, University of Washington, Box 351580, Seattle, WA 98195-1580, USA}

\author[0000-0001-6443-5570]{Elisa Toloba}
\email{}
\affiliation{Department of Physics, University of the Pacific, 3601 Pacific Avenue, Stockton, CA 95211, USA}

\author[0000-0002-6442-6030]{Daniel R. Weisz}
\email{}
\affiliation{Department of Astronomy, University of California, Berkeley, CA 94720, USA}

\author[0000-0003-0603-8942]{Andrew Wetzel}
\email{}
\affiliation{Department of Physics \& Astronomy, University of California, Davis, Davis, CA 95616, USA}

\begin{abstract}
We present deep HST imaging of one of the nearest ultra-diffuse galaxies (UDGs) outside of the Local Group: F8D1, a satellite of M81 known to be tidally disrupting. UDGs are an enigmatic and diverse population, with evolutionary pathways ranging from tidal processing to bursty feedback and high initial angular momentum. To determine F8D1's evolutionary drivers, we resolve stars in F8D1's central $\sim$1\,kpc and in a parallel field $\sim$6\,kpc along its major axis to deep photometric limits, reaching below the Red Clump. We also image eight shallower fields along F8D1's major and minor axes. We calculate the star formation history (SFH) in the two deep fields, finding that while currently quiescent, both regions experienced a substantial burst $\sim$2\,Gyr ago and a smaller burst $\sim$500\,Myr ago, which likely formed F8D1's nuclear star cluster. In the shallow fields, using the ratio of evolved Asymptotic Giant Branch and Red Giant Branch stars out to $\sim$13\,kpc along F8D1's known stellar stream, we confirm that F8D1 was globally star-forming until at least $\sim$2\,Gyr ago. We estimate a total progenitor stellar mass, including the stream, of $\sim$1.3$\times$10$^8\,M_{\odot}$, with an average [M/H]\,$\sim$\,$-$0.8. We compare F8D1's properties to those of Local Group galaxies with similar initial stellar mass. We find that F8D1 is consistent with a progenitor star-forming galaxy similar to NGC 6822 that is in the midst of a transition to a Sagittarius-like system. Notably, this evolutionary sequence can be accomplished through tidal processing alone in galaxies that have experienced sufficiently bursty feedback to have created cored profiles.
\end{abstract}

\section{Introduction}
\label{sec:intro}

\subsection{Ultra Diffuse Galaxies: Extreme End of the LSB Population}
\label{sec:intro-udgs}

In the last decade, a newly-labeled class of low surface brightness (LSB) galaxy has been discovered in unexpectedly large numbers in both nearby galaxy clusters and throughout the local field: so-called ultra-diffuse galaxies \citep[UDGs;][]{vandokkum2015,koda2015,mihos2015}. Though LSB galaxies have been studied for decades \citep[with their own rich literature, e.g.,][and others]{dalcanton1997,deblok&mcgaugh1997}, and there is no ironclad set of UDG selection criteria, these UDGs have rapidly garnered intense interest owing to their extreme properties, including relatively low stellar masses ($M_{\star}\,{=}\,10^7{-}10^{8.5}\,M_{\odot}$), yet radii ($R_{\rm eff}\,{=}\,2{-}4$\,kpc) rivaling much more massive galaxies, and extremely faint central surface brightness (SB; $\mu_{0,V}\,{\gtrsim}\,23{-}24$\,\magsqarc). The discovery of these galaxies has challenged the field to account for their existence in the standard paradigm of galaxy formation. Do these galaxies follow unique evolutionary channels, or are they an expected edge case of the overall galaxy distribution? 

Extreme examples of these UDGs have been discovered in the Local Group (e.g., the Sagittarius dwarf galaxy, \citealt{ibata1995}; Andromeda XIX, \citealt{mcconnachie2008}; Antlia 2, \citealt{torrealba2019}); the faintest of which were effectively hidden prior to the discerning power of nearby stellar motions provided by \textit{Gaia}. Evidence suggests that these Local Group UDGs are exclusively in the final throes of tidal disruption. It is unclear whether these nearby systems are analogous to the emerging populations of more distant UDGs, which are, in contrast, largely symmetric and exhibit no discernible evidence of disruption to deep SB limits \citep{vandokkum2015b,burkert2017,Lim2020}. Furthermore, many UDGs have been found in the field, far from the tidal influence of a massive central galaxy \citep{papastergis2017,barbosa2020}. With the discovery of enigmatically rich globular cluster populations in some (but not all) UDGs \citep[e.g.,][]{peng&lim2016,vandokkum2016,Amorisco2018,Lim2020}, unusual inferred mass-to-light ratios in several systems \citep[e.g.,][]{vandokkum2018,toloba2018,danieli2019,vandokkum2019,Toloba2023}, and high neutral gas fractions in the field \citep[e.g.,][]{papastergis2017,mancera-pina2019,jones2023}, the mystery of UDGs has only deepened. How do these unique galaxies form? 

\begin{figure*}[t]
    \centering
    \includegraphics[width=\linewidth]{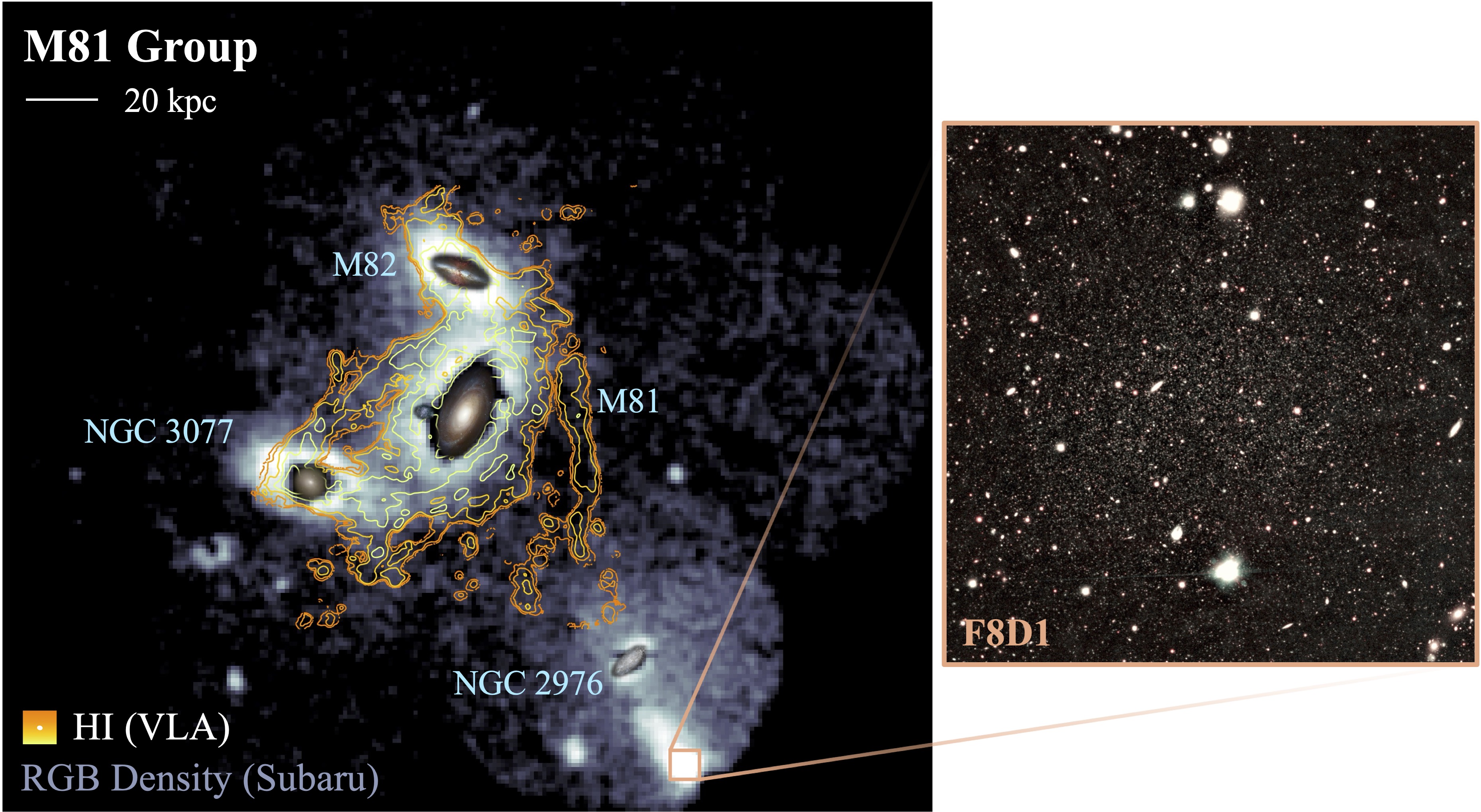}
    \caption{Density of RGB stars in the M81 Group, revealing the global structure of its stellar halo and providing group context for F8D1 (lower right) and its extended tidal stream. North is up and East is left. RGB stars have been selected according to the \cite{smercina2020} criteria for $gri$\ photometry, including size $<$0\farcs7, colors consistent with the $g{-}r$/$r{-}i$\ stellar locus, and within 1.8 magnitudes of the tip of the RGB. The density map features 1\farcm1 pixels, smoothed with a $\sigma\,{=}\,0.8$\,pixel Gaussian kernel, and is scaled logarithmically (minimum value of 1, maximum value of 25 star per arcmin$^2$). Stars were detected and identified in multi-epoch three-filter Subaru HSC imaging. Also overlaid are contours of \textsc{H\,i} 21\,cm emission, from the \cite{deBlok2018} natural-weighted D-array VLA Moment 0 map. Contours are shown in logarithmic steps of 0.5 from 0.01--3 Jy\,beam\,km\,s$^{-1}$. The VLA beam is shown as the white filled ellipse at the bottom left. A zoomed $r/i$\ Subaru HSC color-image of F8D1 is shown in the right panel.}
    \label{fig:m81-group}
\end{figure*}

Like their observational properties, a theoretical consensus on the origins of UDGs has yet to be reached. Four primary ideas have emerged from theoretical studies: (1) tidal processing in cluster or group environments \citep{yozin2015,safarzadeh2017,Carleton2019}, (2) unusually bursty star formation histories (SFHs), resulting in stellar feedback-driven oscillation of galaxy sizes \citep{el-badry2016,dicintio2017,chan2018}, (3) early accretion of gas with high angular momentum, resulting in a naturally shallower central potential and more diffuse star formation for a high-spin tail of normal galaxies \citep{amorisco&loeb2016}, and (4) `failed' galaxies, who experienced violent early star-forming episodes that completely expelled their gas reservoirs before they could form a dense baryonic component \citep{vandokkum2015,peng&lim2016,ferre-mateu2023,Toloba2023}. Some of these mechanisms are mutually exclusive, yet particularly for UDGs in cluster or group environments, most of these mechanisms could work in concert. Further complicating this picture, how UDGs are observationally defined plays an important role in pinning down distinct physical origins \citep[e.g.,][]{vannest2022}, as with any unusual galaxy population. 

It is clear that no single galaxy will encompass all of the unusual properties of the UDG population. However, thus far only the UDGs in the Local Group have been studied to the level of their individual resolved stellar populations \citep[e.g., And XIX;][]{collins2022}. The flavor of evidence is therefore very different when comparing Local Group UDGs to their more distant counterparts. An equally in-depth study of a UDG at intermediate distance, in a different group environment, would go a long way to helping improve understanding of the evolutionary trajectories of these exceptional galaxies. Enter: F8D1. 

\subsection{F8D1: A Prototypical UDG}
\label{sec:intro-f8d1}

\begin{figure*}[t]
    \centering
    \includegraphics[width=0.8\linewidth]{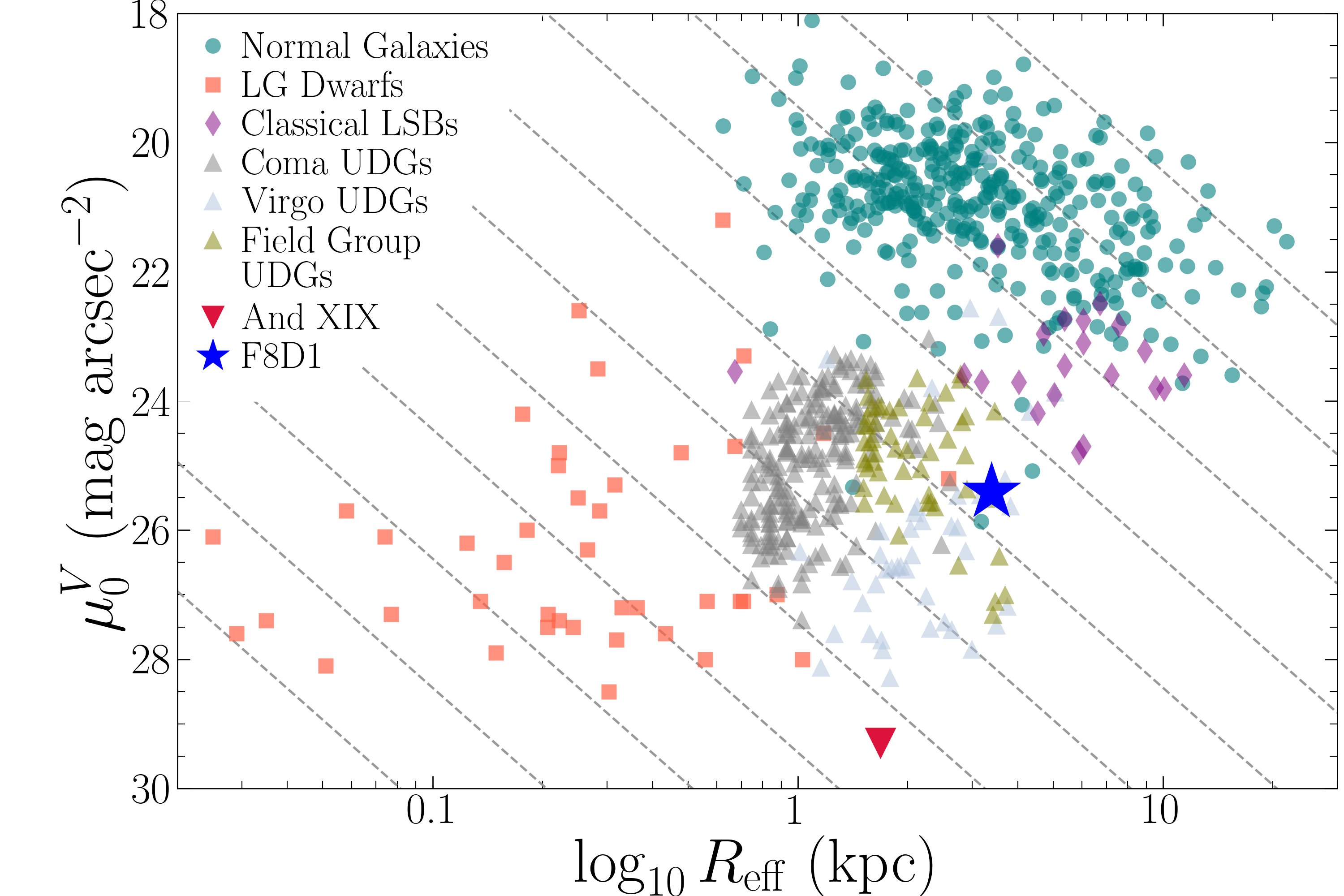}
    \caption{Effective (half-light) radius of different galaxy populations plotted against their central surface brightness in $V$-band ($\mu_0^V$). Lines of constant absolute $V$-band magnitude ($M_V$), assuming an exponential surface brightness profile, are shown as gray dashed lines, ranging from $-$2 (lowest; faintest dwarf galaxies) to $-$22 (highest; brightest galaxies). ``Normal'' galaxies are taken from the CALIFA survey \citep{mendez-abreu2017} and \cite{mcgaugh2005}, shown as teal circles. Local Group dwarf galaxies are taken from \cite{mcconnachie2012} and are shown as red squares. Classical LSBs are taken from \cite{dalcanton1997} and \cite{deblok&mcgaugh1997}, shown as purple diamonds. UDGs identified in the Coma Cluster are shown as gray triangles, taken from \cite{vandokkum2015} and \cite{yagi2016}. UDGs in the Virgo Cluster are shown as light blue triangles, taken from \cite{Lim2020}, with $\mu_{0,V}$\ provided via private communication by Sungsoon Lim, and radii calculated assuming a distance of 16.5\,Mpc \citep{Mei2007,Blakeslee2009}. Several samples of UDGs discovered in galaxy groups in low-density environments are also shown as olive triangles, taken from \cite{merritt2016} and \cite{marleau2021}. Andromeda XIX, the tidally disrupted extremely diffuse satellite of M31 \citep[e.g.,][]{collins2022}, is denoted by a downward red triangle. Finally, F8D1 is denoted by the large blue star, using the original values measured by \cite{caldwell1998}. Note that we derive updated parameters for F8D1 in \S\,\ref{sec:profile}.}
    \label{fig:udgs}
\end{figure*}

F8D1, discovered by \cite{caldwell1998} in a CCD survey with the Burrell Schmidt telescope on Kitt Peak and confirmed with the \textit{Hubble Space Telescope} (\textit{HST}), is a nearby low-surface brightness dwarf satellite of the M81 Group (M81 distance is 3.6\,Mpc, \citealt{radburn-smith2011}). 

We give a visual summary of the M81 Group in Figure \ref{fig:m81-group}, showing the positions of the major galaxies overlayed on the density of RGB stars detected in Subaru Hyper Suprime-Cam (HSC) imaging, using data presented in \cite{smercina2020}, \cite{zemaitis2023}, and new observations from the recent Subaru/Gemini exchange proposal S23A-TE088-GQ (PI: Bell). Nearly invisible in typical ground-based imaging, such as SDSS, F8D1 is the lowest surface brightness ($\mu_{V,0}\,{\simeq}\,25$\ mag\,arcsec$^{-2}$; \citealt{caldwell1998}) known satellite of the M81 Group of galaxies. The M81 Group is a rich environment, with three satellites at least as massive as the LMC: the starburst disk galaxy M82, the compact elliptical NGC 3077, and the M33-analog flocculent spiral NGC 2976. Both M82 and NGC 3077 appear to be in an advanced gravitational interaction with M81 \citep{yun1994,okamoto2015,smercina2020}, while NGC 2976 may or may not have yet experienced a close pass to M81 \citep{williams2010,Drzazga2016}. F8D1 is $\sim$120\,kpc to M81's southwest (in projection) --- much nearer to NGC 2976 in projected distance. 

We show the properties of F8D1, derived by \cite{caldwell1998}, in the context of other galaxy populations in Figure \ref{fig:udgs}, including ``normal'' galaxies from the CALIFA survey\footnote{The CALIFA sample are relatively nearby galaxies, with morphological types spanning the Hubble Diagram, masses $M_{\star}\,{>}\,10^9\,M_{\odot}$, and a median distance of 67\,Mpc.} \citep{mendez-abreu2017}, Local Group dwarf galaxies \citep{mcconnachie2012}, and known UDGs at larger distances, both in the Coma Cluster \citep{vandokkum2015,yagi2016} and near known groups/associations in low-density environments \citep{merritt2016,marleau2021}. This is not an exhaustive comparison, and selection functions of the comparison samples have not been considered. Figure \ref{fig:udgs} is simply illustrative of the unique properties of UDGs among other galaxy populations and shows that, indeed, F8D1 is prototypical among known UDGs. This makes F8D1 the nearest `garden variety' UDG that has been resolved into its constituent stars. 

As is readily apparent in Figure \ref{fig:m81-group}, access to its stellar populations on large scales has revealed an extended tidal stream associated with F8D1, first described by \cite{zemaitis2023}. The distance to F8D1 was first estimated by \cite{caldwell1998}, using the Tip of the Red Giant Branch (TRGB), at 4.0$\pm$0.2\,Mpc. A total of seven distance measurements to F8D1 have now been made, including more recent Galactic reddening maps and distance fitting techniques, with a median distance of 3.8$^{+0.09}_{-0.08}$\,Mpc (${\mu}\,{=}\,27.9{\pm}0.05$\,mag) \citep{caldwell1998,Ferrarese2000,Karachentsev2000,Dalcanton2009,jacobs2009,weisz2011,Anand2021}, which we adopt here. \cite{caldwell1998} revealed that F8D1 has likely not experienced a quiet history. They discovered a substantial population of luminous Asymptotic Giant Branch (AGB) stars, suggesting a more recent episode of star formation ($\sim$3--5\,Gyr) in F8D1 than typical satellites of this mass around the MW. They also measured a significant metallicity dispersion in F8D1 (average [Fe/H]\,=\,$-$1), further hinting at a likely extended star formation history (SFH).

With hints of a more active recent evolutionary history, F8D1 is a powerful probe of UDG evolution. In order to disentangle in even greater detail how F8D1's evolution has been shaped by the various possible UDG formation scenarios, we obtained deep HST imaging of its center and extended stream, and of a number of additional flanking fields along the stream. We present the results of this survey in the remainder of this paper. We describe the data in \S\,\ref{sec:data}, including the observations (\S\,\ref{sec:obs}) and stellar photometry (\S\,\ref{sec:phot}), followed by our results in \S\,\ref{sec:results}. In \S\,\ref{sec:sfh}, and its subsections, we describe the process of fitting observed color--magnitude diagrams (CMDs) of F8D1's stars to reconstruct its detailed SFH. In \S\,\ref{sec:agb-rgb}, we present a larger-scale analysis of the ``shutdown time'' of star formation in F8D1 using the ratio of luminous AGB-to-RGB stars in the shallower flanking fields, and compare to the SFH fits. In \S\,\ref{sec:profile}, we use the full HST dataset, in conjunction with ground-based archival data, to model the radial structure of F8D1 and estimate the total stellar content. \S\,\ref{sec:discussion} contains a detailed discussion of F8D1's evolution and its relevance to the evolution of other ultra-diffuse systems. We end with our conclusions in \S\,\ref{sec:conclusions}. 




\section{Data}
\label{sec:data}

\subsection{Observations}
\label{sec:obs}
The observations presented in this paper were taken through the Hubble Space Telescope Cycle 28 program GO-16191 (PI: Smercina). The observations were taken between 09 December 2020 and 21 February 2022. 19 of the 31 orbits were executed during the Cycle 28 schedule (late 2020--early 2021), the remaining 12 orbits were rescheduled to the following cycle due to single guide star acquisition failures during the initial visits. 

\newcolumntype{t}{!{\extracolsep{16pt}}l!{\extracolsep{0pt}}}
\newcolumntype{p}{!{\extracolsep{15pt}}c!{\extracolsep{0pt}}}

\begin{deluxetable*}{tppppppp}
\tablecaption{\textnormal{Observations}\label{tab:obs}}
\tablecolumns{8}
\setlength{\extrarowheight}{2pt}
\tabletypesize{\footnotesize}
\tablehead{%
\nocolhead{} &
\colhead{R.A.} &
\colhead{Dec.} &
\nocolhead{} &
\nocolhead{} &
\nocolhead{} & 
\colhead{Exp.} &
\colhead{50\% Comp.} \vspace{-2mm}\\
\colhead{\hspace{-27pt}Field Name} &
\colhead{(hh:mm:ss)} &
\colhead{(dd:mm:ss)} &
\colhead{Camera} &
\colhead{Filter} &
\colhead{Date} & 
\colhead{(s)} &
\colhead{(mag)} \vspace{-5mm}\\
}
\startdata
\multirow{2}{*}{NAME-F8D1} & \multirow{2}{*}{09:44:46.6} & \multirow{2}{*}{+67:26:18.9} & \multirow{2}{*}{WFC3} & F814W & 2021-01-31 & 46,059 & 28.34 \\
& & & & F606W & 2021-02-01 & 27,108 & 28.77 \\[3pt]
\multirow{2}{*}{F8D1-257ne-35908} & \multirow{2}{*}{09:45:30.1} & \multirow{2}{*}{+67:30:17.9} & \multirow{2}{*}{ACS} & F814W & 2021-02-01 & 33,281 & 28.37 \\
& & & & F606W & 2021-01-31 & 33,301 & 29.07 \\[3pt]
\multirow{2}{*}{F8D1-FIELD1} & \multirow{2}{*}{09:44:55.3} & \multirow{2}{*}{+67:29:40.2} & \multirow{2}{*}{WFC3} & F814W & 2021-02-06 & 1,392 & \dots \\
& & & & F606W & 2021-02-06 & 1,200 & \dots \\[3pt]
\multirow{2}{*}{F8D1-FIELD2} & \multirow{2}{*}{09:45:07.3} & \multirow{2}{*}{+67:31:54.3} & \multirow{2}{*}{WFC3} & F814W & 2021-02-05 & 1,392 & \dots \\
& & & & F606W & 2021-02-05 & 1,200 & \dots \\[3pt]
\multirow{2}{*}{F8D1-FIELD3} & \multirow{2}{*}{09:45:34.8} & \multirow{2}{*}{+67:22:22.0} & \multirow{2}{*}{WFC3} & F814W & 2022-02-11 & 1,392 & \dots \\
& & & & F606W & 2022-02-11 & 1,200 & \dots \\[3pt]
\multirow{2}{*}{F8D1-FIELD4} & \multirow{2}{*}{09:45:10.8} & \multirow{2}{*}{+67:23:43.1} & \multirow{2}{*}{WFC3} & F814W & 2020-12-09 & 1,392 & \dots \\
& & & & F606W & 2020-12-09 & 1,200 & \dots \\[3pt]
\multirow{2}{*}{F8D1-419ne-34689} & \multirow{2}{*}{09:45:29.1} & \multirow{2}{*}{+67:34:26.6} & \multirow{2}{*}{ACS} & F814W & 2021-02-06 & 1,153 & \dots \\
& & & & F606W & 2021-02-06 & 1,103 & \dots \\[3pt]
\multirow{2}{*}{F8D1-566ne-34999} & \multirow{2}{*}{09:45:43.9} & \multirow{2}{*}{+67:36:29.8} & \multirow{2}{*}{ACS} & F814W & 2021-02-05 & 1,153 & \dots \\
& & & & F606W & 2021-02-05 & 1,103 & \dots \\[3pt]
\multirow{2}{*}{F8D1-448se-34499} & \multirow{2}{*}{09:46:06.8} & \multirow{2}{*}{+67:27:14.8} & \multirow{2}{*}{ACS} & F814W & 2022-02-11 & 1,165 & \dots \\
& & & & F606W & 2022-02-11 & 1,115 & \dots \\[3pt]
\multirow{2}{*}{F8D1-632se-7787} & \multirow{2}{*}{09:45:59.8} & \multirow{2}{*}{+67:20:24.4} & \multirow{2}{*}{ACS} & F814W & 2020-12-09 & 1,153 & \dots \\
& & & & F606W & 2020-12-09 & 1,103 & \dots \\
\enddata
\tablecomments{Summary of observations, including coordinates, camera, filter, and exposure information. Deep fields are listed first, followed by the shallower flanking fields. In each case, the ACS parallels that correspond to each WFC3 primary observation are ordered the same as their WFC3 counterparts. For example, F8D1-FIELD1 and F8D1-419ne-34689 were taken in parallel. For the deep fields, we also give the 50\% completeness depth in each camera, determined from ASTs (\S\,\ref{sec:phot}). See \S\,\ref{sec:data} for a more detailed description of the observations.}
\end{deluxetable*}

\begin{figure*}[t]
    \centering
    \includegraphics[width=\linewidth]{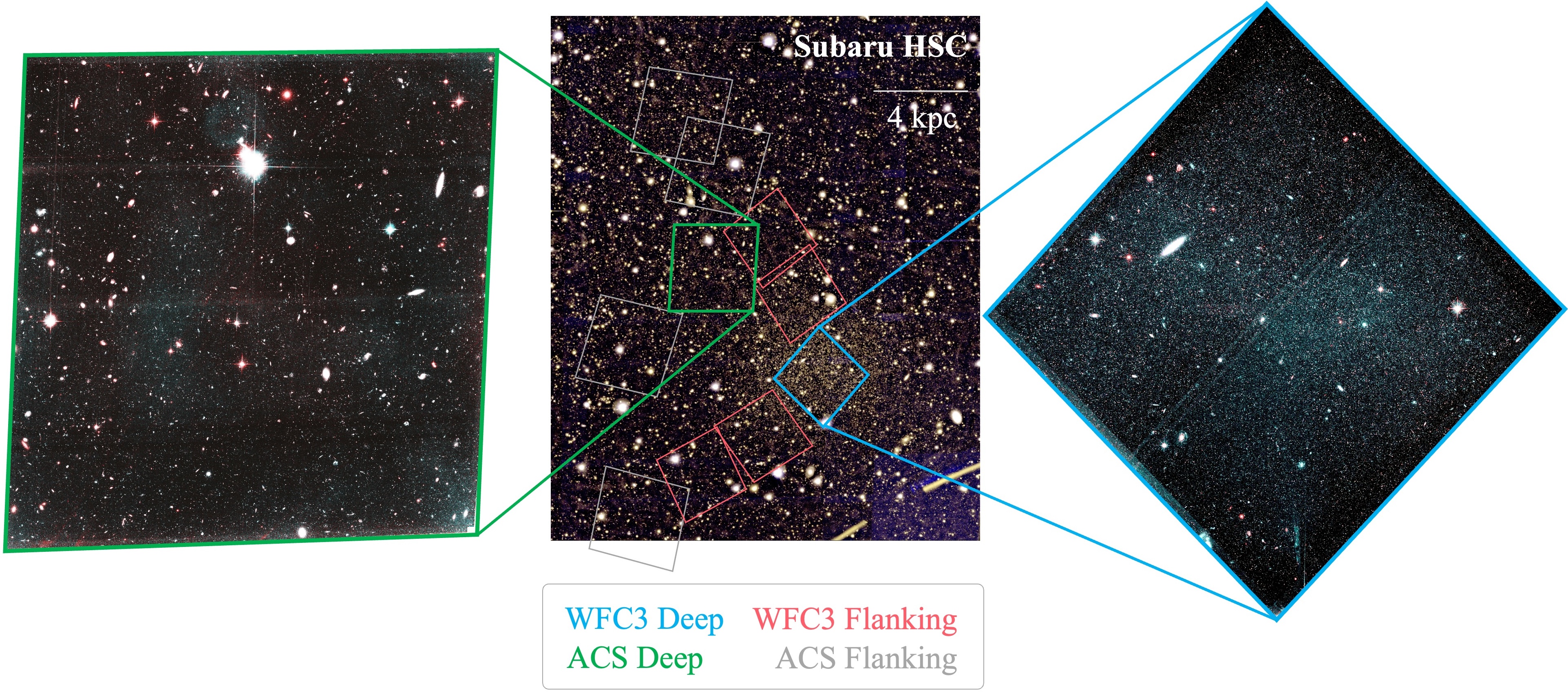}
    \caption{Overview of the layout of our HST survey of F8D1, with a $g$,$r$,$i$\ image from Subaru HSC at the center. The two deep fields are shown, with WFC3 in blue and ACS in green, as well as the shallower flanking fields, with WFC3 pointings in red and ACS in gray. We also show the drizzled F606W/F814W images for the two deep fields as zoomed insets. Images are oriented with North up and East to the left. The central WFC3 field covers F8D1's approximate effective radius.}
    \label{fig:survey}
\end{figure*}

We targeted the cataloged center of F8D1 with the Wide Field Camera 3 (WFC3), as well as coordinated parallel Advanced Camera for Surveys (ACS) observations for a field to the north east, for a total of 27 orbits. Each single-orbit visit was divided into three equal-length WFC3 sub-pixel dithers, using a linear 3-point pattern with 0\farcs135 spacing, for a target of 2,715 seconds of total exposure time per orbit. Corresponding parallel observations comprised three \textit{approximately} equal-length corresponding ACS exposures, for a target of 2,567 seconds of total exposure time per orbit. The visits were also manually mapped to a 4-point box pattern, achieved by using \texttt{POS TARG} to shift the target positions from visit-to-visit, designed to cover the chip gap. The box was designed with vertices ($-$2.042, $-$2.173), (2.042, $-$2.173), (2.042, 2.173), ($-$2.042, 2.173); each four visits therefore completed a full box pattern. Of the 27 orbits, the WFC3 observations were divided into 17 orbits for the F814W filter and 10 orbits for the F606W filter, while the ACS observations were divided into 14 orbits for F814W and 13 orbits for F606W. We designed this observing strategy to detect resolved individual stars at the distance of F8D1 to approximately one magnitude below the predicted core helium-burning phase for old stellar populations, the `Red Clump' (RC; \citealt{alves&sarajedini1999,girardi2016}), in both cameras. 

We also targeted four single-orbit ``flanking'' fields with WFC3 as primary and coordinated ACS parallels, aimed at detecting the presence of possible tidal debris or extended stellar outskirts in F8D1. Each of these four flanking fields was observed in both F606W and F814W, with two dithered exposures each. To maximize the exposure time possible in a single orbit, we split the total F606W exposure time per orbit by $\sim$1/3 in the first exposure and $\sim$2/3 in the second, and the opposite for F814W ($\sim$2/3 and $\sim$1/3, respectively). Executed exposure times sometimes varied by several seconds relative to the Phase-II program submitted in the Astronomer's Proposal Tool (APT), particularly for the rescheduled observations, to adjust for HST's evolving orbital visibility window. The total executed exposure times for each field in this program are given in Table \ref{tab:obs}. All images can be obtained from the MAST archive: \dataset[doi:10.17909/zry4-hr90]{https://dx.doi.org/10.17909/zry4-hr90}.

\subsection{Photometry}
\label{sec:phot}

We performed point spread function (PSF)-fitting on the HST pipeline-calibrated images (\texttt{flc} extension) using the most up-to-date version of the \texttt{DOLPHOT} software package \citep{dolphin2000,Dolphin2016}. We ran artificial star tests (ASTs) in the two deep fields to assess the photometric quality, completeness, and bias of our observations. For each field, we used the Bayesian Extinction and Stellar Tool (\texttt{BEAST}) package \citep{gordon2016} to generate input lists of 170,000 artificial stars, sampled from a realistic range of stellar population models. We then used \texttt{DOLPHOT} to run these input lists as ASTs, following the method developed by the Panchromatic Hubble Andromeda Treasury (PHAT) program \citep{williams2014}. 

Using the results of the ASTs, we compute completeness curves in each filter, for each target field. We used these completeness curves to optimize the selection criteria for stellar sources --- i.e.\ balancing loosening criteria for greater depth and tightening them to reduce contamination. We consider sources to be `good stars' (GST), if they pass our criteria for \texttt{DOLPHOT}'s signal-to-noise ratio (\texttt{SNR}), crowding (\texttt{CROWD}), and sharpness (\texttt{SHARP}) parameter outputs. Due to the depth of our imaging, and very little foreground or background contamination, we choose slightly lower \texttt{SNR} criteria than are typically used in the deep fields. In depth-limited, rather than crowding-limited, fields such as these the \texttt{CROWD} parameter is particularly useful for identifying spurious sources; for example, the diffraction spikes of bright foreground stars, which are often resolved into many individual sources. In all fields, we choose the smallest \texttt{CROWD} parameter that is able to remove the majority of these contaminating sources without significantly changing the measured 50\% completeness depth relative to more standard values, such as used by the PHAT Survey. For example, adopting the \texttt{CROWD} and \texttt{SHARP} values of \citep{williams2014} yields changes of $<$0.05\,mag to our completeness limits, for the same \texttt{SNR}. We give our GST criteria for each field and camera combination in Table \ref{tab:gst}. 

\begin{figure*}[!ht]
    \centering
    \includegraphics[width=0.9\linewidth]{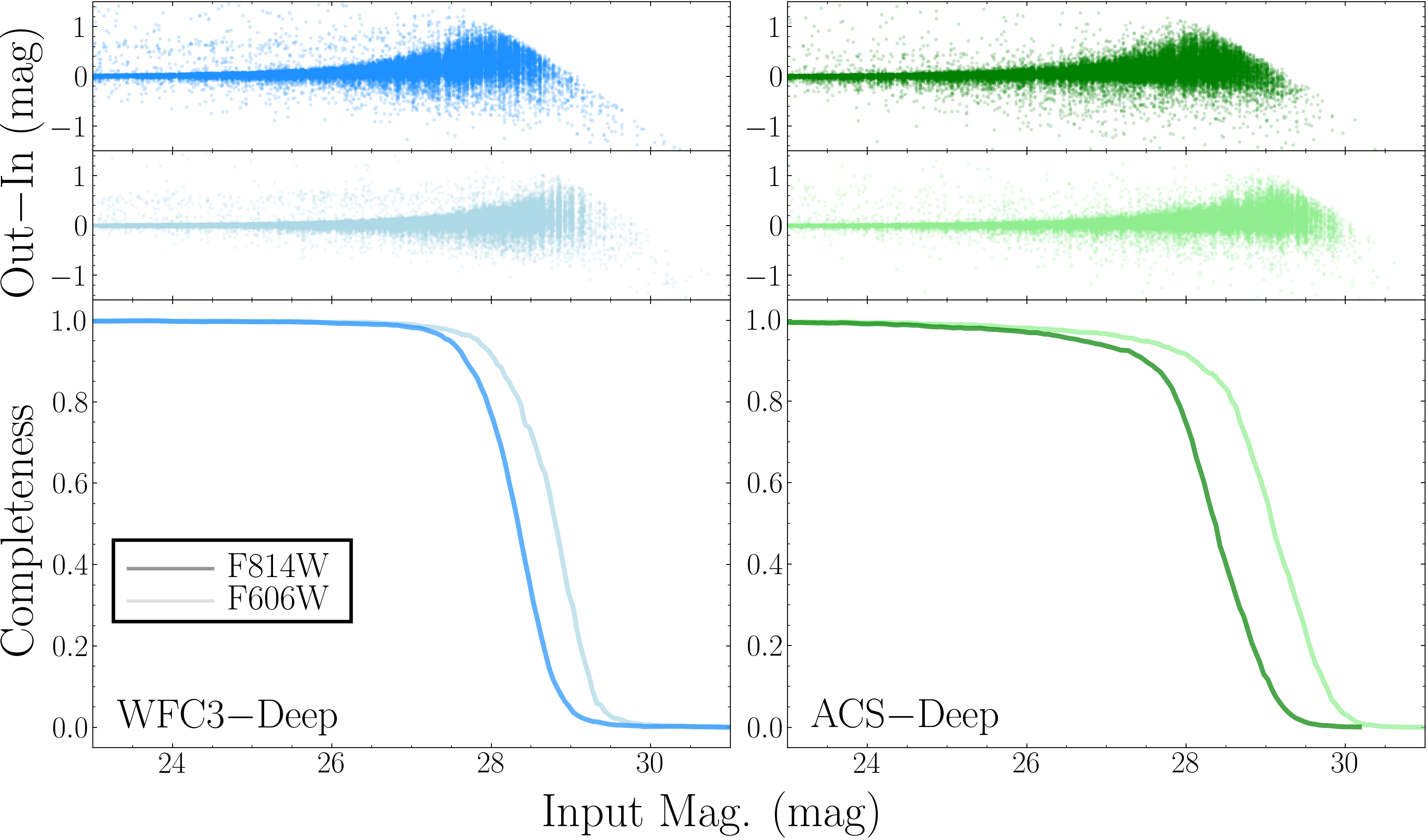}
    \caption{Summary of the data quality in the two WFC3 (left column; blue) and ACS (right column; green) deep fields, for GST sources selected in F8D1, assessed from the AST results. Results for the F606W filter are shown in the lighter shade, and those for F814W in the darker shade. The top two rows show the difference in the measured magnitudes of GST-selected artificial stars compared to their input magnitudes. As is commonly seen near the completeness limit, there is a slight bias for the faintest stars, where they are measured to be 0.1--0.2 magnitudes fainter than expected on average. This suggests the data are not crowding-limited, as blends in crowded fields would typically show biases to brighter recovered magnitudes. The bottom row shows the detection completeness of recovered artificial GST-selected sources, as a function of magnitude.
    }
    \label{fig:completeness}
\end{figure*}

In Figure \ref{fig:completeness}, we show photometry residuals and the completeness curves for GST sources in F606W and F814W, for the deep WFC3 and ACS fields. In Figure \ref{fig:deep-cmds}, we show the GST color--magnitude diagrams (CMDs) and density maps for the two deep fields. These observations are very deep, with 50\% completeness limits of 28.77 in F606W and 28.34 in F814W for WFC3, and 29.07 in F606W and 28.37 in F814W for ACS. 

\section{Results}
\label{sec:results}

\begin{deluxetable*}{lcccccc}[!ht]
\tablecaption{\textnormal{GST Selection Criteria}\label{tab:gst}}
\tablecolumns{7}
\setlength{\extrarowheight}{2pt}
\tabletypesize{\small}
\tablehead{%
\colhead{} &
\colhead{\texttt{SNR}$_{\rm F606W}$} &
\colhead{\texttt{SHARP}$^2_{\rm F606W}$} &
\colhead{\texttt{CROWD}$_{\rm F606W}$} & 
\colhead{\texttt{SNR}$_{\rm F814W}$} &
\colhead{\texttt{SHARP}$^2_{\rm F814W}$} &
\colhead{\texttt{CROWD}$_{\rm F814W}$} \vspace{-2mm}\\
\colhead{Field+Camera} & 
\colhead{($>$)} & 
\colhead{($<$)} & 
\colhead{($<$)} & 
\colhead{($>$)} & 
\colhead{($<$)} & 
\colhead{($<$)} \vspace{-5mm}\\
}
\startdata
Deep WFC3 & 3.3 & 0.1 & 1 & 3.3 & 0.1 & 1 \\
Deep ACS & 3.3 & 0.3 & 0.2 & 3.3 & 0.3 & 0.2 \\
Shallow WFC3 & 4 & 0.1 & 0.2 & 4 & 0.1 & 0.2 \\
Shallow ACS & 4 & 0.1 & 0.2 & 4 & 0.1 & 0.2 \\
\enddata
\tablecomments{Selection criteria for `Good Star' (GST) sources, using the \texttt{DOLPHOT} output parameters.}
\end{deluxetable*}

In the following sections, we present the results of our analysis of these observations. First, in \S\,\ref{sec:sfh} we fit our photometry to infer detailed star formation histories (SFHs) over the past 6\,Gyr for both of the central WFC3 and outer ACS deep fields in F8D1. Next, in \S\,\ref{sec:agb-rgb} we use the ratio of the number of luminous Asymptotic Giant Branch (AGB) stars to stars on the upper Red Giant Branch (RGB) to estimate the `quenching time' for star formation across F8D1, and compare these results to the full calculated SFHs. Finally, we use these and other results from the literature to study the structure of F8D1 using resolved star counts. 

\subsection{A Detailed History of Star Formation in F8D1}
\label{sec:sfh}

\begin{figure*}[t]
    \centering
    \includegraphics[width=\linewidth]{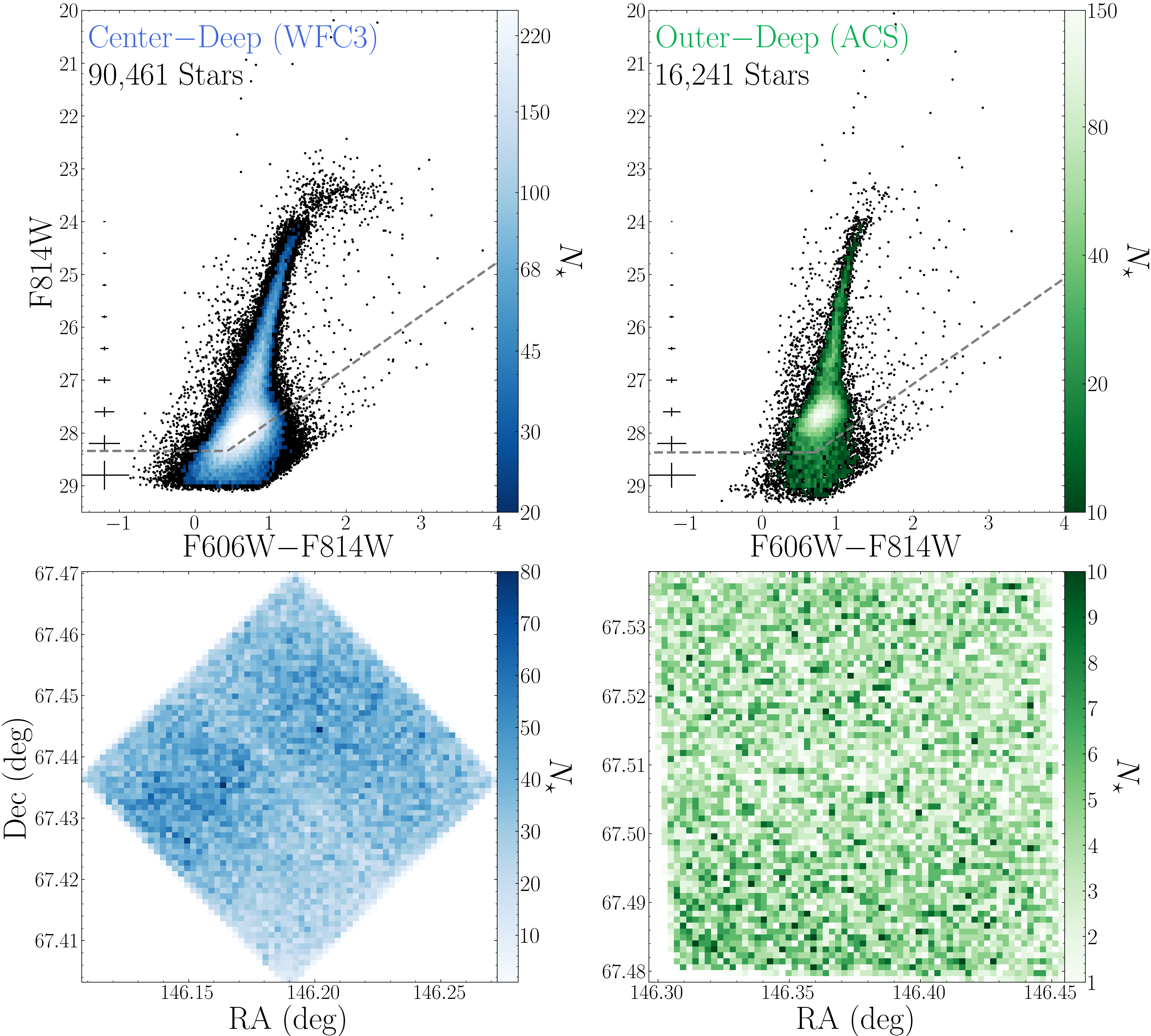}
    \caption{\textit{Top row}: Color--magnitude diagrams (CMDs) of GST-selected stars in the central deep WFC3 field (left) and outer deep ACS field (right). Stellar density is shown above a level of $\sim$3,000 $\star$/mag$^2$\ ($N_{\star}\,{=}$\,10/bin). Below this density individual stars are shown. Median errors for the photometry are shown to the left, from F814W\,=\,24--29\,mag at 0.6\,mag increments. The peak of the Red Clump is visible above the 50\% completeness curve (shown as a gray curve) in both fields, with a center at F814W\,$\sim$\,27.6--28.0. We note that the different RC morphology visible between the two fields is an early hint at the differences in SFH of the two populations. As shown in Figures \ref{fig:cmd-burst} \&\ \ref{fig:RC-LFs}, the variable morphology of this feature encodes important information about F8D1's SFH. \textit{Bottom row}: Spatial density of GST sources in each field, shown at a scale of $\sim$3\arcsec\ per pixel.}
    \label{fig:deep-cmds}
\end{figure*}

\begin{figure*}[t]
    \centering
    \includegraphics[width=\linewidth]{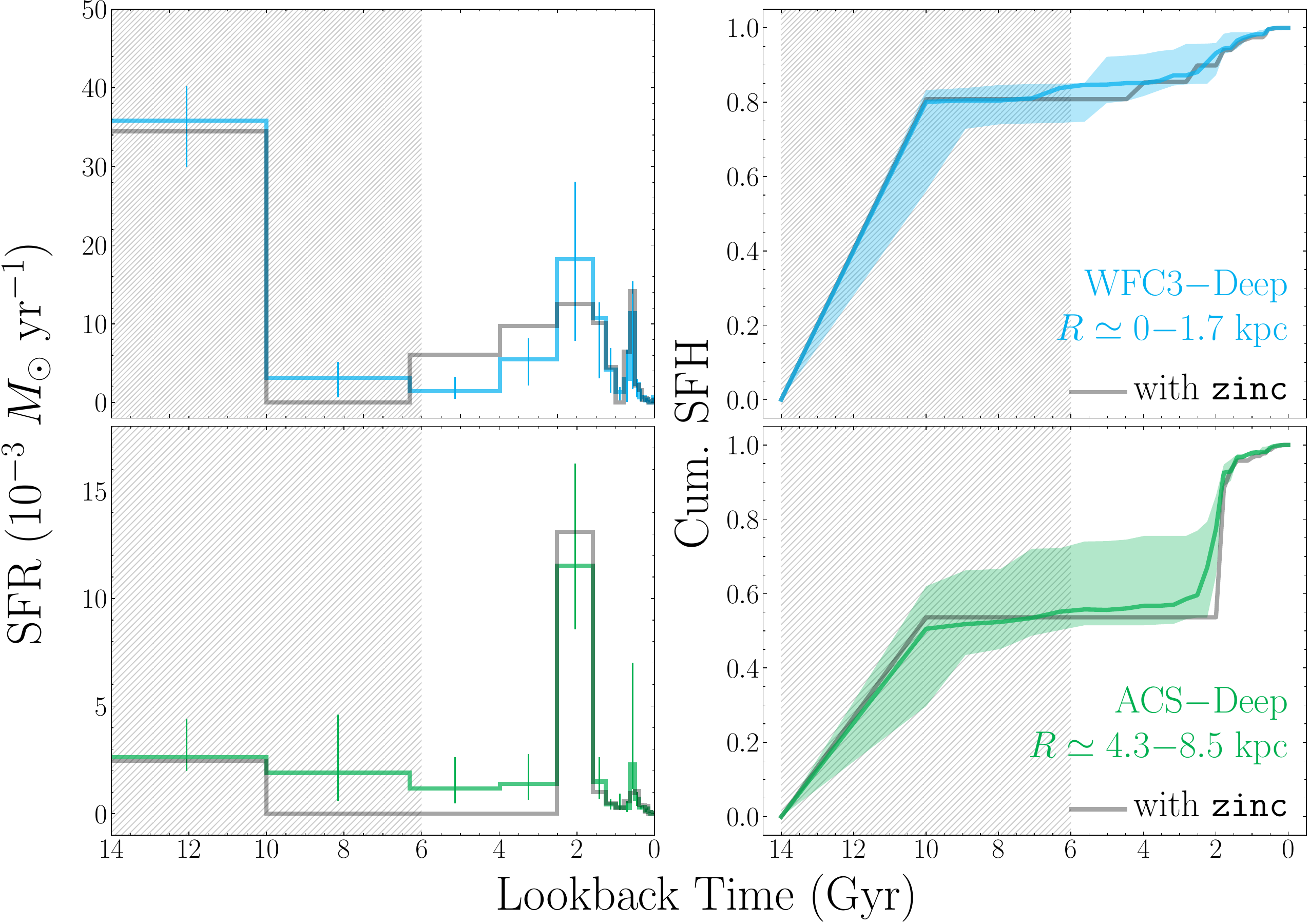}
    \caption{F8D1's star formation history (SFH) calculated from the deep central WFC3 (top row, blue) and outer ACS (bottom row, green) fields. The instantaneous SFHs (SFR over time) are shown in the left column, while the cumulative SFHs, displayed as a fraction of the present-day stellar mass, are shown in the right column. The 1$\sigma$\ combined random and systematic uncertainties on the SFR in each bin are shown, and the corresponding 1$\sigma$\ envelopes for the cumulative SFHs are shown as shaded regions in the same color. Note the different $y$-axis scales for the two instantaneous SFHs. As discussed in \S\,\ref{sec:age-mh}, we show results for both a free age--metallicity relation (color) and a strictly increasing age--metallicity relation (MATCH's \texttt{zinc} option; dark gray). The light hatched region denotes the age range where SFH results are highly uncertain for the photometric depth of these data. Note that these results correspond only to the FOVs shown in Figure \ref{fig:survey}.}
    \label{fig:sfhs}
\end{figure*}

Our observations in the deep WFC3 and ACS fields reach $\sim$1.0\,mag below the RC, providing reliable access to F8D1's ancient stellar populations and SFH. To calculate the SFHs of these populations, we use the well-tested color--magnitude fitting software MATCH \citep{dolphin2002}. Using stellar evolutionary models, MATCH forward models the observed CMD of a stellar population as a composite of stars formed at different ages and with different metallicities (see \citealt{weisz2014} for an updated description). It is a widely used tool for deriving the SFHs of stellar populations in dwarf galaxies throughout the local universe \citep[e.g.,][and others]{weisz2008,mcquinn2009,mcquinn2010,weisz2011,weisz2012,weisz2014,mcquinn2015,albers2019,gallart2021,rusakov2021,collins2022,mcquinn2023,savino2023}. When comparing the results of observations reaching below the RC and much deeper observations reaching the oldest main sequence turnoff (oMSTO), \cite{weisz2014} found that photometry resolving the RC is sufficient for MATCH to recover accurate SFHs back to $\sim$6\,Gyr. We show a more detailed breakdown of the sensitivity of our data to the RC in F8D1 in \textsc{Appendix \ref{sec:rc-lfs}} (Figure \ref{fig:RC-LFs}).

\begin{figure*}[t]
\centering
\includegraphics[width=\linewidth]{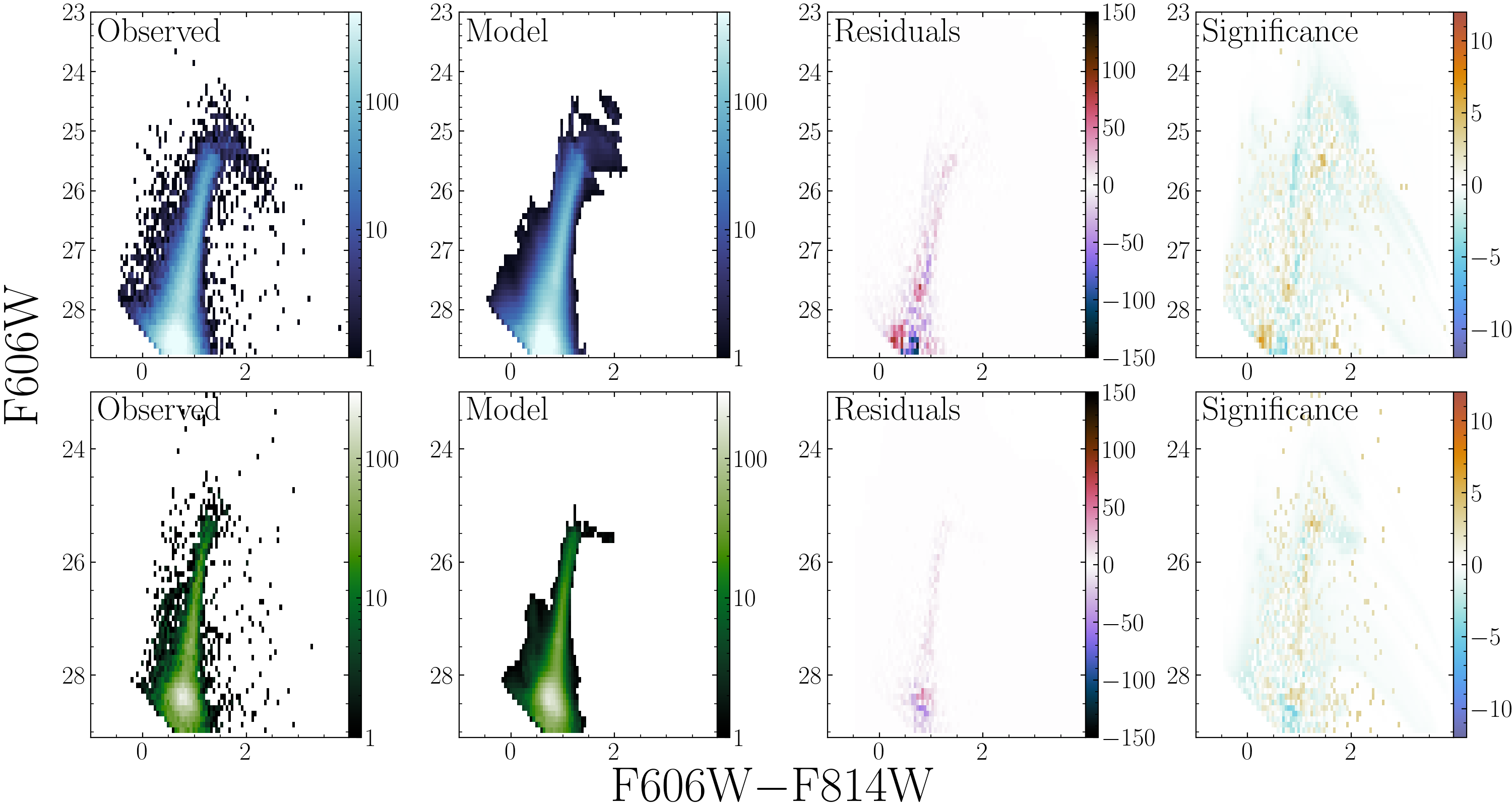}
\caption{Hess Diagram fitting results for our fiducial SFH in the deep WFC3 (top row, blue) and ACS (bottom row, green). The leftmost column shows the observed Hess Diagram, the second column shows the best-fit model, the third column shows the raw residuals (data--model), and the fourth column shows the residual ``significance'' (the sum of the significance diagram returns the overall goodness-of-fit value), which can be interpreted as the residual weighted by the number of stars in bin. Some structure is expected in the raw residuals in regions of high stellar density (such as the RC), but the weighted residuals are largely unstructured in both cases, indicating good fits to the data.\footnote{Note that as the models draw from continuous tracks, some features with very low model densities ($<$0.1 star per pixel) are visible in the significance plots, such as the faint AGB `waterfall' features extending to red colors.}}
\label{fig:match-cmds}
\end{figure*}

Fitting was performed on the portions of the CMDs lying above the gray completeness curve in Figure \ref{fig:deep-cmds}, which corresponds to 50\% completeness for our GST selection criteria (see also Table \ref{tab:obs}), up to 20$^{\rm th}$\,magnitude in each filter and from $-$0.5 to 4 in F606W$-$F814W color. We use the Padova stellar libraries \citep{girardi2000,cioni2006,marigo2008,girardi2010,girardi2016} to fit the deep CMDs of F8D1's stellar populations and calculate their SFHs. The updated Padova model suite (distinct from the related PARSEC suite; \citealt{bressan2012}) supports a broad range of metallicities, and includes model tracks for thermally-pulsating (TP)-AGB stars, of which F8D1 appears to have a substantial population \citep{caldwell1998}. We fit the SFH using 71 logarithmically-spaced time bins, with bin size 0.05, from $\log[{\rm Age\,(yr)}]\,{=}\,6.6{-}10.15$. We fit each field in \texttt{MATCH} assuming a \cite{kroupa2001} initial mass function, a binary fraction of 35\% (the default for \texttt{MATCH}), and a fixed distance modulus of 27.9 \citep{weisz2011,Anand2021}. Using our 71 time bins, we fit our observed CMDs with a grid of models with varying reddening. We assumed simple foreground reddening, as F8D1 does not host any detectable \textit{in situ} ISM. We allowed for a range of foreground reddening values, as the Galactic cirrus in the vicinity of the M81 Group is notably variable, from $A_V\,{=}\,0.2{-}0.5$\ in steps of 0.05\,mag. We find best-fit extinction values of $A_V\,{=}\,0.25$\,mag for the WFC3 field, and $A_V\,{=}\,0.45$\,mag for the ACS field. In \textsc{Appendix} \ref{sec:degen} we perform a example analysis of how reasonable uncertainties on distance and $A_V$\ impact SFH inference. The magnitude of these effects is small and well-captured by our treatment of systematic uncertainties (see \S\,\ref{sec:uncertainties}).

\begin{figure*}[t]
    \centering
    \includegraphics[width=\linewidth]{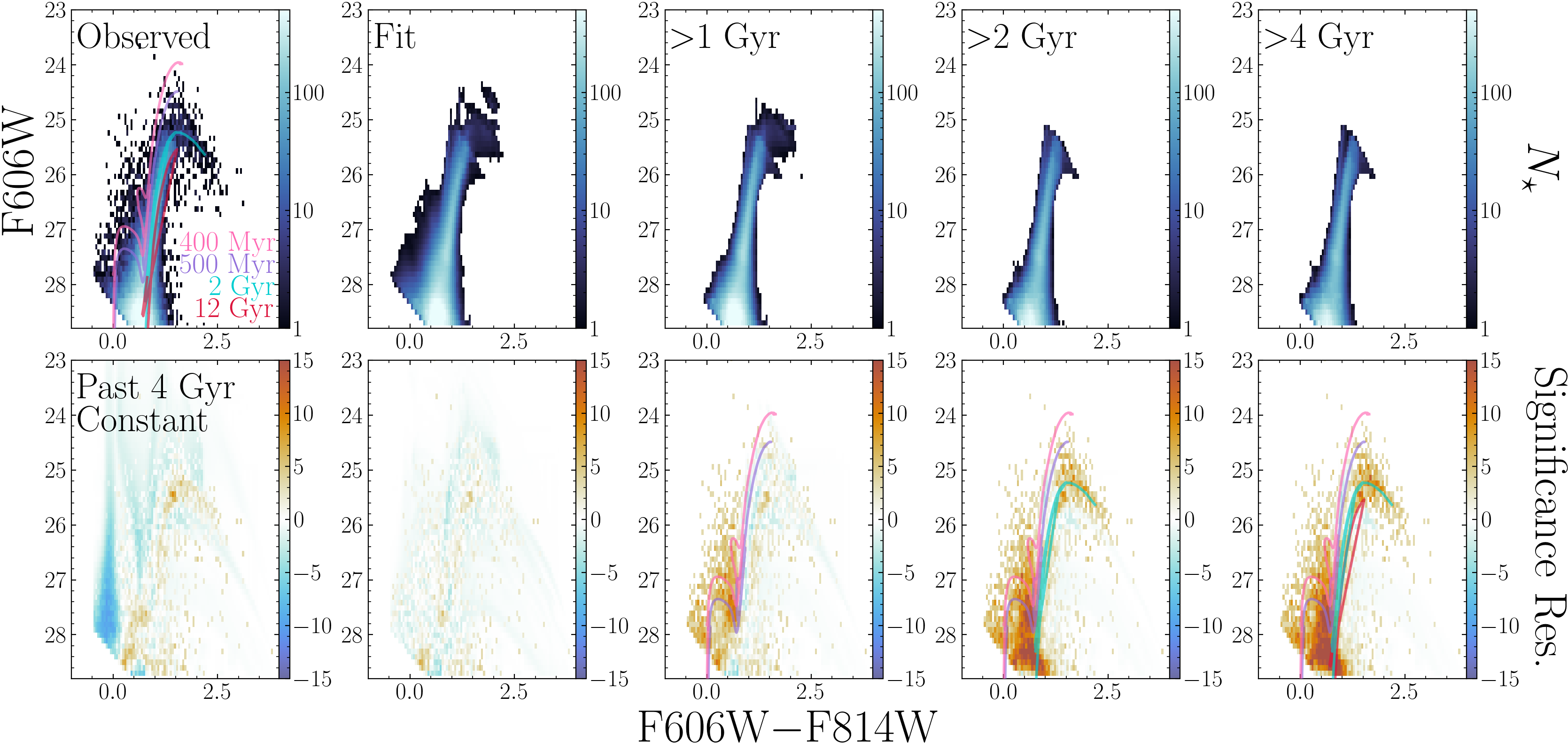}
    \caption{A deconstruction of the best-fit CMD model (second from left) of the central deep WFC3 field. From left to right, we remove progressively older populations from the best-fit model: first removing only the youngest populations ($<$1\,Gyr), then the populations formed at the peak of the 2\,Gyr burst, and finally the intermediate populations from 2--4\,Gyr, leaving only the oldest stars. The top row shows the Hess Diagram for each of these cases, while the bottom row shows the significance (weighted) residuals. We also show the observed CMD (left) for reference, with PARSEC isochrone tracks \citep{bressan2012} overlaid at 400\,Myr (pink; [M/H]\,=\,$-$0.75), 500\,Myr burst (purple; [M/H]\,=\,$-$0.5), 2\,Gyr burst (turquoise; [M/H]\,=\,$-$0.6), and 12\,Gyr (red; [M/H]\,=\,$-$0.85). Metallicities are taken from the SFH fit. The changes when including the 2--4\,Gyr population are subtle and only visible by scrutinizing the residuals. The inclusion of the 1--2\,Gyr population reproduces the remainder of the RC, the RGB, and the bright AGB populations. The remaining features at this point are all blueward of the RC and RGB. Including the $<$1\,Gyr bins reproduces these remaining populations very well. Comparing to the isochrones in the leftmost panel, there is clear evidence for strong contributions from the MSTO and Helium-burning tracks of stars 400-500\,Myr in age. In the bottom left panel, we also show the residuals for a fit with the same fraction of stellar mass formed over the past 4\,Gyr, but with a constant SFH. The residuals are notably worse for the entire CMD, highlighting the necessity of two bursts in explaining the observed CMD.}
    \label{fig:cmd-burst}
\end{figure*}

We show the resulting SFHs in Figure \ref{fig:sfhs}, and the Hess Diagram model fits in Figure \ref{fig:match-cmds}. There is some patterning in the raw fit residuals, as is expected for any instance of CMD fitting for regions of such high stellar density, however the ``significance'' map, which weights the residuals based on the number of stars in each bin, shows very few areas with significant residuals -- particularly around the RC. Notably, both fields show strong bursts of star formation around 2--2.5\,Gyr ago, with a sharp decline at 1--1.5\,Gyr. This burst is consistent with the inference of an intermediate-age population in the HST imaging of \cite{caldwell1998}, based on the presence of a large number of luminous AGB stars, and is broadly consistent with the recent SFH of F8D1 inferred by \cite{weisz2011} -- which also showed an enhancement around 1.5--2\,Gyr ago -- from these much shallower existing HST data. Both fields also show a second distinct and statistically-significant burst of star formation $\sim$500\,Myr ago. In the central field, these bursts represent enhancements of $\sim$4$\times$\ and $\sim$3$\times$\ the lifetime average SFR, respectively. The burst at 2\,Gyr in the outer field represented a 25$\times$\ enhancement over the average, while the 500\,Myr burst was similar to the central field at $\sim$3$\times$. The burst at 500\,Myr formed a stellar mass of approximately $8{\times}10^5\,M_{\odot}$\ within the central WFC3 field, of which ${\sim}5{\times}10^5\,M_{\odot}$\ remains. This burst coincides with the spectroscopic age estimate of F8D1's nuclear star cluster (480\,Myr; \citealt{forbes2024}), which has a stellar mass of ${\sim}3{\times}10^{4}\,M_{\odot}$\ \citep{caldwell1998}. 

In Figure \ref{fig:cmd-burst}, we deconstruct the model that best fits the central WFC3 deep field, showing how the different populations in the CMD are explained by stars of different ages. It is readily apparent that the observed morphology of the RC, AGB Bump, and bright AGB populations can only be explained by a substantial population $\sim$2\,Gyr and that faint blue populations can only be reproduced by MSTO and helium-burning stars with ages $\sim$500\,Myr. Further, we compare the past 4\,Gyr of star formation to a constant SFH forming the same stellar mass as the best-fit SFH from \texttt{MATCH}. The fit is significantly visually and statistically poorer than the recovered SFH, particularly around the RC and young MS -- indicating that specific enhancements are uniquely required at ages of $\sim$2\,Gyr and $\sim$500\,Myr to explain the observed CMD. We note that the brightest (youngest) subset of the red and blue helium-burning stars are visible in the original HST imaging of \cite{caldwell1998}. Our observations' gain of $\sim$2 magnitudes in depth has now confirmed that these stars indeed belong to a population formed in the past 500\,Myr.

In the following two sub-sections, we describe in greater detail our treatment of the age--metallicity degeneracy in the fits (\S\,\ref{sec:age-mh}), as well as the estimation of uncertainties on the fits (\S\,\ref{sec:uncertainties}). 

\begin{figure*}[t]
    \centering
    \includegraphics[width=0.9\linewidth]{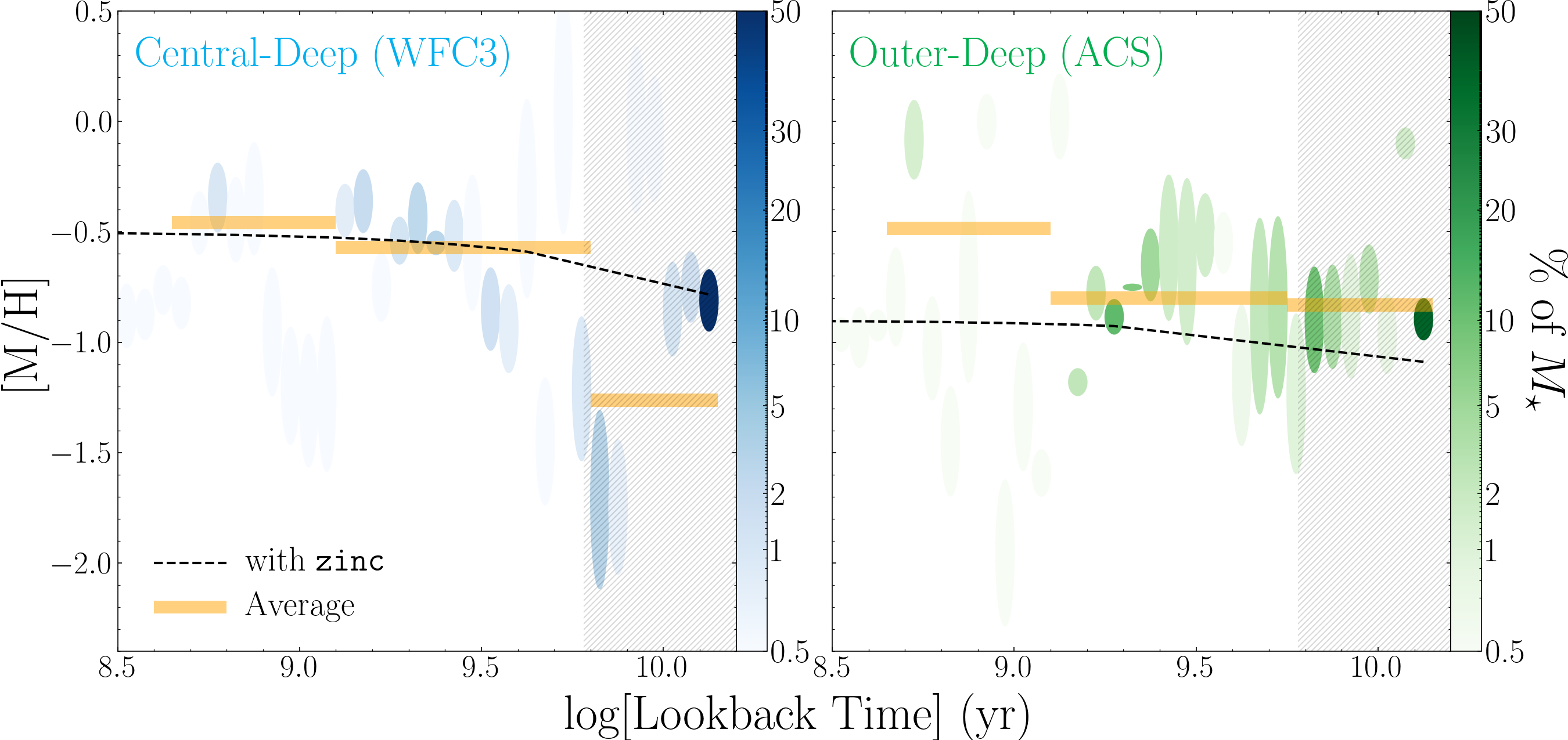}
    \caption{Recovered age--metallicity relations from MATCH for the two SFH fits to F8D1's stellar populations. The axes of each ellipse denote the bin width (0.05\,dex) and the combined uncertainty on [M/H] for star formation within the bin. Ellipses are color-mapped to the fraction of total stellar mass formed in each. We have chosen a colormap that gives less visual weight to bins with a smaller contribution to the stellar mass. The average metallicity in three broad age bins is shown in orange. This is meant to illustrate the overall trends in the metallicity of the stellar populations over time. We note that the metallicity estimates that MATCH returns should not be taken as precise metallicity measurements (as can be inferred from the large uncertainty envelopes), but are useful as a rough guide. We also show the output when assuming the age--metallicity relation to be monotonically increasing (set using MATCH's \texttt{zinc} argument). As in Figure \ref{fig:sfhs}, the most uncertain age range falls within the hatched region.}
    \label{fig:age-MH}
\end{figure*}

\subsubsection{Age--Metallicity Relation}
\label{sec:age-mh}

For our fiducial SFHs, we do not assume an age--metallicity relation, instead allowing the metallicity of stars formed in a given time bin to vary freely with age. This approach results in better fits to the observed CMD, but introduces more degeneracy in the age--metallicity--reddening parameter space, and can sometimes produce unphysical metallicity values or changes over a galaxy's lifetime. As the reddening in these fields is limited to the foreground cirrus, and it should be very common for the age--metallicity relation to be complex for all but the smallest galaxies, we forgo an assumption of an age--metallicity relation. We find that \texttt{MATCH} produces a reasonable age--metallicity relation in both fields, with an average change of $\lesssim$0.1\,dex between most adjacent time bins. For comparison, we also ran \texttt{MATCH} with a monotonically-increasing age--metallicity relation, as is often standard for dwarf galaxy SFHs \citep[e.g.,][]{weisz2014,savino2023,mcquinn2024b}, using the \texttt{zinc} flag (i.e. `\textit{Z} increasing'). In Figure \ref{fig:age-MH} we show the resulting age--metallicity relation for our fiducial run, including the average over three broad age bins, and compare to the results from the \texttt{zinc} run. For the non-\texttt{zinc} run, we allowed [M/H] between $-$2.3 and 0. For the \texttt{zinc} run, we allowed $-$2.3 to 0 as the minimum/maximum initial metallicity, and $-$1.8 to 0 as the minimum/maximum present-day metallicity. 

We see very good agreement in both fields between the average of the non-\texttt{zinc} and \texttt{zinc} runs, and observe an average metallicity increase of $\sim$0.15-0.3\,dex over from the oldest to youngest time bins. For the non-\texttt{zinc} runs, metallicity behavior appears fairly stable; bins with very low metallicities are typically insignificant to the overall fit, contributing $<$1\% of the total stellar mass. There is one region of note: between 5--8\,Gyr ago ($\sim$10$^{9.7}$--10$^{9.9}$\,yr), there is a visible drop in metallicity in the central region, and a slight drop in the outer field at the same ages. While the contribution to the total stellar mass formed in each field is negligible, which means there are few stars at that age present for constraining the metallicity, the consistent drop across fields, and across adjacent time bins within each field, could hint at a legitimate feature in the metallicity evolution of the galaxy.  However, we also note that this is at the limit in age (5--6\,Gyr) for which reasonable inferences of the SFH can be made (see \S\,\ref{sec:uncertainties}). A more accurate inference of F8D1's age--metallicity relation will require deeper imaging or independent metallicity information (e.g., spectroscopy). 

Overall, while allowing the metallicity in each time bin to be fit freely gives more room for physical variation in the metallicity evolution of the galaxy, the metallicity evolution for the significant star-forming bins reasonably reproduces a monotonically-increasing age--metallicity relation. Thus, the choice of an open or monotonically-increasing age--metallicity relation prior has no significant impact on the shape of F8D1's SFH. While the remaining degeneracy in age and metallicity should preclude interpreting these age--metallicity relations as precise inferences, it is at least as accurate as the commonly used technique of estimating stellar metallicities from the color of the RGB. We therefore report the average metallicities in each field, for our fiducial SFH fit. For the central WFC3 field, the lifetime average metallicity is [M/H]\,=\,$-$0.73, with an average of [M/H]\,=\,$-$0.53 for star formation in the past 3\,Gyr. The fit to the outer ACS field yields a slightly lower metallicity, with a lifetime average of [M/H]\,=\,$-$0.84, and an average of [M/H]\,=\,$-$0.80 for star formation in the past 3\,Gyr. We note that this implies a remarkably flat radial metallicity gradient in F8D1. 

\subsubsection{SFH Uncertainties}
\label{sec:uncertainties}

The calculation of uncertainties in the inference of SFH from resolved CMD fitting is not trivial. The procedure for calculating both systematic uncertainties and statistical uncertainties with \texttt{MATCH} are described in detail in \cite{dolphin2012} and \cite{dolphin2013}, respectively. We only provide a brief summary of these procedures here. 

Statistical uncertainties are calculated using the \texttt{hybridMC} routine in MATCH, based on the Hybrid (or Hamiltonian) Monte Carlo (HMC) algorithm \citep{duane1987}. \texttt{hybridMC} creates samples of SFHs, based on the fiducial fit, whose density is proportional to the probability density function. The 1$\sigma$\ spread around the weighted median is recorded. 

Systematic uncertainties are more insidious, dominated by intrinsic variation in the calculation of stellar properties in different stellar model grids. One of the most common approaches for use with \texttt{MATCH} is to introduce random shifts in stellar parameters that are designed to mimic differences in atmosphere models, evolutionary tracks, and bolometric corrections between large numbers of available model suites \citep[e.g.,][]{weisz2011,dolphin2012,weisz2014,skillman2017}. Other works take the empirical approach of comparing the results for several model suites directly \citep[e.g.,][]{williams2017,lazzarini2022,mcquinn2024b}. The random shift approach was calibrated against differences in multiple model suites, and has been shown to be fully consistent with, while often more conservative than, comparing against any two chosen evolutionary model suites \citep[e.g.,][]{skillman2017}. Given the presence of certain populations, such as the TP-AGB, that are not included in most evolutionary models, and our lack of ancient MSTO depth, we employ this method of randomly shifting stellar parameters to estimate our systematic uncertainties as conservatively as possible.

We adopt depth-dependent values for shifts in $\log(T_{\rm eff})$\ (Effective Temperature) and $M_{\rm bol}$\ (Bolometric Magnitude) following \cite{dolphin2012}. Systematic uncertainties were estimated by running 50 Monte Carlo SFH fits, with each realization applying a single shift in $\log\,{T_{\rm eff}}$\ and $M_{\rm bol}$\ for all stars, randomly drawn from normal distributions with $\sigma[\log(T_{\rm eff})]\,{=}\,0.022$\ and $\sigma[M_{\rm bol}]\,{=}\,0.27$. These shift values are estimated by interpolating the depth of our observations to the optimized values suggested in Table 3 of \cite{weisz2014}. We adopt the 68\% range of these realizations around the median as our systematic uncertainty. As a check on this systematic uncertainty estimate, we also fit the SFH with the MIST evolutionary tracks and found the difference between MIST and Padova to be completely consistent with the estimated systematic uncertainties using the Gaussian perturbation method.

The statistical and systematic uncertainties are combined at the end of these steps (using the MATCH task \texttt{zcmerge}) into a single set of upper and lower 1$\sigma$\ uncertainty bounds on the SFH at each timestep. As is apparent in the SFHs shown in Figure \ref{fig:sfhs}, while the uncertainties on the SFR in each time bin may be significant, the corresponding uncertainties on the cumulative SFH, accounting for the covariance between time bins, is much better constrained. At the depth of our data (and with the numbers of stars detected), both the instantaneous and cumulative SFHs are well measured to ages of approximately 5--6\,Gyr. At ages older than this, while difficult to fully account for in the systematic uncertainties, it has been well-documented that the age--metallicity degeneracy is essentially insurmountable without access to the oMSTO \citep[e.g.,][]{weisz2014}. Readers should take this limitation into account when interpreting the age--metallicity and SFH fits. 

We give results for the cumulative SFHs in Table \ref{tab:sfh}, including the calculated lookback times at which the most recent 50\% of the stellar mass formed. It is worth noting, as the percentiles given in Table \ref{tab:sfh} show, that the systematic uncertainties on the cumulative SFH should be interpreted as not just an uncertainty on star formation in a given time bin, but also uncertainty on the \textit{timing} of features in the SFH. $\tau_{90}$, for example, which is dominated by the enhancement at $\sim$2\,Gyr, has an asymmetric uncertainty of several Gyr at the oldest end in the central field. This should be interpreted as an uncertainty on the timing of the 2\,Gyr feature (see also \textsc{Appendix}\,\ref{sec:mist}), which is typical for SFHs inferred from CMDs of this depth \citep[e.g.,][]{williams2017}. The outer deep field, however, is much more tightly constrained to a $\tau_{90}\,{\sim}\,2$\,Gyr.

\subsection{Spatially Resolving the Quenching Time in F8D1 with Luminous AGB Stars}
\label{sec:agb-rgb}

\begin{figure*}[t]
    \centering
    \includegraphics[width=0.88\linewidth]{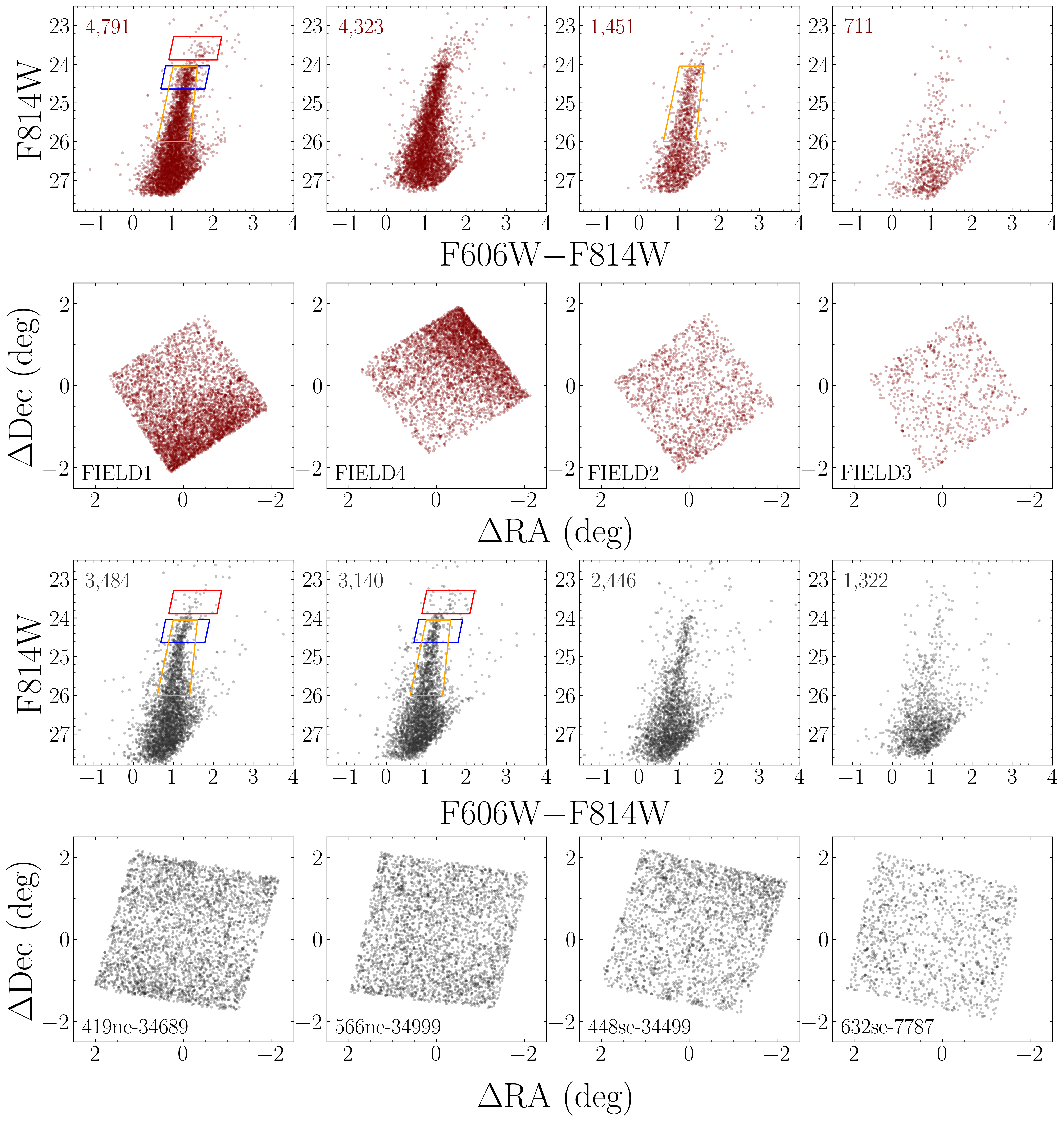}
    \caption{CMDs and maps of RGB-selected sources for our flanking fields. Following Figure \ref{fig:survey}, WFC3 fields are red and ACS fields are gray. Each camera is ordered by decreasing RGB density from left to right. We give the total number of stars in each CMD in the upper left corner. We show the selection regions used for AGB/RGB ratio as red and blue regions, each 0.5\,magnitudes tall. We also show the RGB selection region for measuring F8D1's density profile in orange, covering the brightest 2\,magnitudes of the RGB. The unreddened TRGB is at F814W\,$\simeq$\,24.0. Note that we do not calculate an AGB/RGB ratio for FIELD2, as it overlaps the ACS Deep field in radius. We do, however, use it in our density profile analysis.}
    \label{fig:outer-fields}
\end{figure*}

The photometry in the WFC3/ACS flanking fields is much shallower than the central deep fields, only reaching approximately 3\,magnitudes below the TRGB, and shy of the RC. As such, they do not provide enough information to reliably overcome the age--metallicity degeneracy and reconstruct their SFHs as far back as 6\,Gyr using \texttt{MATCH}. However, the populations of thermally-pulsating AGB (TP-AGB) stars brighter than the TRGB, as well as the most luminous populations of RGB stars, are both resolved at high-S/N. We show the CMDs of the flanking fields in Figure \ref{fig:outer-fields}. 

\cite{harmsen2023} derived an empirical relation between the ratio of AGB-to-upper RGB stars and the time at which 90\% of the stars formed in a population ($\tau_{90}$), using SFHs derived from deep CMDs of quiescent Local Group dwarf satellite galaxies. The efficacy of this relation depends on the assumption that star formation was halted $\gtrsim$1\,Gyr ago, and therefore that $\tau_{90}$\ corresponds to the approximate `shutdown' or `quenching' time of star formation \citep[e.g.,][]{weisz2015,skillman2017}. It is not appropriate for use on galaxies with star formation occurring within the past $\sim$1\,Gyr (noting that the $\tau_{90}$\ of a constant SFH is $\sim$1.3\,Gyr). \cite{harmsen2023} found the following best-fit relation between the logarithm of the ratio between the number of TP-AGB to RGB stars and $\tau_{90}$: 
\begin{equation}
\begin{split}
\tau_{90}\,{=}\, & 4.4{\pm}0.3\,{\rm Gyr}\,{-} \\
& 11.3{\pm}1.6 \big[\log_{10}(N_{\star,AGB}/N_{\star,TRGB})+1\big]. 
\end{split}
\end{equation} 
AGB and RGB stars are selected from the CMD in 0.5\,magnitude regions, identical to the \cite{harmsen2023} selections, which are shown in Figure \ref{fig:outer-fields}. The relation has an intrinsic scatter of 1.45$\pm$0.18\,Gyr, which in practice limits this technique's usefulness to measuring $\gtrsim$1.5\,Gyr variations in $\tau_{90}$. 

\begin{figure*}[t]
    \centering
    \includegraphics[width=0.7\linewidth]{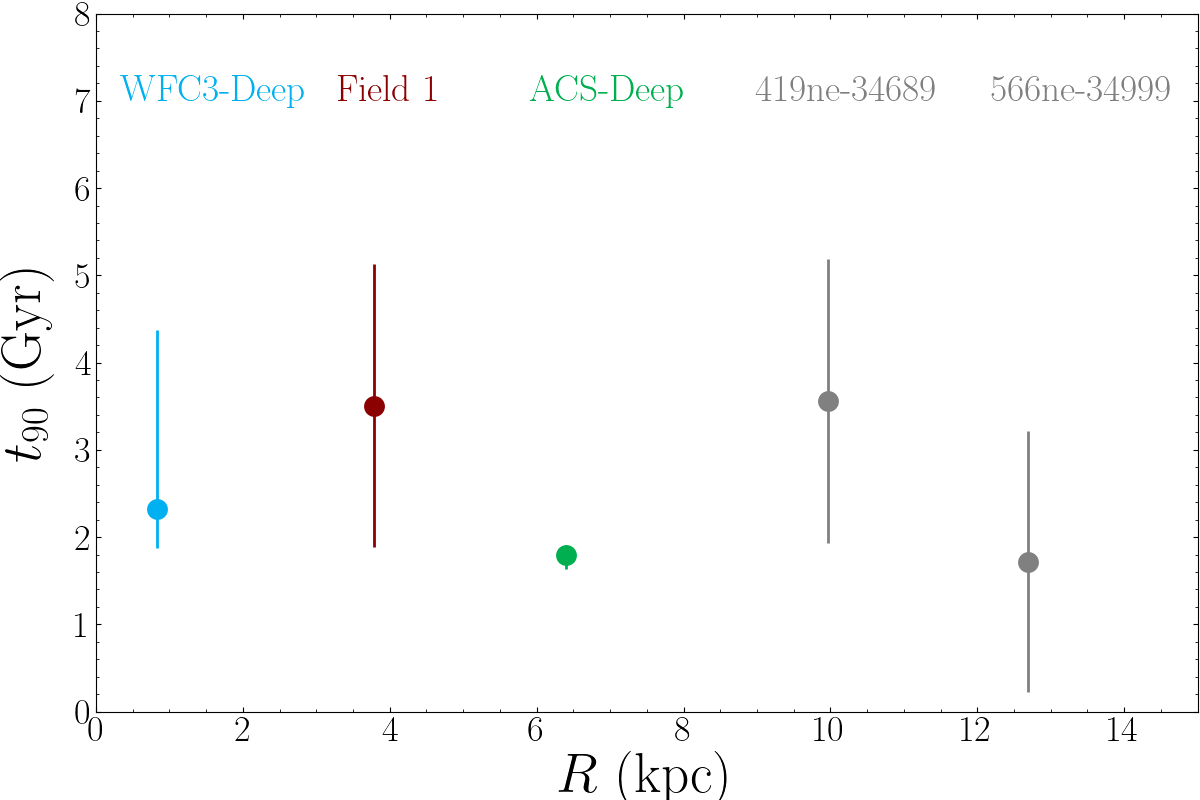}
    \caption{Radial profile of estimated $\tau_{90}$\ (i.e. `quenching time') in F8D1, in the direction of its tidal stream, combining our deep SFH analysis and inferences from the AGB-to-RGB ratio in several of the flanking fields. While individual points have substantial uncertainties, a universal quenching time within the past 3\,Gyr, for the entire galaxy, appears to be a good fit to the data.}
    \label{fig:t90-profile}
\end{figure*}

The placement of the flanking fields relative to the 2 deep fields allows us to use this technique to estimate the quenching time of F8D1's populations out to $>$10\,kpc -- well into the region visibly dominated by its stellar stream. Specifically, we use Field 2 ($R\,{=}$\,3.8\,kpc), Field 2-parallel (419ne-34689; $R\,{\simeq}$\,10\,kpc), and Field 4-parallel (566ne-34999; $R\,{\simeq}$\,12.7\,kpc), as they cover radii along F8D1's stream axis that are distinct from the more precisely-measured deep fields. We use a set of eight fields from the GHOSTS survey, identified by \cite{radburn-smith2011} as containing only foreground stars and background galaxies, to correct for contaminant sources in these sparser flanking fields. We find that both the AGB and RGB selection regions host, on average, 0.35 and 0.74 contaminants per ACS field (0.031 and 0.065 per arcmin$^2$) using identical photometric cuts, with a scatter of 0.2. We give the ratios with Poisson and contaminant-subtraction uncertainties, as well as the estimated $\tau_{90}$\ values with combined uncertainties, in Table \ref{tab:t90}.

Figure \ref{fig:t90-profile} shows the inferred $\tau_{90}$\ as a function of radius. The three fields show fairly consistent ratios of $\sim$0.12--0.17, corresponding to $\tau_{90}$s of 1.7--3.5\,Gyr. These are consistent within the uncertainties with $\tau_{90}$\ for the two deep fields of 2.3\,Gyr and 1.79\,Gyr (Table \ref{tab:sfh}), suggesting that F8D1 was globally star-forming $\sim$2\,Gyr ago, with the inner regions of the galaxy and the outer regions currently contained in the tidal stream quenching relatively close to one another. 

\newcolumntype{g}{!{\extracolsep{5pt}}c!{\extracolsep{0pt}}}

\begin{deluxetable}{ggg}[t]
\tablecaption{SFH Results\label{tab:sfh}}
\tablecolumns{3}
\tabletypesize{\small}
\setlength{\extrarowheight}{5pt}
\tablehead{
\colhead{Cum. SFH} & 
\colhead{$\tau_{\rm WFC3}$} & 
\colhead{$\tau_{\rm ACS}$} \vspace{-5pt}\\
%
\colhead{(\%)} & 
\colhead{(Gyr)} & 
\colhead{(Gyr)} \\
\cmidrule(l{0pt}r{0pt}){1-3} 
\colhead{$M^{\rm form}_{\star}\,{=}$} & 
\colhead{$5.19^{+0.58}_{-1.30}{\times}10^7\,{M_{\odot}}$} & 
\colhead{$5.82^{+0.46}_{-1.64}{\times}10^6\,{M_{\odot}}$} \\
\colhead{$\langle$[M/H]$\rangle_{\rm life}$\,=} & 
\colhead{$-$0.81} & 
\colhead{$-$0.84} \\
\colhead{$\langle$[M/H]$\rangle_{\rm 3\,Gyr}$\,=} & 
\colhead{$-$0.53} & 
\colhead{$-$0.80}
}
\startdata
50  & $>$6 & $>$6 \\
60  & $>$6 & $2.50^{+7.63}_{-0.28}$ \\
70  & $>$6 & $2.17^{+5.24}_{-0.22}$ \\
80  & $>$6 & $1.96^{+0.26}_{-0.10}$ \\
90  & $2.32^{+2.89}_{-0.42}$  & $1.81^{+0.09}_{-0.18}$ \\
100 & $0.13^{+0.00}_{-0.07}$  & $0.10^{+0.01}_{-0.04}$ \\
\vspace{-10pt}
\enddata
\tablecomments{Results from the best-fit \texttt{MATCH} SFH. We give the lookback times ($\tau$) at which fractions of the total stellar mass formed, from 50--100\% in steps of 10\%, for the WFC3 and ACS deep fields. We also give the total \textit{formed} stellar mass,$^{\rm a}$\ lifetime average metallicity, and average metallicity over the past 3\,Gry for the best fit in each field. Note that we estimate F8D1's total \textit{present-day} stellar mass to be $1.3{\times}10^{8}\,M_{\odot}$\ (see \S\,\ref{sec:profile}). Times are not given for fractions $<$50\%, as they are all at ancient times below the time sensitivity of our observations. A reminder that results for ages $>$6\,Gyr are highly uncertain. Uncertainties at younger ages are typical of SFHs inferred for Local Group dwarfs.\vspace{5pt}\\
$^{\rm a}$\texttt{MATCH} calculates formed stellar mass by integrating from 0--$\infty$, rather than the range of 0.1--100\,$M_{\odot}$\ assumed for a \cite{kroupa2001} IMF. Following \cite{telford2020}, we correct the masses reported by \texttt{MATCH} using a factor of 0.76 that brings these masses into agreement with a standard \cite{kroupa2001} IMF.}
\end{deluxetable}

\newcolumntype{k}{!{\extracolsep{8pt}}c!{\extracolsep{0pt}}}

\begin{deluxetable}{lkkk}[t]
\vspace{-38pt}
\tablecaption{AGB/RGB Ratio\label{tab:t90}}
\tablecolumns{4}
\tabletypesize{\small}
\setlength{\extrarowheight}{5pt}
\tablehead{
\colhead{} &
\colhead{Radius} & 
\multirow{2}{*}{$\rm log_{10}{\bigg(\dfrac{AGB}{RGB}\bigg)}$} & 
\colhead{$\tau_{\rm 90}$} \\[-5pt] 
\colhead{Field} & 
\colhead{(kpc)} &
\colhead{} & 
\colhead{(Gyr)}
}
\startdata
FIELD1 & 3.79 & $-$0.92\,$\pm$\,0.09 & 3.51\,$\pm$\,1.62 \\
419ne-34689 & 9.97 & $-$0.93\,$\pm$\,0.13 & 3.56\,$\pm$\,1.63 \\
566ne-34999 & 12.68 & $-$0.76\,$\pm$\,0.14 & 1.72\,$\pm$\,1.50 \\
\enddata
\tablecomments{AGB/RGB ratios and corresponding $\tau_{90}$\ for the three flanking fields used in this analysis (\S\,\ref{sec:agb-rgb}).}
\end{deluxetable}

\subsection{The Structure of F8D1 in Resolved Stars}
\label{sec:profile}

Finally, we combined the deep fields and shallow flanking fields to investigate the radial structure of F8D1 and its stream. We selected stars from the top 2\,magnitudes of the RGB (orange selection in Figure \ref{fig:outer-fields}). To increase the number of data points for the radial profiles, we divided each HST field exactly in half. Uncertainties on the surface density of RGB stars in the flanking fields are the result of the quadrature sum of Poisson error and the uncertainty in the TRGB of each field due to differential extinction, at the level of $\pm$0.07\,mag in E($B{-}V$). As in \S\,\ref{sec:agb-rgb}, for the AGB/RGB analysis, we use the `empty' fields identified by \cite{radburn-smith2011} to characterize foreground and background contamination. We find an average of 4.0 contaminants per ACS field (0.36 per arcmin$^2$), using our photometric criteria and deeper RGB selection window, with a scatter of 0.9. We subtract this source density from all fields used in the density analysis. 

We convert RGB density to stellar mass density by calibrating to the SFH fits for the two deep fields (\S\,\ref{sec:sfh}; Table \ref{tab:sfh}) -- by taking the ratio of stellar mass to number of stars in the top 2\,magnitudes of the RGB (where completeness is 100\%). It is important to note that \texttt{MATCH} calculates the \textit{formed} stellar mass (which we correct to the equivalent for a Kroupa IMF; see Table \ref{tab:sfh} and \citealt{telford2020}) and does not account for the mass lost by stellar death and evolved star winds. We correspondingly assume a 40\% mass loss to present-day, consistent with a \cite{kroupa2001} IMF. Due to the difference in SFH between the two fields and our ignorance of the SFH of the more distant stream populations, we average the two fields to obtain a single conversion value of 5.23$\times$10$^3\,M_{\odot}$\ per RGB star within our selection region. The uncertainty on this conversion factor is $\sim$17\% and includes in quadrature the difference in stellar mass per RGB star between the two fields, the uncertainty on the total stellar mass formed in each field, and the Poisson error on the total number of stars in each of the two calibration fields. The total uncertainty on the stellar mass surface density combines the uncertainty on the RGB surface density and stellar mass conversion. 

\cite{zemaitis2023} used star counts from Subaru HSC to detect and measure the profile of F8D1's stream from $\sim$9--60\,kpc. We combine these measurements with our HST measurements to provide a complete view of the galaxy from its center to 60\,kpc along its stream, including a more accurate view of the inner structure. One of the \cite{zemaitis2023} data points overlaps with our HST field at 9\,kpc. We scale this data point to match the value obtained for our HST data: their value of $\sim$18 RGB stars\,arcmin$^{-2}$\ at 9\,kpc corresponds to an estimated stellar mass density of 3.54$\times$10$^5\,M_{\odot}$\,kpc$^{-2}$\ at the same radius in our HST field. We scale the remaining data points from \cite{zemaitis2023} by the ratio of these two values. 

\begin{figure*}[t]
    \centering
    \includegraphics[width=0.85\linewidth]{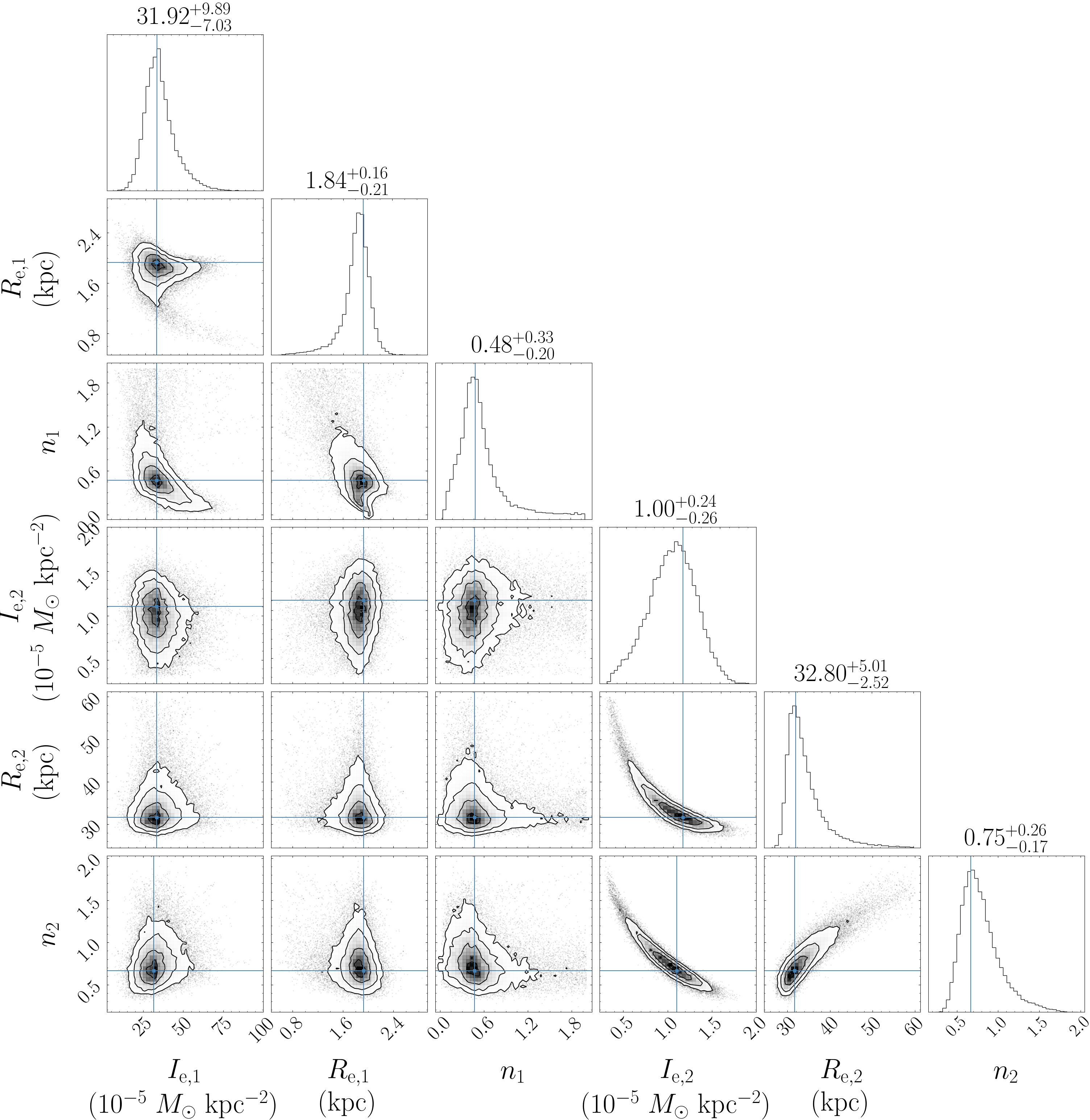}
    \caption{Corner plot showing the 1-dimensional and joint posterior probability distributions for the parameters in our two-component S\'{e}rsic model. See \S\,\ref{sec:profile} for details. Best-fit values are shown at the top of each column, corresponding to the parameter along the $x$-axis at the base of the column.}
    \label{fig:corner}
\end{figure*}

\begin{figure*}[t]
    \centering
    \includegraphics[width=0.92\linewidth]{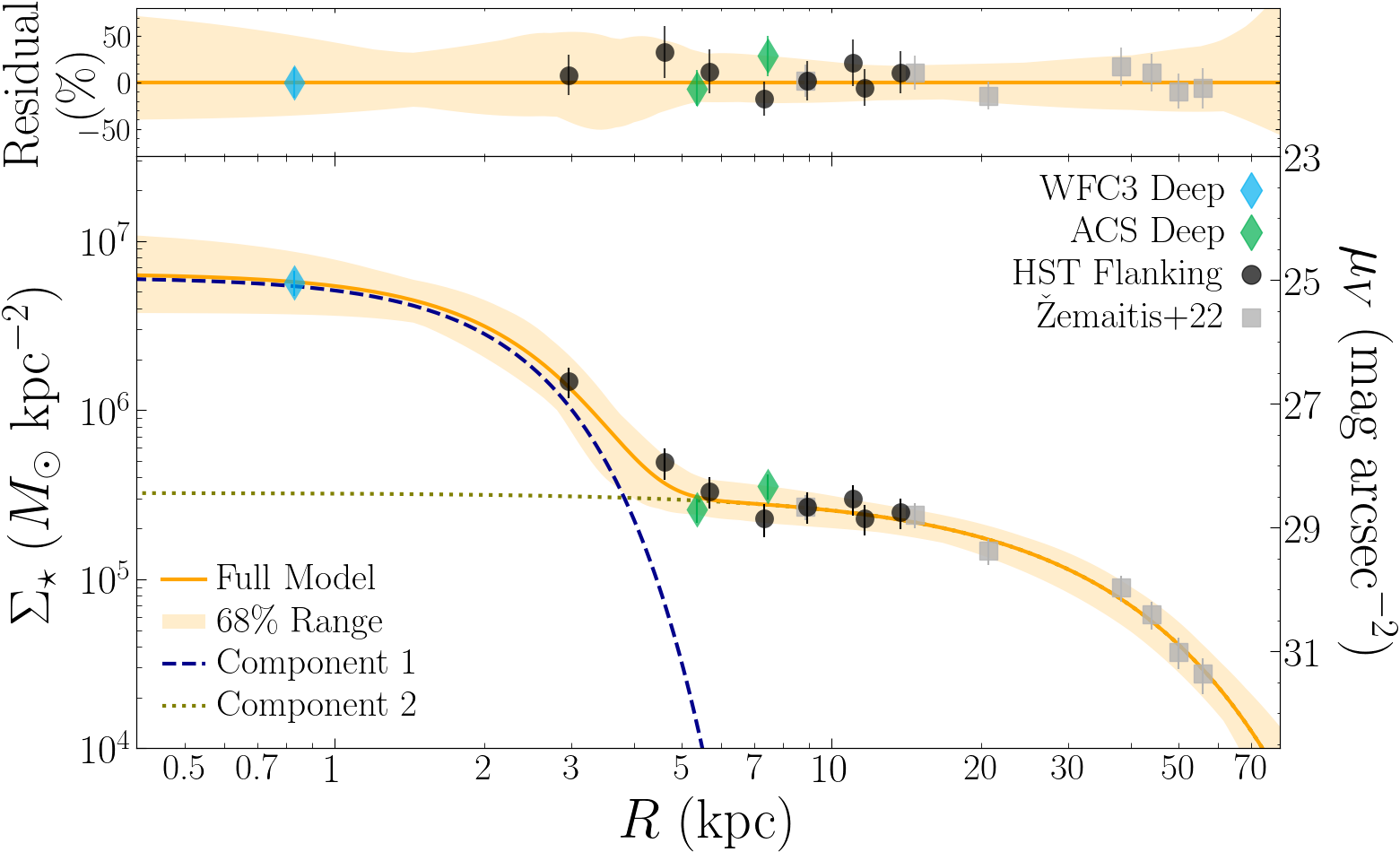}
    \caption{Radial stellar surface mass density, ($\Sigma_{\star}$) profile of F8D1, calculated as described in \S\,\ref{sec:profile}. Equivalent $V$-band surface brightness ($\mu_V$) is shown on the right axis, assuming $M/L_V\,{=}\,3.0$\ (see \S\,\ref{sec:mlv}). The central and outer deep fields are shown as diamonds, colored blue and green, respectively. Only the northern half of the central WFC3 field is used, as we do not have extended radial coverage in the south at any other radii. All other HST fields are split into two equal-area measurements. The scaled measurements from \cite{zemaitis2023} are shown as gray squares. We show the most probable double-S\'{e}rsic model from our MCMC analsis in orange, with the 68\% range shown as the light orange envelope. Component 1 (representing F8D1's remnant spheroid) is shown as a dashed dark blue curve, while Component 2 (representing the tidal stream) is shown as a dotted dark green curve. In the topmost panel, we show the fractional residuals of each measurement relative to the error envelope from the best model. The data are very well fit, within the uncertainties, by the double-S\'{e}rsic model.}
    \label{fig:density-profile}
\end{figure*}

In the remainder of this section, we analyze the radial stellar mass surface density of F8D1 along the northern stream direction. Only northern half of the inner deep WFC3 field is considered as well as the remaining field halves from the ACS deep field and 4 WFC3+ACS flanking fields, and the 7 scaled points from \cite{zemaitis2023}. The profile appears to be well described by two components: 1) an inner component, which dominates at radii $<$5\,kpc; and 2) an outer component, which results in a dramatic flattening of the density profile and dominates at radii $>$5\,kpc. To physically describe this profile, we apply a two component S\'{e}rsic fit \citep{sersic1968} using the Python Markov Chain Monte Carlo (MCMC) fitting module \texttt{emcee}. We allow each S\'{e}rsic component to vary freely, parameterized using the half-light radius, $R_e$, as
\begin{equation}
    I(r) = I_e \exp\Bigg\{-b_n\Bigg[\bigg(\frac{R}{R_e}\bigg)^{1/n} - 1\Bigg]\Bigg\}, 
\end{equation}
where $n$\ is the S\'{e}rsic index and $b_n$\ is the inverse of the regularized lower half incomplete gamma function, such that the incomplete gamma function ($\gamma$) and the complete gamma function ($\Gamma$) are related by $\Gamma(2n)\,{=}\,2\gamma(2n; b_n)$.\footnote{This is accomplished using the SciPy special function \texttt{gammaincinv}.}. We adopt the log-likelihood function of \cite{hogg2010} for a linear model fit to data: 
\begin{equation}
    \ln \mathcal{L} = -\frac{1}{2}\sum_{i=1}^n \Bigg[\frac{1}{\sigma_i^2}(x_i-x^\prime_i)^2 + \ln \sigma_{x,i}^2\Bigg], 
\end{equation}
where $x_i$\ is the data, $x_i^\prime$\ is the model, and $\sigma_{x,i}$\ is the uncertainty on each point. 5,000 realizations, with a discarded burn-in set of 200, was more than sufficient to obtain stability in the posterior probability density functions for each of the 6 parameters, which are shown in Figure \ref{fig:corner}. We show the radial profile, along with the best-fit components in Figure \ref{fig:density-profile}. 

\newcolumntype{f}{!{\extracolsep{62pt}}l!{\extracolsep{0pt}}}
\newcolumntype{k}{!{\extracolsep{62pt}}r!{\extracolsep{0pt}}}

\begin{deluxetable}{fk}
\tablecaption{\textnormal{Global Properties of F8D1}\label{tab:prop}}
\tablecolumns{2}
\setlength{\extrarowheight}{5pt}
\tabletypesize{\small}
\tablehead{%
\colhead{Parameter} &
\colhead{\hspace{85pt}Value} 
}
\startdata
$I_{\rm e,1}$ & $31.92^{+9.89}_{-7.03}{\times}10^{5}\,M_{\odot}\,{\rm kpc^{-2}}$ \\
$R_{\rm e,1}$ & $1.84^{+0.16}_{-0.21}$\ kpc \\
$n_{\rm 1}$ & $0.48^{+0.33}_{-0.20}$ \\
$M_{\rm \star,1}$ & $9.67^{+1.78}_{-1.99}{\times}10^7\,M_{\odot}$ \\
$I_{\rm e,2}$ & $1.00^{+0.24}_{-0.26}{\times}10^{5}\,M_{\odot}\,{\rm kpc^{-2}}$ \\
$R_{\rm e,2}$ & $32.80^{+5.01}_{-2.52}$\ kpc \\
$n_{\rm 2}$ & $0.75^{+0.26}_{-0.17}$ \\
$M_{\rm \star,2}^{\rm north}$ & $3.46^{+0.44}_{-0.25}{\times}10^7\,M_{\odot}$ \\
$M_{\rm \star,total}$ & $1.31^{+0.39}_{-0.19}{\times}10^8\,M_{\odot}$ \\
\cmidrule(l{0pt}r{0pt}){1-2} 
\multicolumn{2}{c}{Assuming $M/L_V\,{=}\,3.0$} \\
\cmidrule(l{0pt}r{0pt}){1-2}
$M_{V,{\rm inner}}$ & ${-}14.0$ \\
$M_{V,{\rm total}}$ & ${-}14.3$ \\
$\mu_{V,0}$ & 25.6 \\
$\mu_{V,{\rm e}}$ & 26.3 \\
$\langle{\mu_{V}\rangle}_{\rm e}$ & 25.9 \\
\vspace{-10pt}
\enddata
\tablecomments{Derivation of fit parameters is described in \S\,\ref{sec:profile}. Components 1 and 2 are taken to be F8D1's bound central spheroid and its tidal stream, respectively. Stellar masses include a typical lifetime mass loss of 40\% for an evolved population. For quick comparison to the observed properties of other nearby dwarf galaxies, we also estimate total magnitudes and surface brightnesses in the $V$-band, assuming a mass-to-light ratio $M/L_V\,{=}\,3.0$, calculated as a weighted average of the integrated $M/L_V$\ from the SFHs of the two deep fields (see \S\,\ref{sec:mlv}).} 
\end{deluxetable}

We integrate the two components to obtain estimates of the stellar mass in each. We assume the central component of F8D1 to be spherical. In this case, the second component, which we attribute to the stream, can be modeled as an ellipse with a semi-major axis in the stream direction and a semi-minor axis (i.e. width) equivalent to the effective radius of the inner component, where the two components meet. This gives us an approximate axis ratio for the stream of $b/a = R_{\rm e,1}/R_{\rm e,2}$, which we use to integrate the radial profile of the stream as an ellipse. We estimate  uncertainties on the stellar mass of each component by integrating all 5,000 of the MCMC realizations and taking the corresponding 16--84\% range. We estimate total stellar masses of $9.67^{+1.78}_{-1.99}{\times}10^7\,M_{\odot}$\ for the main galaxy and $3.46^{+0.44}_{-0.25}{\times}10^7\,M_{\odot}$\ for the northern stream. We combine the two components to estimate a total \textit{progenitor} stellar mass for F8D1 of $1.31^{+0.39}_{-0.19}{\times}10^8\,M_{\odot}$. This includes an additional uncertainty to the upper bound on F8D1's total stellar mass, reflecting our complete uncertainty about the presence of a southern stellar stream component. Whether or not we expect a Southern component to the stream is entirely dependent on its orbital history (see \S\,\ref{sec:tides} for a more in-depth discussion); both are possible, for example if F8D1 is disrupting along a southwest path versus near a current apocenter, where the visible stream may comprise multiple distance components along the line-of-sight. We therefore add a 100\% uncertainty (equal mass in the north and south) on the total mass of the stream component. 

We give the parameters for the best-fit S\'{e}rsic models, as well as total stellar mass estimates in Table \ref{tab:prop}. We also investigated whether the stars produced in the most recent 500\,Myr burst differ in their spatial distribution relative to the RGB stars. We found no spatially-coherent structure in the distributions of either sources blue-ward of the RGB, or the TP-AGB population. In both cases, sources are distributed essentially evenly across both fields.

\subsubsection{Comparison to Integrated Light}
\label{sec:mlv}

As a check on our density profile and stellar mass estimation, we convert our stellar mass density ($\Sigma_{\star}$) profile to $V$-band surface brightness ($\mu_V$) and compare against the profile measured by \cite{caldwell1998}. Converting our SFH-derived stellar mass results into equivalent surface brightness in the traditional $V$-band (Table \ref{tab:prop}) requires an estimate of the $V$-band mass-to-light ratio, $M/L_V$. Among the satellites of the MW in the mass range $M_{\star}\,{=}\,10^7{-}10^9\,M_{\odot}$, there is an enormous range in $M/L_V$\ -- from $<$1 to $\gg$10 \citep[e.g.,][]{Mateo1998}. For F8D1, we use the SFHs calculated in \S\,\ref{sec:sfh} to infer $M/L_V$\ directly from the models. Taking a grid of PARSEC isochrones (using the integrated magnitudes output) in the standard UBVRI filter set, with log-age from 8.3--10.1 and [M/H] from $-$1.7 to $-$0.2, we calculated the weighted total luminosity in the $V$-band contributed by each of our SFH bins, in each of the two fields. We estimate $M/L_V\,{=}\,3.1$\ and 1.8 for the WFC3 and ACS fields, respectively. As we do not have detailed SFH information for the outer fields beyond 7\,kpc, other than the AGB/RGB-based inference that the populations are consistent with the F8D1's main body (\S\,\ref{sec:agb-rgb}), we take the mass-weighted average of these two estimates for an mean $M/L_V\,{=}\,3.0$. This is similar to the $M/L_V$\ for populations in other dwarf galaxies hosting intermediate-age populations, including the Sculptor dSph \citep{Mateo1998}.

\begin{figure}[t]
    \centering
    \includegraphics[width=\linewidth]{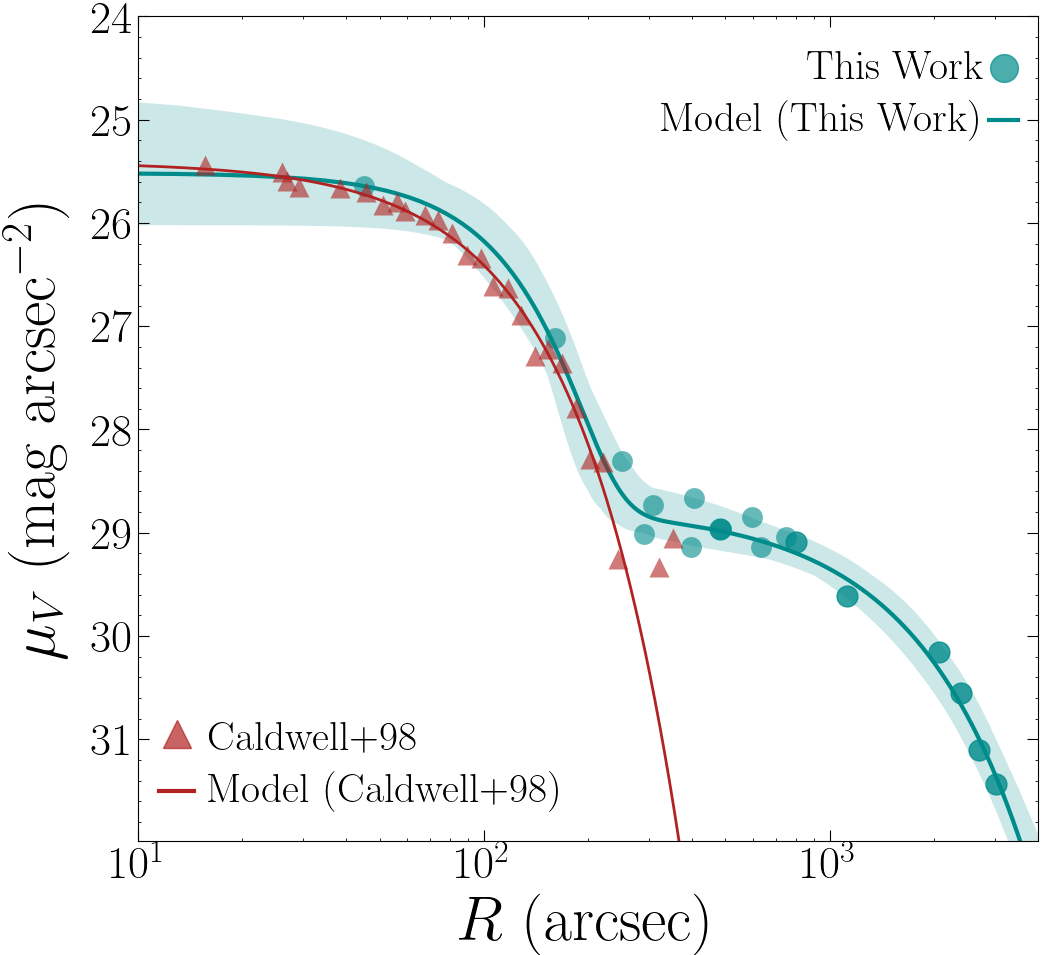}
    \caption{A comparison of the $V$-band surface brightness profile measured by \cite{caldwell1998} and our converted stellar mass surface density profile, assuming $M/L_V\,{=}\,3.0$\ -- obtained by integrating over the SFHs calculated for F8D1.}
    \label{fig:vprof}
\end{figure}

We compare our converted $\mu_V$\ profile to the profile of \cite{caldwell1998} in Figure \ref{fig:vprof}. In general, we find excellent agreement between the two profiles, as has been the case in other works comparing star counts and surface brightness \citep[e.g.,][]{harmsen2017,smercina2020}. The \cite{caldwell1998} profile fully overlaps our two innermost data points and lies almost entirely inside the error envelope from our MCMC analysis. We infer a slightly lower central surface brightness: 25.7 vs.\ 25.4, and slightly lower absolute magnitude of the inner component: $-$14.0 vs.\ $-$14.25. These small differences are likely due to the different parameters of our S\'{e}rsic profile -- we find a lower index of the central component, 0.43 vs. 1.45. We note that, while \cite{caldwell1998} did not explicitly infer the existence of an extended stream structure, their measurements include two points beyond 200\arcsec\ where the profile seems to flatten, that their S\'{e}rsic fit does not run through. This is remarkably consistent with the behavior of our extended stream profile, suggesting that this was likely the first evidence of the existence of this structure, seen in integrated light. These two points lie a bit further below our profile than the rest; this could be due to the difficulty of background-subtraction at such low surface brightnesses. However, the \cite{caldwell1998} profile is azimuthally-averaged, suggesting that this could also hint at asymmetry in the presence of the stream from one side of the galaxy to the other.

\section{Reflections on the evolution of F8D1 and UDGs}
\label{sec:discussion}

In summary, our results indicate that F8D1 experienced multiple episodes of enhanced star formation: the first $\sim$2\,Gyr ago, followed by a period of sharp decline (i.e. initial `quenching'); the second, much smaller burst occurred only $\sim$500\,Myr ago, representing a brief period of rejuventation that was followed by complete cessation of star formation to the present day (\S\,\ref{sec:sfh}). Stars formed in the burst 1--3\,Gyrs ago represent a much higher fraction of the current stellar mass in F8D1's outskirts ($\sim$39\%), relative to its center ($\sim$11\%). This indicates that F8D1 hosts an age gradient, rather than a purely phase-mixed population. Quenching appears to have occurred on similar timescales for the entire galaxy, including its tidal stream (\S\,\ref{sec:agb-rgb}). Updated fits to F8D1's global structure indicate that it is a reasonably massive satellite, with $M_{\star}\,{\simeq}\,10^{8}\,M_{\odot}$, and a large fraction of that mass is currently contained in its stream. 

In this discussion, we reflect on F8D1's evolutionary history and future as a tidally disrupting galaxy (\S\,\ref{sec:tides}), how it can help us understand the evolution of UDGs (\S\,\ref{sec:udgs}), and how consideration of comparable progenitors illuminates the physical processes driving galaxy evolution at this mass scale (\S\,\ref{sec:prog}). 


\subsection{F8D1 as a Product of Tidal Disruption}
\label{sec:tides}

The most obvious feature of F8D1 is its ongoing tidal disruption \citep{zemaitis2023}. The HST-based measurements of F8D1's SFH and global structure we have presented in this paper place new constraints on F8D1's orbital history and state of tidal disruption. 

Both the deep central WFC3 field and the deep parallel ACS field show two significant recent bursts of star formation 2\,Gyr and 500\,Myr ago. Because our structural fits to F8D1 show that the ACS deep parallel field is dominated by F8D1's stream, we can say that at least the inner part of F8D1's tidal stream is $<$500\,Myr old, implying that its last pericenter from its main disturber was no more than 500\,Myr ago. From our structural fits, the stellar mass of the central spheroidal component, which we take to signify the main body of F8D1, is $9.67^{+1.78}_{-1.99}{\times}10^8\ M_{\odot}$. Our shallow HST fields and the profiles from \citet{zemaitis2023} cover only the Northern part of F8D1's debris field; that Northern part of the stream contains $0.35^{+0.44}_{-0.25}\times10^8 M_{\odot}$ in stars. In the case that the Northern stream is the only stream (i.e., if F8D1 is projected close to the stream apocenter), then the stream is 26\% of the total system mass. If, instead, we assume that there is a symmetric tail to the South that neither the Subaru nor HST data have imaged, the stream is 42\% of the total system mass. 

\subsubsection{What is Responsible for F8D1's Tidal Disruption?}
\label{sec:orbit}

Despite these new insights, the origin of F8D1’s tidal disruption remains uncertain. Its position in the M81 Group means that, without a bespoke modeling effort, its tidal features could equally likely result from interaction with M81, NGC 2976, or a combination of both. To investigate the tidal influence from each system, we first calculate the tidal index \citep[see e.g.,][]{Karachentsev2014,Pearson2016}, ignoring velocity and line-of-sight separation, and taking into account only the stellar mass of M81 and NGC 2976 respectively ($M_{\star,M81} = 6.0{\times}10^{10}\,M_{\odot}$ and $M_{\star,\rm NGC~2976}\,{=}\,1.9{\times}10^9\,M_{\odot}$), and their projected separation from F8D1 ($R_{\rm proj, M81}\,{=}\,120$\,kpc and $R_{\rm proj, NGC~2976}\,{=}\,33$\,kpc). We find that each system yields comparable and large tidal influence on F8D1, where $\Theta_{\rm M81}\,{=}\,2.54$ and $\Theta_{\rm NGC~2976}\,{=}\,2.72$. 
Note that the line-of-sight velocity separation between F8D1 and M81 is a factor of 3 smaller than between F8D1 and NGC 2976, $-$103\,km\,s$^{-1}$\ \citep{forbes2024} vs.\ $-$34\,km\,s$^{-1}$, but that the potential line-of-sight separation between F8D1 (which is assumed to be at $D_{\rm los}\,{=}\,3.8$\,Mpc in this paper) and M81 (at $D_{\rm los}\,{=}\,3.6$\,Mpc; \citealt{radburn-smith2011}) is greater than the projected separation between F8D1 and NGC 2976. This distance difference is the primary uncertainty in establishing F8D1's orbital history. Establishing distance moduli to galaxies in the M81 Group to better than $\sim$0.2\,mag is particularly difficult due to the substantially variable Galactic cirrus in this region. See \cite{smercina2020} and \cite{zemaitis2023} for deep, wide-field images of the M81 Group that show the variable cirrus throughout this region. 

If the most recent bursts of star formation, around 500 Myr ago, was triggered by pericentric passages that helped produce the observed stream from F8D1, we can estimate the orbital velocity required for F8D1 to have passed near M81 or NGC 2976 at that time. To have passed close to M81 500\,Myr ago, which is at least 120\,kpc away (projected distance), F8D1 would need to move tangentially with a velocity of at least $\sim$235\,km\,s$^{-1}$. To have passed close to NGC 2976 500\,Myr ago, which is only 33\,kpc away (projected distance), F8D1 would need to move tangentially with a velocity of at least$\sim$65\,km\,s$^{-1}$. The line-of-sight velocity of F8D1 is ${-}103{\pm}36$\,km\,s$^{-1}$\ \citep{forbes2024}, while M81 and NGC 2976 have $v_r\,{=}\,{-}34$\,km\,s$^{-1}$\ and $v_r\,{=}\,3$\,km\,s$^{-1}$, respectively \citep{deVaucouleurs1991}. Thus, the total velocity of F8D1 with respect to M81 and NGC 2976 would be $\sim$\,$-$245\,km\,s$^{-1}$\ and $\sim$\,$-$124\,km\,s$^{-1}$, respectively. 

In stellar mass, M81 is a MW analog and NGC 2976 is similar to the LMC. Assuming that their dark matter halo masses also follow estimates of the MW and LMC halo masses, this means that at distance of 120\,kpc, F8D1 is likely bound to M81, and at a distance of 33\,kpc F8D1 can also be bound to NGC 2976 --- similar to the MW/LMC/SMC system. Here we ignore the line-of-sight distances as the errors on these are large. Thus, it seems likely that F8D1 could be bound to either host and that its stream could be produced from an interaction with either system. 

To explore whether the earlier burst around 2\,Gyr ago could correspond to a previous pericenter, it is helpful to compare to orbits of known systems. For example, the Sagittarius dwarf in the Milky Way shows how orbital phases can be connected to repeated star formation and stripping events. However, reaching a galactocentric radius of at least 120\,kpc in 500\,Myr, preceded by a pericenter pass 2\,Gyr ago, might be difficult to explain for a Sagittarius-like dwarf galaxy system around a MW-like host \citep[see orbital decays in][and Bell et al, in prep.]{laporte2018,hunt2021,vasiliev2021}. If an interaction with NGC 2976 is responsible for the formation of F8D1's tidal stream and its two most recent bursts in star formation, the scenario resembles a dwarf-dwarf interaction, similar to the LMC--SMC or NGC 4490--4485 systems \citep[][]{Besla2012,Pearson2018}. The orbits of these dwarf-pair analogs and orbits of dwarf pairs analyzed from TNG100 in \citet{chamberlain2024b} show that a pericenter of a few $\times$100\,Myr, preceded by a pericenter $\sim$2\,Gyr ago, is plausible. It is also possible that the stream is formed due to a combination of `pre-processing' due to gravitational interaction with NGC 2976 and tidal forces from M81. Future detailed modeling is necessary to determine which explanation is more likely, and to better constrain the nature of the interaction.  

\subsubsection{Comparison With Other Nearby Disrupting Satellites}
\label{sec:sag-compare}

Here we compare F8D1's current state of disruption against other known examples of disrupting satellites. Two galaxies in nearby groups are at somewhat earlier stages in their disruption, but already show important effects from tides on their main bodies. NGC 147 is a satellite in the M31 group with $M_{\star}\,{\sim}\,6{\times}10^8\,M_{\odot}$\ \citep{Geha2010} with tidal arms that may contain $\sim10$\% of its total mass \citep{Arias2016}. NGC 147's main body exhibits isophote twists, showing that its main body is significantly affected by tides \citep{Crnojevic2014}. NGC 3077, a roughly LMC-mass member of the M81 group, is at a similar stage in its disruption as NGC 147, with ${\sim}$13\% of its stellar mass in its tidal envelope \citep{smercina2020} and a similar $S$-shaped morphology \citep[e.g.,][]{Okamoto2023}. In NGC 3077's case,  almost all of its \textsc{H\,i} gas has already been stripped from it in its interaction with M81 \citep{Walter2002,deBlok2018}, and the remaining gas is being rapidly consumed in a modest starburst with $t_{\rm dep}\,{\sim}\,10$\,Myr \citep{Meier2001}. 

In contrast, the Sagittarius dwarf spheroidal (Sgr) and its stream appear to be a similar system to F8D1, just at a much later stage in its disruption. We note that while Sgr is the closest example, it several satellites in nearby galaxy groups have been identified that closely resemble Sagittarius in their bulk properties and state of disruption, such as Scl-MM-Dw2 around NGC 253 \citep{toloba2016b}, and CenA-MM-Dw3 around Centaurus A \citep{crnojevic2016}. While the total mass, half-light radius, and metallicity of Sgr are all very similar to F8D1 (see \citealt{mcconnachie2012} and Figure \ref{fig:udgs}), there are some important differences. First, Sgr is more disrupted (70\% of the mass is in the tails; \citealt{NO_Sgr_2010} vs.\ 26--42\% in the tails for F8D1). The SFH of Sgr's tidal tails and core \citep{Siegel2007} show a major drop in SFR $\sim$3--5\,Gyr ago \citep{Siegel2007,weisz2014,deBoer2015}, although Sgr's core experienced a significant episode of more recent star formation $\sim$2.5--3\,Gyr ago, with some possibly younger stars \citep{Siegel2007,weisz2014}. The metallicity of the most recently-formed stars appear quite similar to F8D1 ([M/H]\,${\sim}\,{-}0.5$; compare Figure \ref{fig:age-MH} with e.g., \citealt{Hasselquist2021}). Because Sgr is so close to the MW, its chemical evolution is more richly constrained, showing $\alpha$/Fe trends lower than the Milky Way for most metallicities (consistent with its lower total mass and star formation efficiency), showing a flattening in $\alpha$/Fe for the most recent star formation as $\alpha$ element production kept pace with Iron production from Type Ia Supernovae \citep{Hasselquist2021}. The overall picture is that F8D1 is quite consistent with a less evolved version of Sgr.

\subsection{How Does F8D1 Inform Our Ideas About UDG Formation?}
\label{sec:udgs}

UDGs are a heterogeneous group of galaxies, brought together by virtue of very low central surface brightness ($\mu_{0,V}\,{\gtrsim}\,23{-}24$\,\magsqarc) and extended half-light radii ($R_{\rm eff}\,{=}\,2{-}4$\,kpc; e.g., \citealt{vandokkum2015}, \citealt{Zaritsky2023}). While originally explored in the Coma galaxy cluster \citep{vandokkum2015}, UDGs have been discovered in a wide range of environments. UDGs occur as satellites in groups all the way from the Milky Way's group mass scale to large galaxy clusters \citep[e.g.,][]{vanderBurg2017,Lim2020,Karunakaran2023,Zaritsky2023}. Most of these satellite UDGs in clusters and groups lack ongoing star formation, while those in isolated environments can have very high gas content, relative to their stars, and host current star formation \citep[e.g.,][]{papastergis2017,He2019,Prole2019,Kadowaki2021}. Some UDGs have exceptionally high inferred mass-to-light ($M/L$) ratios, indicating a high degree of dark matter domination (see e.g., \citealt{Kravtsov2024}), while others have $M/L$\ values in the range consistent with stellar populations alone \citep[e.g.,][]{vandokkum2018,toloba2018,vandokkum2019,Toloba2023}. It is virtually certain that there are a number of mechanisms that lead to UDGs, with various mechanisms being more or less important depending on the particular types of UDG being studied. Indeed, many theoretical studies find that multiple mechanisms --- internal to the galaxy and external to the galaxy --- should operate in concert to give a realistic range of UDGs.  Accordingly, we do not attempt to `solve' the issue of UDG formation using F8D1. Instead, here we reflect on how F8D1, as a case study, illuminates certain UDG formation scenarios. 

\subsubsection{Star Formation Histories of UDGs}

Many of the formation mechanisms for UDGs are connected to important signatures in their SFHs. F8D1's early star formation history is relatively uncertain, with $\sim$80\% of its stars forming before $\sim$6\,Gyr ago. There was a prominent episode of star formation 2--2.5\,Gyr ago in which $\sim$15--25\% of its stars likely formed ($\sim$11\% in the central field, $\sim$39\% in the deep ACS inner stream field). An additional episode of star formation happened $\sim$500\,Myr ago, in which a few percent of its stars formed. In this section, we will briefly discuss how F8D1's SFH compares with the signatures of possible UDG formation models. This discussion is necessarily tentative, in great part because F8D1's UDG nature may be dominated by the impacts of tides (and other environmental processes). 

Several papers emphasize that intense feedback from bursty star formation, which is pivotal in DM core creation, has the consequence of lowering the central density of $M_{\star}\,{\sim}\,10^8\,M_{\odot}$\ galaxies, and can lead to the formation of UDGs \citep[e.g.,][]{el-badry2016,dicintio2017,chan2018}. F8D1 shows some features consistent with this picture. Where the SFH is well-resolved, F8D1 clearly shows multiple bursts of SF at $\sim$2\,Gyr and $\sim$500\,Myr ago. Yet, there are some areas where the F8D1's properties appear to differ from the theoretical expectations. F8D1's SFH before $\sim$2\,Gyr ago is concentrated at early times (with substantial SFH uncertainties), but not outwardly similar to the kind of $\sim$300\,Myr--2\,Gyr starburst cadence characteristic of the simulations from \citet{el-badry2016} or \citet{dicintio2017}. In addition, the models suggest that the oldest stars should have experienced the most radial migration, resulting in preferentially older outer parts. This is in contrast to F8D1's age gradient, which shows outskirts that are clearly younger than the center ($\tau_{80}$\ of $<$2\,Gyr vs.\ $>$6\,Gyr, respectively; see Figure \ref{fig:sfhs} and Table \ref{tab:sfh}). Obviously, the interpretation of the observations is complicated by F8D1's extensive tidal processing, and a DM core sculpted by bursty star formation would strongly change the impact of tides on F8D1. We conclude that F8D1's SFH shows some consistency with a bursty SF picture, but with some differences (e.g., early star formation and younger outskirts) that are not trivially interpreted in such a picture. 

The SFH implications of UDGs as the high angular momentum tail of the dwarf galaxy population \citep[e.g.,][]{dalcanton1997,amorisco&loeb2016,liao2019} are somewhat unclear. The naive expectation of such a picture might be that star formation should proceed more or less continuously with low efficiency \citep[e.g.,][]{Benavides2024}, in diffuse \textsc{H\,i} disks over long timescales, similar to their star-forming low surface brightness galaxy siblings \citep{deBlok1996,Bell2000,Schombert2014}. In contrast, F8D1 experienced much of its SF at relatively early times, but did retain the gas necessary to make $\gtrsim$20\% of its stellar mass in late star formation $\lesssim$4\,Gyr ago. F8D1's more recent star formation was likely entirely driven by tidal interaction, however, and does not necessarily invalidate a `high-spin' formation scenario. 

F8D1's properties do not strongly align with the expectations of a `failed galaxy' picture of UDG formation. In this picture, these `failed' UDGs stop forming stars early, $z\sim 2$, leaving a remnant with an unusually high number of globular clusters for their stellar mass and no current SF \citep{vandokkum2015,peng&lim2016,ferre-mateu2018}. In this picture, environmental processes (e.g., ram pressure) are usually credited with causing the early cessation of star formation. Visual inspection of our HST images shows 6 globular cluster candidates, including the GC already discovered by \citet{caldwell1998}. This number is similar to the number of globular clusters in galaxies of a similar mass (e.g., Fornax, \citealt{Pace2021}; Sagittarius dwarf spheroidal and stream, \citealt{Bellazzini2020}; and many of the UDGs found in clusters, e.g., \citealt{Lim2020,Toloba2023}), indicating no dramatic deficit of stellar mass compared to its globular cluster number or inferred dark matter mass. On the balance, F8D1's late-time star formation and reasonably typical globular cluster number argue against a `failed galaxy' origin for F8D1. 

One can see that F8D1 shows aspects of many of the factors that are thought to shape UDG formation and evolution. It had an initial episode of star formation that was followed by a fallow period, consistent broadly with a bursty formation scenario (or a failed `failure'!). Yet, F8D1 retained a reservoir of gas (consistent with bursty or high angular momentum scenario) that fueled further bursts of star formation $\sim$2\,Gyr and $\sim$500\,Myr ago. Importantly, the dominant driver of F8D1's recent evolution is tidal processing, which may have contributed to (or triggered) F8D1's most recent SF bursts, and created tidal tails containing 26--42\% of its total stellar mass. 

\subsubsection{The Importance of Tides in UDG Formation}

One point that is exceptionally clear, from both this work and \citet{zemaitis2023}, is the importance of tides in shaping F8D1, and by extension, in shaping at least some UDGs. Isolated UDGs are, obviously, not expected to have been affected by tides. Deep imaging studies in groups do not find tidal tails around some UDGs (e.g., NGC 1052's UDGs; \citealt{Mueller2019}, although see also \citealt{Montes2020} who claim tidal debris around NGC1052-DF4), but find the impacts of environmental processing in the forms of ram pressure stripping and/or tidal tails around other group UDGs \citep[e.g.,][]{Bennet2018,Gannon2021,Fielder2024,Watts2024}, and even more exotic dwarf merger scenarios \citep[e.g.,][]{vanDokkum2022,Keim2025}. 

\begin{figure*}[t]
    \centering
    \includegraphics[width=\linewidth]{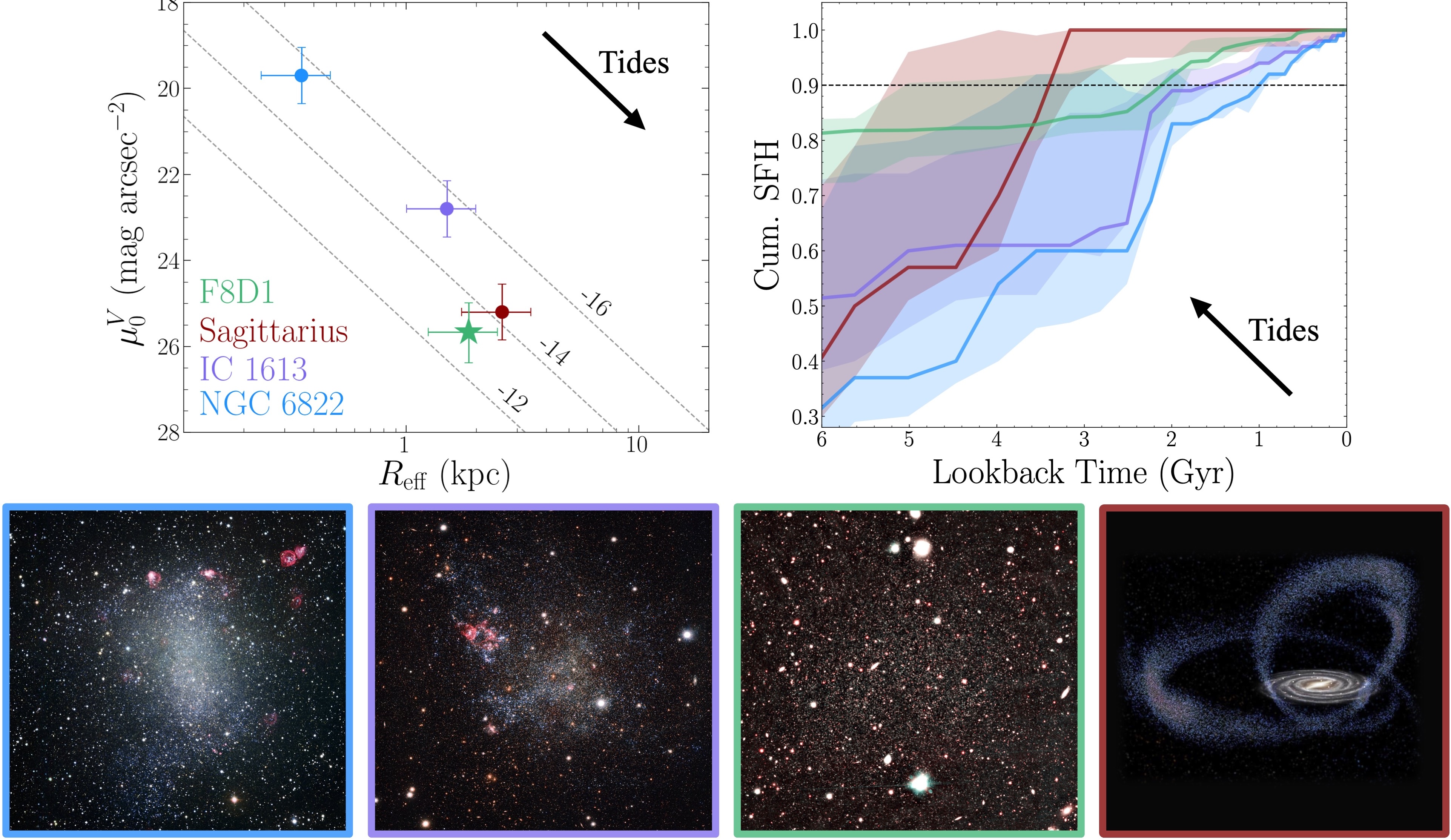}
    \caption{Comparison of F8D1's (green) properties measured in this work to those of local galaxies with analogous progenitor stellar mass: NGC 6822 (blue), IC 1613 (purple), and Sagittarius (dark red). \textit{Top Left}: Effective radius vs.\ central $V$-band surface brightness for F8D1 and the three Local Group galaxies. As in Figure \ref{fig:udgs}, lines of constant $M_V$\ have been drawn, assuming exponential profiles. Note that none of the four galaxies' light profiles follow a true exponential, and therefore these are only guides. We assume a consistent 33\% uncertainty on the effective radii. \textit{Top Right}: Combined cumulative SFHs of the four galaxies over the past 6\,Gyr (this work and \citealt{weisz2014}). Note that Sagittarius combines fields along the stream and the remnant dwarf spheroidal. In both spaces, the four galaxies form a plausible evolutionary sequence, where tidal processing similarly impacts both the morphology and star formation in dwarf galaxies at this mass scale. NGC 6822 and IC 1613 are currently star-forming and are plausible analogs of F8D1's progenitor, while Sagittarius is in a more advanced stage of tidal disruption. \\
    \textit{Images}: From left to right --- $B/V/R/{\rm H\alpha}$\ image of NGC 6822 from the 2.2\,m MPG/ESO telescope at La Silla Observatory; $U/G/R/{\rm H\alpha}$\ image of IC 1613 from OmegaCAM on ESO's VLT Survey Telescope; $r/i$\ color image of F8D1 from Subaru HSC; artistic representation of the interaction between Sagittarius and the Milky Way, from Gabriel Pérez Díaz at the Instituto de Astrof\'{i}sica de Canarias (IAC).
    }
    \label{fig:sfh-compare}
\end{figure*}

An important factor in determining the response of a satellite galaxy to tides is the structure of the central parts of its dark matter halo. Galaxies with cusped DM halo profiles do not substantially change in their central structure during a tidal interation, while galaxies with cored dark matter halos `puff up' substantially as they are tidally stripped \citep[e.g.,][]{Errani2015,Carleton2019}. Velocity profiles indicative of cores are particularly common in galaxies with stellar masses ${\sim}10^8\,M_{\odot}$\ \citep[both non-star-forming spheroidal galaxies and star-forming irregulars; e.g.,][]{Battaglia2008,Walker2011,Oh2011}. Galaxies simulated in a $\Lambda$CDM cosmological framework require strong effects from stellar feedback to produce such strongly cored dark matter density profiles, which on one hand requires enough star formation to inject a large enough amount of energy, and on the other hand a shallow enough potential to allow the outflows to significantly affect the distribution of the ISM \citep[e.g.,][]{Pontzen2012,DiCintio2014,Chan2015,Lazar2020}. This favors core formation in a broad stellar mass range centered on $M_{\star}\,{\sim}\,10^{8}\,M_{\odot}$, similar in mass to F8D1.  

We can estimate the magnitude of this increase in size for a cored progenitor of F8D1, using the framework of \citet{Errani2015}. As discussed in \S\,\ref{sec:sag-compare}, F8D1's tidal streams contain 26--42\% of its stellar mass. \citet{Errani2015} suggest that this degree of stellar mass loss indicates a loss of 80--90\% of its total mass. For this degree of mass loss, negligible changes in $R_{\rm e}$\ would be expected for a cusped DM density profile. In contrast, for a cored profile, \citet{Errani2015} predicts a factor of 2.5--4 increase in $R_{\rm e}$. This suggests that F8D1's progenitor may have had a significant core and been substantially more compact than its current half-light radius ($\sim$1.8\,kpc), with half-light radii between 500\,pc and 1\,kpc. We will use this information in \S\,\ref{sec:prog} where we consider possible progenitor analogs for F8D1. 

\subsection{Reflections on Possible F8D1 Progenitors}
\label{sec:prog}

In the light of these different formation scenarios, it is interesting to reflect on possible progenitors of the F8D1 system. If F8D1 experienced no increase in its half-light radius ($\sim$1.8\,kpc in its main body; this work) owing to tides, then there are a limited number of analogs of `star-forming F8D1s'. The largest samples of analogs are drawn from larger volume catalogs. For example, \citet{Janowiecki19} identify a sample of 71 \textsc{H\,i}-rich UDGs with $M_g\,{\sim}\,{-}16$ and $R_{\rm e}\,{>}\,1.5$\,kpc, with \textsc{H\,i} masses around ${\sim}10^9\,M_{\odot}$, that might represent star-forming progenitors of F8D1 (see also \citealt{Leisman2017}). Although the identification of 71 progenitors seems like such galaxies might be common, these galaxies were selected from a parent sample of more than 15,000 \textsc{H\,i}-bearing galaxies between $25<D/Mpc<120$; star-forming UDGs are reasonably rare. In addition, consideration of their star formation rates and histories suggests that they remain ultra-diffuse by virtue of a low star formation efficiency over much of the lifetime of the Universe \citep{Kado-Fong2022}, perhaps somewhat at variance with F8D1's SFH, in which $\sim$80\% of its stars were formed $>$6\,Gyr ago or more.

If F8D1's half-light radius has increased as part of the process of tidal stripping, owing to a cored mass profile \citep[e.g.,][]{Errani2015}, then possible progenitor (and descendant) systems can be identified more readily, to the point where an `evolutionary sequence' can be envisaged using galaxies in the Local Group alone. Among Local Group members, NGC 6822 and IC 1613 are the closest star-forming galaxies in stellar masses to F8D1, with \textsc{H\,i} masses of $1.3{\times}10^8\,M_{\odot}$\ \citep{deBlok2000} and $5{\times}10^7\,M_{\odot}$\ \citep{Silich2006}, respectively. IC 1613's may be experiencing initial tidal disruption/processing, as evidenced both in its distribution of stars and \textsc{H\,i} \citep{battinelli2007}, and, as discussed in \S\,\ref{sec:sag-compare}, Sagittarius is a plausible descendant system. We illustrate the comparison between F8D1 and its local analogs schematically in Figure \ref{fig:sfh-compare}. We compare the surface brightnesses, sizes, visual appearances and SFHs of NGC 6822, IC 1613, F8D1 and Sagittarius. The galaxies exhibit nearly monotonic trends in all four properties with stage of tidal disruption. We therefore postulate that increasing impacts of tidal processing puffs up the stellar distribution, decreases the central density, and is reflected on the SFH by later $\tau_{90}$\ values, as star formation is quenched by the effects of group ram pressure. In this paradigm, F8D1 represents an intermediate stage of a common evolutionary pathway experienced by satellite galaxies at this mass scale. 

It is worth briefly considering the fate of F8D1's ISM in this scenario. Attributing specific features of the complex \textsc{H\,i} field in the M81 Group to specific galaxies will require a detailed orbital modeling effort (see \S\,\ref{sec:orbit}). However, we note that the Eastern `arm' of \textsc{H\,i} visible in the map of \cite{deBlok2018} (see Figure \ref{fig:m81-group}), which visually overlaps the end of F8D1's stellar stream, contains ${\sim}10^8\,M_{\odot}$\ of gas --- highly comparable to the current reservoir of NGC 6822, and possibly to the expected pre-infall reservoir of F8D1. It is viable that this arm originally resided in the larger NGC 2976, which may also be experiencing ram pressure effects \citep[e.g.,][]{Drzazga2016}. However, NGC 2976 is still star-forming and does not appear to show a substantial environmental impact (either positive or negative) in its recent SFH \citep{williams2010}. Part of our future modeling effort will be to investigate whether this \textsc{H\,i} feature could comprise the ram pressure-stripped remnants of F8D1's former ISM. 

\section{Conclusions} \label{sec:conclusions}

We have used new HST observations to infer the star formation history of the nearby UDG, F8D1, over the past 6\,Gyr, using detections of individual stars to below the Red Clump in a central WFC3 field and an outer ACS field. Using detected evolved RGB and AGB stars in a set of shallower WFC3+ACS fields, in concert with published Subaru HSC observations, we have measured the properties of F8D1's main body and stellar stream. This temporal and morphological study of F8D1 represents one of the most detailed case studies of a UDG to-date. We find: 

\begin{enumerate}[topsep=0pt,leftmargin=12pt,itemsep=0pt]
    \item Three distinct periods of star formation; the first, $>$6\,Gyr ago, when $\sim$80\% of F8D1's stars formed (with considerable uncertainties). This was followed by a burst of star formation $\sim$2--2.5\,Gyr ago, that formed $\sim$11\% of the stars in the central field and $\sim$39\% in the outer field. Lastly, a smaller burst occurred only $\sim$500\,Myr ago, forming a few \% of F8D1's stars, and which also likely formed F8D1's nuclear star cluster. 
    \item A global shutdown time of $<$2\,Gyr in F8D1, including its core and tidal stream out to 13\,kpc, combining the SFH fits with the ratio of AGB-to-upper RGB stars. This indicates that F8D1 was globally star-forming 2\,Gyr ago. 
    \item A two-component morphology, with a total progenitor stellar mass of 1.31$^{+0.39}_{-0.19}{\times}10^8\,M_{\odot}$. An MCMC fitting analysis yields two distinct S\'{e}rsic components to its stellar mass surface density profile: an inner spheroid, with a half-light radius of 1.84\,kpc, and an extended tidal stream, with a half-light radius of 33\,kpc. Our new structural model of F8D1 predicts that 26--42\% of its stellar mass is currently unbound in this extensive tidal stream. 
    \item The outer deep field, which lies on the inner portion of F8D1's stream, was star-forming 500\,Myr ago, constraining the formation time of the inner portion of the stream.  
    \item It is as yet unclear whether M81 or NGC 2976 are responsible for environmentally processing F8D1; both are plausible interaction partners when taking into account projected separations, line-of-sight distances, and velocities. One or both of F8D1's recent bursts of star formation may be associated with pericenter passages, but the recent star formation in F8D1's tidal tail constrains its most recent pericenter to $<$500\,Myr ago. More detailed orbital modeling, deeper mapping of F8D1's tidal streams, and/or more careful consideration of distance constraints and F8D1's SFH, will be necessary to clarify F8D1's tidal history.
    \item F8D1's bursty star formation history illustrates characteristics anticipated by theoretical models that link such histories with the `puffing up' of a galaxy through strong episodic stellar feedback, leading to DM core formation. The presence of significant star formation late in its history, however, alongside its seemingly typical globular cluster population, argues against a `failed galaxy' scenario for F8D1's origin. \\
\end{enumerate} 

While a variety of UDG formation pathways are necessary to account for the full diversity of these low surface brightness galaxies, F8D1 underlines and richly illustrates the importance of environment, and tidal processing particularly, in forming UDGs. Simulations of the tidal disruption of satellites residing in cored potentials, as expected at F8D1's stellar mass scale, predict substantial morphological changes and size increases for tidally-disrupting satellites at F8D1's stage of disruption. In this paradigm, we find that plausible progenitors for F8D1 exist in the local universe, irrespective of how strongly tidal processing affects the central surface brightness of satellites. In the (likely) event that F8D1 has a cored density profile, F8D1's progenitor may have had properties similar to Local Group galaxies NGC 6822 and IC 1613, illustrating a possible sequence from star-forming dwarf irregular, to an ultra-diffuse galaxy undergoing tidal disruption (F8D1), eventually resulting in a more mature stream system like the Sagittarius dwarf spheroidal and its stream at later times.

\facility{Hubble Space Telescope (ACS, WFC3), MAST, Subaru Telescope (HSC)}

\software{\texttt{MATCH} \citep{dolphin2002}, \texttt{DOLPHOT} \citep{Dolphin2016}, \texttt{Matplotlib} \citep{matplotlib}, \texttt{NumPy} \citep{numpy-guide,numpy}, \texttt{Astropy} \citep{astropy}, \texttt{SciPy} \citep{scipy}, \texttt{SAOImage DS9} \citep{ds9}}

\begin{acknowledgments}
We thank the anonymous referee for a thoughtful and careful review that improved this paper.

This research is based on observations made with the NASA/ESA Hubble Space Telescope, associated with program GO-16191, and obtained from the Space Telescope Science Institute. AS is supported by NASA through the NASA Hubble Fellowship grant HST-HF2-51567, also awarded by the Space Telescope Science Institute. The Space Telescope Science Institute is operated by the Association of Universities for Research in Astronomy, Inc., under NASA contract NAS 5–26555. ET is thankful for the support from  NSF-AST- 2206498 grant. AM acknowledges support from the ANID FONDECYT Regular grant 1251882, from the ANID BASAL project FB210003, and funding from the HORIZON-MSCA-2021-SE-01 Research and Innovation Programme under the Marie Sklodowska-Curie grant agreement number 101086388. TKC is supported by the `Improvement on Competitiveness in Hiring New Faculties' Funding Scheme (4937210-4937211-4937212), the Direct Grant project (4053662,4443786,4053719) from the Chinese University of Hong Kong, and the RGC Early Career Scheme (24301524). TKC was supported by the E. Margaret Burbidge Prize Postdoctoral Fellowship from the Brinson Foundation at the Departments of Astronomy and Astrophysics at the University of Chicago.
\end{acknowledgments}

\bibliographystyle{aasjournal}
\bibliography{references}

@ARTICLE{NO_Sgr_2010,
       author = {{Niederste-Ostholt}, M. and {Belokurov}, V. and {Evans}, N.~W. and {Pe{\~n}arrubia}, J.},
        title = "{Re-Assembling the Sagittarius Dwarf Galaxy}",
      journal = {\apj},
         year = 2010,
        month = mar,
       volume = {712},
       number = {1},
        pages = {516-526},
          doi = {10.1088/0004-637X/712/1/516},
archivePrefix = {arXiv},
       eprint = {1002.0266},
 primaryClass = {astro-ph.GA},
       adsurl = {https://ui.adsabs.harvard.edu/abs/2010ApJ...712..516N}
}

@ARTICLE{chamberlain2024b,
       author = {{Chamberlain}, Katie and {Patel}, Ekta and {Besla}, Gurtina and {Torrey}, Paul and {Rodriguez-Gomez}, Vicente},
        title = "{A Physically Motivated Framework to Compare the Merger Timescales of Isolated Low- and High-mass Galaxy Pairs Across Cosmic Time}",
      journal = {\apj},
         year = 2024,
        month = nov,
       volume = {975},
       number = {1},
          eid = {104},
        pages = {104},
          doi = {10.3847/1538-4357/ad7bad},
archivePrefix = {arXiv},
       eprint = {2409.02233},
 primaryClass = {astro-ph.GA},
       adsurl = {https://ui.adsabs.harvard.edu/abs/2024ApJ...975..104C}
}

@ARTICLE{hunt2021,
       author = {{Hunt}, Jason A.~S. and {Stelea}, Ioana A. and {Johnston}, Kathryn V. and {Gandhi}, Suroor S. and {Laporte}, Chervin F.~P. and {B{\'e}dorf}, Jeroen},
        title = "{Resolving local and global kinematic signatures of satellite mergers with billion particle simulations}",
      journal = {\mnras},
         year = 2021,
        month = nov,
       volume = {508},
       number = {1},
        pages = {1459-1472},
          doi = {10.1093/mnras/stab2580},
archivePrefix = {arXiv},
       eprint = {2107.06294},
 primaryClass = {astro-ph.GA},
       adsurl = {https://ui.adsabs.harvard.edu/abs/2021MNRAS.508.1459H}
}

@ARTICLE{vasiliev2021,
       author = {{Vasiliev}, Eugene and {Belokurov}, Vasily and {Erkal}, Denis},
        title = "{Tango for three: Sagittarius, LMC, and the Milky Way}",
      journal = {\mnras},
         year = 2021,
        month = feb,
       volume = {501},
       number = {2},
        pages = {2279-2304},
          doi = {10.1093/mnras/staa3673},
archivePrefix = {arXiv},
       eprint = {2009.10726},
 primaryClass = {astro-ph.GA},
       adsurl = {https://ui.adsabs.harvard.edu/abs/2021MNRAS.501.2279V}
}

@ARTICLE{laporte2018,
       author = {{Laporte}, Chervin F.~P. and {Johnston}, Kathryn V. and {G{\'o}mez}, Facundo A. and {Garavito-Camargo}, Nicolas and {Besla}, Gurtina},
        title = "{The influence of Sagittarius and the Large Magellanic Cloud on the stellar disc of the Milky Way Galaxy}",
      journal = {\mnras},
         year = 2018,
        month = nov,
       volume = {481},
       number = {1},
        pages = {286-306},
          doi = {10.1093/mnras/sty1574},
archivePrefix = {arXiv},
       eprint = {1710.02538},
 primaryClass = {astro-ph.GA},
       adsurl = {https://ui.adsabs.harvard.edu/abs/2018MNRAS.481..286L}
}

@ARTICLE{Pearson2018,
       author = {{Pearson}, Sarah and {Privon}, George C. and {Besla}, Gurtina and {Putman}, Mary E. and {Mart{\'\i}nez-Delgado}, David and {Johnston}, Kathryn V. and {Gabany}, R. Jay and {Patton}, David R. and {Kallivayalil}, Nitya},
        title = "{Modelling the baryon cycle in low-mass galaxy encounters: the case of NGC 4490 and NGC 4485}",
      journal = {\mnras},
         year = 2018,
        month = nov,
       volume = {480},
       number = {3},
        pages = {3069-3090},
          doi = {10.1093/mnras/sty2052},
archivePrefix = {arXiv},
       eprint = {1807.03791},
 primaryClass = {astro-ph.GA},
       adsurl = {https://ui.adsabs.harvard.edu/abs/2018MNRAS.480.3069P}
}

@ARTICLE{Pearson2016,
       author = {{Pearson}, Sarah and {Besla}, Gurtina and {Putman}, Mary E. and {Lutz}, Katharina A. and {Fern{\'a}ndez}, Ximena and {Stierwalt}, Sabrina and {Patton}, David R. and {Kim}, Jinhyub and {Kallivayalil}, Nitya and {Johnson}, Kelsey and {Sung}, Eon-Chang},
        title = "{Local Volume TiNy Titans: gaseous dwarf-dwarf interactions in the Local Universe}",
      journal = {\mnras},
         year = 2016,
        month = jun,
       volume = {459},
       number = {2},
        pages = {1827-1846},
          doi = {10.1093/mnras/stw757},
archivePrefix = {arXiv},
       eprint = {1603.09342},
 primaryClass = {astro-ph.GA},
       adsurl = {https://ui.adsabs.harvard.edu/abs/2016MNRAS.459.1827P}
}

@ARTICLE{Siegel2007,
       author = {{Siegel}, Michael H. and {Dotter}, Aaron and {Majewski}, Steven R. and {Sarajedini}, Ata and {Chaboyer}, Brian and {Nidever}, David L. and {Anderson}, Jay and {Mar{\'\i}n-Franch}, Antonio and {Rosenberg}, Alfred and {Bedin}, Luigi R. and {Aparicio}, Antonio and {King}, Ivan and {Piotto}, Giampaolo and {Reid}, I. Neill},
        title = "{The ACS Survey of Galactic Globular Clusters: M54 and Young Populations in the Sagittarius Dwarf Spheroidal Galaxy}",
      journal = {\apjl},
         year = 2007,
        month = sep,
       volume = {667},
       number = {1},
        pages = {L57-L60},
          doi = {10.1086/522003},
archivePrefix = {arXiv},
       eprint = {0708.0027},
 primaryClass = {astro-ph},
       adsurl = {https://ui.adsabs.harvard.edu/abs/2007ApJ...667L..57S}
}

@ARTICLE{deBoer2015,
       author = {{de Boer}, T.~J.~L. and {Belokurov}, V. and {Koposov}, S.},
        title = "{The star formation history of the Sagittarius stream}",
      journal = {\mnras},
         year = 2015,
        month = aug,
       volume = {451},
       number = {4},
        pages = {3489-3503},
          doi = {10.1093/mnras/stv946},
archivePrefix = {arXiv},
       eprint = {1505.00787},
 primaryClass = {astro-ph.GA},
       adsurl = {https://ui.adsabs.harvard.edu/abs/2015MNRAS.451.3489D}
}

@ARTICLE{Bell2000,
       author = {{Bell}, Eric F. and {Barnaby}, David and {Bower}, Richard G. and {de
        Jong}, Roelof S. and {Harper}, Doyal A. and {Hereld}, Mark and
        {Loewenstein}, Robert F. and {Rauscher}, Bernard J.},
        title = "{The star formation histories of low surface brightness galaxies}",
      journal = {\mnras},
         year = 2000,
        month = Mar,
       volume = {312},
        pages = {470-496},
          doi = {10.1046/j.1365-8711.2000.03180.x},
archivePrefix = {arXiv},
       eprint = {astro-ph/9909401},
 primaryClass = {Astrophysics},
       adsurl = {https://ui.adsabs.harvard.edu/#abs/2000MNRAS.312..470B}
}

@ARTICLE{vandokkum2015,
       author = {{van Dokkum}, Pieter G. and {Abraham}, Roberto and {Merritt}, Allison
        and {Zhang}, Jielai and {Geha}, Marla and {Conroy}, Charlie},
        title = "{Forty-seven Milky Way-sized, Extremely Diffuse Galaxies in the Coma
        Cluster}",
      journal = {\apj},
         year = 2015,
        month = Jan,
       volume = {798},
          eid = {L45},
        pages = {L45},
          doi = {10.1088/2041-8205/798/2/L45},
archivePrefix = {arXiv},
       eprint = {1410.8141},
       adsurl = {https://ui.adsabs.harvard.edu/#abs/2015ApJ...798L..45V}
}

@ARTICLE{caldwell1998,
       author = {{Caldwell}, Nelson and {Armandroff}, Taft E. and {Da Costa}, G.~S. and
        {Seitzer}, Patrick},
        title = "{Dwarf Elliptical Galaxies in the M81 Group: The Structure and Stellar
        Populations of BK5N and F8D1}",
      journal = {\aj},
         year = 1998,
        month = Feb,
       volume = {115},
        pages = {535-558},
          doi = {10.1086/300233},
archivePrefix = {arXiv},
       eprint = {astro-ph/9710286},
       adsurl = {https://ui.adsabs.harvard.edu/#abs/1998AJ....115..535C}
}

@ARTICLE{vandokkum2016,
       author = {{van Dokkum}, Pieter and {Abraham}, Roberto and {Brodie}, Jean and
        {Conroy}, Charlie and {Danieli}, Shany and {Merritt}, Allison
        and {Mowla}, Lamiya and {Romanowsky}, Aaron and {Zhang}, Jielai},
        title = "{A High Stellar Velocity Dispersion and ̃100 Globular Clusters for the
        Ultra-diffuse Galaxy Dragonfly 44}",
      journal = {\apj},
         year = 2016,
        month = Sep,
       volume = {828},
          eid = {L6},
        pages = {L6},
          doi = {10.3847/2041-8205/828/1/L6},
       adsurl = {https://ui.adsabs.harvard.edu/#abs/2016ApJ...828L...6V}
}

@ARTICLE{jacobs2009,
       author = {{Jacobs}, Bradley A. and {Rizzi}, Luca and {Tully}, R. Brent and
        {Shaya}, Edward J. and {Makarov}, Dmitry I. and {Makarova},
        Lidia},
        title = "{The Extragalactic Distance Database: Color-Magnitude Diagrams}",
      journal = {\aj},
         year = 2009,
        month = Aug,
       volume = {138},
        pages = {332-337},
          doi = {10.1088/0004-6256/138/2/332},
archivePrefix = {arXiv},
       eprint = {0902.3675},
       adsurl = {https://ui.adsabs.harvard.edu/#abs/2009AJ....138..332J}
}

@ARTICLE{Carleton2019,
       author = {{Carleton}, Timothy and {Errani}, Rapha{\"e}l and {Cooper}, Michael and {Kaplinghat}, Manoj and {Pe{\~n}arrubia}, Jorge and {Guo}, Yicheng},
        title = "{The formation of ultra-diffuse galaxies in cored dark matter haloes through tidal stripping and heating}",
      journal = {\mnras},
         year = 2019,
        month = may,
       volume = {485},
       number = {1},
        pages = {382-395},
          doi = {10.1093/mnras/stz383},
archivePrefix = {arXiv},
       eprint = {1805.06896},
 primaryClass = {astro-ph.GA},
       adsurl = {https://ui.adsabs.harvard.edu/abs/2019MNRAS.485..382C}
}

@ARTICLE{Errani2015,
       author = {{Errani}, R. and {Penarrubia}, J. and {Tormen}, G.},
        title = "{Constraining the distribution of dark matter in dwarf spheroidal galaxies with stellar tidal streams.}",
      journal = {\mnras},
         year = 2015,
        month = apr,
       volume = {449},
        pages = {L46-L50},
          doi = {10.1093/mnrasl/slv012},
archivePrefix = {arXiv},
       eprint = {1501.04968},
 primaryClass = {astro-ph.GA},
       adsurl = {https://ui.adsabs.harvard.edu/abs/2015MNRAS.449L..46E}
}

@ARTICLE{Battaglia2008,
       author = {{Battaglia}, G. and {Helmi}, A. and {Tolstoy}, E. and {Irwin}, M. and {Hill}, V. and {Jablonka}, P.},
        title = "{The Kinematic Status and Mass Content of the Sculptor Dwarf Spheroidal Galaxy}",
      journal = {\apjl},
         year = 2008,
        month = jul,
       volume = {681},
       number = {1},
        pages = {L13},
          doi = {10.1086/590179},
archivePrefix = {arXiv},
       eprint = {0802.4220},
 primaryClass = {astro-ph},
       adsurl = {https://ui.adsabs.harvard.edu/abs/2008ApJ...681L..13B}
}

@ARTICLE{deBlok2018,
       author = {{de Blok}, W.~J.~G. and {Walter}, Fabian and {Ferguson}, Annette M.~N. and {Bernard}, Edouard J. and {van der Hulst}, J.~M. and {Neeleman}, Marcel and {Leroy}, Adam K. and {Ott}, J{\"u}rgen and {Zschaechner}, Laura K. and {Zwaan}, Martin A. and {Yun}, Min S. and {Langston}, Glen and {Keating}, Katie M.},
        title = "{A High-resolution Mosaic of the Neutral Hydrogen in the M81 Triplet}",
      journal = {\apj},
         year = 2018,
        month = sep,
       volume = {865},
       number = {1},
          eid = {26},
        pages = {26},
          doi = {10.3847/1538-4357/aad557},
archivePrefix = {arXiv},
       eprint = {1808.02840},
 primaryClass = {astro-ph.GA},
       adsurl = {https://ui.adsabs.harvard.edu/abs/2018ApJ...865...26D}
}

@ARTICLE{Crnojevic2014,
       author = {{Crnojevi{\'c}}, D. and {Ferguson}, A.~M.~N. and {Irwin}, M.~J. and {McConnachie}, A.~W. and {Bernard}, E.~J. and {Fardal}, M.~A. and {Ibata}, R.~A. and {Lewis}, G.~F. and {Martin}, N.~F. and {Navarro}, J.~F. and {No{\"e}l}, N.~E.~D. and {Pasetto}, S.},
        title = "{A PAndAS view of M31 dwarf elliptical satellites: NGC 147 and NGC 185}",
      journal = {\mnras},
         year = 2014,
        month = dec,
       volume = {445},
       number = {4},
        pages = {3862-3877},
          doi = {10.1093/mnras/stu2003},
archivePrefix = {arXiv},
       eprint = {1409.7065},
 primaryClass = {astro-ph.GA},
       adsurl = {https://ui.adsabs.harvard.edu/abs/2014MNRAS.445.3862C}
}

@ARTICLE{Bennet2018,
       author = {{Bennet}, P. and {Sand}, D.~J. and {Zaritsky}, D. and {Crnojevi{\'c}}, D. and {Spekkens}, K. and {Karunakaran}, A.},
        title = "{Evidence for Ultra-diffuse Galaxy {\textquotedblleft}Formation{\textquotedblright} through Galaxy Interactions}",
      journal = {\apjl},
         year = 2018,
        month = oct,
       volume = {866},
       number = {1},
          eid = {L11},
        pages = {L11},
          doi = {10.3847/2041-8213/aadedf},
archivePrefix = {arXiv},
       eprint = {1809.01145},
 primaryClass = {astro-ph.GA},
       adsurl = {https://ui.adsabs.harvard.edu/abs/2018ApJ...866L..11B}
}

@ARTICLE{Montes2020,
       author = {{Montes}, Mireia and {Infante-Sainz}, Ra{\'u}l and {Madrigal-Aguado}, Alberto and {Rom{\'a}n}, Javier and {Monelli}, Matteo and {Borlaff}, Alejandro S. and {Trujillo}, Ignacio},
        title = "{The Galaxy ``Missing Dark Matter'' NGC 1052-DF4 is Undergoing Tidal Disruption}",
      journal = {\apj},
         year = 2020,
        month = dec,
       volume = {904},
       number = {2},
          eid = {114},
        pages = {114},
          doi = {10.3847/1538-4357/abc340},
archivePrefix = {arXiv},
       eprint = {2010.09719},
 primaryClass = {astro-ph.GA},
       adsurl = {https://ui.adsabs.harvard.edu/abs/2020ApJ...904..114M}
}

@ARTICLE{Leisman2017,
       author = {{Leisman}, Lukas and {Haynes}, Martha P. and {Janowiecki}, Steven and {Hallenbeck}, Gregory and {J{\'o}zsa}, Gyula and {Giovanelli}, Riccardo and {Adams}, Elizabeth A.~K. and {Bernal Neira}, David and {Cannon}, John M. and {Janesh}, William F. and {Rhode}, Katherine L. and {Salzer}, John J.},
        title = "{(Almost) Dark Galaxies in the ALFALFA Survey: Isolated H I-bearing Ultra-diffuse Galaxies}",
      journal = {\apj},
         year = 2017,
        month = jun,
       volume = {842},
       number = {2},
          eid = {133},
        pages = {133},
          doi = {10.3847/1538-4357/aa7575},
archivePrefix = {arXiv},
       eprint = {1703.05293},
 primaryClass = {astro-ph.GA},
       adsurl = {https://ui.adsabs.harvard.edu/abs/2017ApJ...842..133L}
}

@ARTICLE{Kado-Fong2022,
       author = {{Kado-Fong}, Erin and {Greene}, Jenny E. and {Huang}, Song and {Goulding}, Andy},
        title = "{Ultra-diffuse Galaxies as Extreme Star-forming Environments. I. Mapping Star Formation in H I-rich UDGs}",
      journal = {\apj},
         year = 2022,
        month = dec,
       volume = {941},
       number = {1},
          eid = {11},
        pages = {11},
          doi = {10.3847/1538-4357/ac9964},
archivePrefix = {arXiv},
       eprint = {2209.05492},
 primaryClass = {astro-ph.GA},
       adsurl = {https://ui.adsabs.harvard.edu/abs/2022ApJ...941...11K}
}

@ARTICLE{Janowiecki19,
       author = {{Janowiecki}, Steven and {Jones}, Michael G. and {Leisman}, Lukas and {Webb}, Andrew},
        title = "{The environment of H I-bearing ultra-diffuse galaxies in the ALFALFA survey}",
      journal = {\mnras},
         year = 2019,
        month = nov,
       volume = {490},
       number = {1},
        pages = {566-577},
          doi = {10.1093/mnras/stz1868},
archivePrefix = {arXiv},
       eprint = {1906.11543},
 primaryClass = {astro-ph.GA},
       adsurl = {https://ui.adsabs.harvard.edu/abs/2019MNRAS.490..566J}
}

@ARTICLE{Bellazzini2020,
       author = {{Bellazzini}, M. and {Ibata}, R. and {Malhan}, K. and {Martin}, N. and {Famaey}, B. and {Thomas}, G.},
        title = "{Globular clusters in the Sagittarius stream. Revising members and candidates with Gaia DR2}",
      journal = {\aap},
         year = 2020,
        month = apr,
       volume = {636},
          eid = {A107},
        pages = {A107},
          doi = {10.1051/0004-6361/202037621},
archivePrefix = {arXiv},
       eprint = {2003.07871},
 primaryClass = {astro-ph.GA},
       adsurl = {https://ui.adsabs.harvard.edu/abs/2020A&A...636A.107B}
}

@ARTICLE{Pace2021,
       author = {{Pace}, Andrew B. and {Walker}, Matthew G. and {Koposov}, Sergey E. and {Caldwell}, Nelson and {Mateo}, Mario and {Olszewski}, Edward W. and {Bailey}, III, John I. and {Wang}, Mei-Yu},
        title = "{Spectroscopic Confirmation of the Sixth Globular Cluster in the Fornax Dwarf Spheroidal Galaxy}",
      journal = {\apj},
         year = 2021,
        month = dec,
       volume = {923},
       number = {1},
          eid = {77},
        pages = {77},
          doi = {10.3847/1538-4357/ac2cd2},
archivePrefix = {arXiv},
       eprint = {2105.00064},
 primaryClass = {astro-ph.GA},
       adsurl = {https://ui.adsabs.harvard.edu/abs/2021ApJ...923...77P}
}

@ARTICLE{Schombert2014,
       author = {{Schombert}, James and {McGaugh}, Stacy},
        title = "{Stellar Populations and the Star Formation Histories of LSB Galaxies: III. Stellar Population Models}",
      journal = {\pasa},
         year = 2014,
        month = sep,
       volume = {31},
          eid = {e036},
        pages = {e036},
          doi = {10.1017/pasa.2014.32},
archivePrefix = {arXiv},
       eprint = {1407.6778},
 primaryClass = {astro-ph.GA},
       adsurl = {https://ui.adsabs.harvard.edu/abs/2014PASA...31...36S}
}

@ARTICLE{deBlok1996,
       author = {{de Blok}, W.~J.~G. and {McGaugh}, S.~S. and {van der Hulst}, J.~M.},
        title = "{HI observations of low surface brightness galaxies: probing low-density galaxies}",
      journal = {\mnras},
         year = 1996,
        month = nov,
       volume = {283},
       number = {1},
        pages = {18-54},
          doi = {10.1093/mnras/283.1.18},
archivePrefix = {arXiv},
       eprint = {astro-ph/9605069},
 primaryClass = {astro-ph},
       adsurl = {https://ui.adsabs.harvard.edu/abs/1996MNRAS.283...18D}
}

@ARTICLE{Gannon2021,
       author = {{Gannon}, Jonah S. and {Dullo}, Bililign T. and {Forbes}, Duncan A. and {Rich}, R. Michael and {Rom{\'a}n}, Javier and {Couch}, Warrick J. and {Brodie}, Jean P. and {Ferr{\'e}-Mateu}, Anna and {Alabi}, Adebusola and {Mould}, Jeremy},
        title = "{A photometric and kinematic analysis of UDG1137+16 (dw1137+16): Probing ultradiffuse galaxy formation in a group environment}",
      journal = {\mnras},
         year = 2021,
        month = apr,
       volume = {502},
       number = {3},
        pages = {3144-3157},
          doi = {10.1093/mnras/stab277},
archivePrefix = {arXiv},
       eprint = {2102.00598},
 primaryClass = {astro-ph.GA},
       adsurl = {https://ui.adsabs.harvard.edu/abs/2021MNRAS.502.3144G}
}

@ARTICLE{Mueller2019,
       author = {{M{\"u}ller}, Oliver and {Rich}, R. Michael and {Rom{\'a}n}, Javier and {Y{\i}ld{\i}z}, Mustafa K. and {B{\'\i}lek}, Michal and {Duc}, Pierre-Alain and {Fensch}, J{\'e}r{\'e}my and {Trujillo}, Ignacio and {Koch}, Andreas},
        title = "{A tidal tale: detection of several stellar streams in the environment of NGC 1052}",
      journal = {\aap},
         year = 2019,
        month = apr,
       volume = {624},
          eid = {L6},
        pages = {L6},
          doi = {10.1051/0004-6361/201935463},
archivePrefix = {arXiv},
       eprint = {1903.07285},
 primaryClass = {astro-ph.GA},
       adsurl = {https://ui.adsabs.harvard.edu/abs/2019A&A...624L...6M}
}

@ARTICLE{Fielder2024,
       author = {{Fielder}, Catherine and {Jones}, Michael G. and {Sand}, David J. and {Bennet}, Paul and {Crnojevi{\'c}}, Denija and {Karunakaran}, Ananthan and {Mutlu-Pakdil}, Bur{\c{c}}in and {Spekkens}, Kristine},
        title = "{All Puffed Up: Exploring Ultra-diffuse Galaxy Origins Through Galaxy Interactions}",
      journal = {\aj},
         year = 2024,
        month = nov,
       volume = {168},
       number = {5},
          eid = {212},
        pages = {212},
          doi = {10.3847/1538-3881/ad74f6},
archivePrefix = {arXiv},
       eprint = {2401.01931},
 primaryClass = {astro-ph.GA},
       adsurl = {https://ui.adsabs.harvard.edu/abs/2024AJ....168..212F}
}

@ARTICLE{Kravtsov2024,
       author = {{Kravtsov}, Andrey},
        title = "{On the dark matter content of ultra-diffuse galaxies}",
      journal = {The Open Journal of Astrophysics},
         year = 2024,
        month = dec,
       volume = {7},
          eid = {117},
        pages = {117},
          doi = {10.33232/001c.127487},
       adsurl = {https://ui.adsabs.harvard.edu/abs/2024OJAp....7E.117K}
}

@ARTICLE{Karunakaran2023,
       author = {{Karunakaran}, Ananthan and {Zaritsky}, Dennis},
        title = "{Extending Ultra-Diffuse Galaxy abundances to Milky Way analogues}",
      journal = {\mnras},
         year = 2023,
        month = feb,
       volume = {519},
       number = {1},
        pages = {884-890},
          doi = {10.1093/mnras/stac3622},
archivePrefix = {arXiv},
       eprint = {2210.00009},
 primaryClass = {astro-ph.GA},
       adsurl = {https://ui.adsabs.harvard.edu/abs/2023MNRAS.519..884K}
}

@ARTICLE{He2019,
       author = {{He}, Min and {Wu}, Hong and {Du}, Wei and {Wicker}, James and {Zhao}, Pingsong and {Lei}, Fengjie and {Liu}, Jifeng},
        title = "{Edge-on H I-bearing Ultra-diffuse Galaxy Candidates in the 40\% ALFALFA Catalog}",
      journal = {\apj},
         year = 2019,
        month = jul,
       volume = {880},
       number = {1},
          eid = {30},
        pages = {30},
          doi = {10.3847/1538-4357/ab2710},
archivePrefix = {arXiv},
       eprint = {1907.10438},
 primaryClass = {astro-ph.GA},
       adsurl = {https://ui.adsabs.harvard.edu/abs/2019ApJ...880...30H}
}

@ARTICLE{Prole2019,
       author = {{Prole}, D.~J. and {van der Burg}, R.~F.~J. and {Hilker}, M. and {Davies}, J.~I.},
        title = "{Observational properties of ultra-diffuse galaxies in low-density environments: field UDGs are predominantly blue and star forming}",
      journal = {\mnras},
         year = 2019,
        month = sep,
       volume = {488},
       number = {2},
        pages = {2143-2157},
          doi = {10.1093/mnras/stz1843},
archivePrefix = {arXiv},
       eprint = {1907.01559},
 primaryClass = {astro-ph.GA},
       adsurl = {https://ui.adsabs.harvard.edu/abs/2019MNRAS.488.2143P}
}

@ARTICLE{Kadowaki2021,
       author = {{Kadowaki}, Jennifer and {Zaritsky}, Dennis and {Donnerstein}, R.~L. and {RS}, Pranjal and {Karunakaran}, Ananthan and {Spekkens}, Kristine},
        title = "{On the Properties of Spectroscopically Confirmed Ultra-diffuse Galaxies across Environments}",
      journal = {\apj},
         year = 2021,
        month = dec,
       volume = {923},
       number = {2},
          eid = {257},
        pages = {257},
          doi = {10.3847/1538-4357/ac2948},
archivePrefix = {arXiv},
       eprint = {2110.00015},
 primaryClass = {astro-ph.GA},
       adsurl = {https://ui.adsabs.harvard.edu/abs/2021ApJ...923..257K}
}

@ARTICLE{vanderBurg2017,
       author = {{van der Burg}, Remco F.~J. and {Hoekstra}, Henk and {Muzzin}, Adam and {Sif{\'o}n}, Crist{\'o}bal and {Viola}, Massimo and {Bremer}, Malcolm N. and {Brough}, Sarah and {Driver}, Simon P. and {Erben}, Thomas and {Heymans}, Catherine and {Hildebrandt}, Hendrik and {Holwerda}, Benne W. and {Klaes}, Dominik and {Kuijken}, Konrad and {McGee}, Sean and {Nakajima}, Reiko and {Napolitano}, Nicola and {Norberg}, Peder and {Taylor}, Edward N. and {Valentijn}, Edwin},
        title = "{The abundance of ultra-diffuse galaxies from groups to clusters. UDGs are relatively more common in more massive haloes}",
      journal = {\aap},
         year = 2017,
        month = nov,
       volume = {607},
          eid = {A79},
        pages = {A79},
          doi = {10.1051/0004-6361/201731335},
archivePrefix = {arXiv},
       eprint = {1706.02704},
 primaryClass = {astro-ph.GA},
       adsurl = {https://ui.adsabs.harvard.edu/abs/2017A&A...607A..79V}
}

@ARTICLE{Zaritsky2023,
       author = {{Zaritsky}, Dennis and {Donnerstein}, Richard and {Dey}, Arjun and {Karunakaran}, Ananthan and {Kadowaki}, Jennifer and {Khim}, Donghyeon J. and {Spekkens}, Kristine and {Zhang}, Huanian},
        title = "{Systematically Measuring Ultra-diffuse Galaxies (SMUDGes). V. The Complete SMUDGes Catalog and the Nature of Ultradiffuse Galaxies}",
      journal = {\apjs},
         year = 2023,
        month = aug,
       volume = {267},
       number = {2},
          eid = {27},
        pages = {27},
          doi = {10.3847/1538-4365/acdd71},
archivePrefix = {arXiv},
       eprint = {2306.01524},
 primaryClass = {astro-ph.GA},
       adsurl = {https://ui.adsabs.harvard.edu/abs/2023ApJS..267...27Z}
}

@ARTICLE{Pontzen2012,
       author = {{Pontzen}, Andrew and {Governato}, Fabio},
        title = "{How supernova feedback turns dark matter cusps into cores}",
      journal = {\mnras},
         year = 2012,
        month = apr,
       volume = {421},
       number = {4},
        pages = {3464-3471},
          doi = {10.1111/j.1365-2966.2012.20571.x},
archivePrefix = {arXiv},
       eprint = {1106.0499},
 primaryClass = {astro-ph.CO},
       adsurl = {https://ui.adsabs.harvard.edu/abs/2012MNRAS.421.3464P}
}

@ARTICLE{DiCintio2014,
       author = {{Di Cintio}, Arianna and {Brook}, Chris B. and {Macci{\`o}}, Andrea V. and {Stinson}, Greg S. and {Knebe}, Alexander and {Dutton}, Aaron A. and {Wadsley}, James},
        title = "{The dependence of dark matter profiles on the stellar-to-halo mass ratio: a prediction for cusps versus cores}",
      journal = {\mnras},
         year = 2014,
        month = jan,
       volume = {437},
       number = {1},
        pages = {415-423},
          doi = {10.1093/mnras/stt1891},
archivePrefix = {arXiv},
       eprint = {1306.0898},
 primaryClass = {astro-ph.CO},
       adsurl = {https://ui.adsabs.harvard.edu/abs/2014MNRAS.437..415D}
}

@ARTICLE{Oh2011,
       author = {{Oh}, Se-Heon and {de Blok}, W.~J.~G. and {Brinks}, Elias and {Walter}, Fabian and {Kennicutt}, Jr., Robert C.},
        title = "{Dark and Luminous Matter in THINGS Dwarf Galaxies}",
      journal = {\aj},
         year = 2011,
        month = jun,
       volume = {141},
       number = {6},
          eid = {193},
        pages = {193},
          doi = {10.1088/0004-6256/141/6/193},
archivePrefix = {arXiv},
       eprint = {1011.0899},
 primaryClass = {astro-ph.CO},
       adsurl = {https://ui.adsabs.harvard.edu/abs/2011AJ....141..193O}
}

@ARTICLE{Hasselquist2021,
       author = {{Hasselquist}, Sten and {Hayes}, Christian R. and {Lian}, Jianhui and {Weinberg}, David H. and {Zasowski}, Gail and {Horta}, Danny and {Beaton}, Rachael and {Feuillet}, Diane K. and {Garro}, Elisa R. and {Gallart}, Carme and {Smith}, Verne V. and {Holtzman}, Jon A. and {Minniti}, Dante and {Lacerna}, Ivan and {Shetrone}, Matthew and {J{\"o}nsson}, Henrik and {Cioni}, Maria-Rosa L. and {Fillingham}, Sean P. and {Cunha}, Katia and {O'Connell}, Robert and {Fern{\'a}ndez-Trincado}, Jos{\'e} G. and {Mu{\~n}oz}, Ricardo R. and {Schiavon}, Ricardo and {Almeida}, Andres and {Anguiano}, Borja and {Beers}, Timothy C. and {Bizyaev}, Dmitry and {Brownstein}, Joel R. and {Cohen}, Roger E. and {Frinchaboy}, Peter and {Garc{\'\i}a-Hern{\'a}ndez}, D.~A. and {Geisler}, Doug and {Lane}, Richard R. and {Majewski}, Steven R. and {Nidever}, David L. and {Nitschelm}, Christian and {Povick}, Joshua and {Price-Whelan}, Adrian and {Roman-Lopes}, Alexandre and {Rosado}, Margarita and {Sobeck}, Jennifer and {Stringfellow}, Guy and {Valenzuela}, Octavio and {Villanova}, Sandro and {Vincenzo}, Fiorenzo},
        title = "{APOGEE Chemical Abundance Patterns of the Massive Milky Way Satellites}",
      journal = {\apj},
         year = 2021,
        month = dec,
       volume = {923},
       number = {2},
          eid = {172},
        pages = {172},
          doi = {10.3847/1538-4357/ac25f9},
archivePrefix = {arXiv},
       eprint = {2109.05130},
 primaryClass = {astro-ph.GA},
       adsurl = {https://ui.adsabs.harvard.edu/abs/2021ApJ...923..172H}
}

@ARTICLE{Meier2001,
       author = {{Meier}, David S. and {Turner}, Jean L. and {Beck}, Sara C.},
        title = "{Molecular Gas and Star Formation in NGC 3077}",
      journal = {\aj},
         year = 2001,
        month = oct,
       volume = {122},
       number = {4},
        pages = {1770-1781},
          doi = {10.1086/323136},
archivePrefix = {arXiv},
       eprint = {astro-ph/0107031},
 primaryClass = {astro-ph},
       adsurl = {https://ui.adsabs.harvard.edu/abs/2001AJ....122.1770M}
}

@ARTICLE{Arias2016,
       author = {{Arias}, Veronica and {Guglielmo}, Magda and {Fernando}, Nuwanthika and {Lewis}, Geraint F. and {Bland-Hawthorn}, Joss and {Bate}, Nicholas F. and {Conn}, Anthony and {Irwin}, Mike J. and {Ferguson}, Annette M.~N. and {Ibata}, Rodrigo A. and {McConnachie}, Alan W. and {Martin}, Nicolas},
        title = "{NGC 147, NGC 185 and CassII: a genetic approach to orbital properties, star formation and tidal debris}",
      journal = {\mnras},
         year = 2016,
        month = feb,
       volume = {456},
       number = {2},
        pages = {1654-1665},
          doi = {10.1093/mnras/stv2781},
archivePrefix = {arXiv},
       eprint = {1511.08289},
 primaryClass = {astro-ph.GA},
       adsurl = {https://ui.adsabs.harvard.edu/abs/2016MNRAS.456.1654A}
}

@ARTICLE{Geha2010,
       author = {{Geha}, M. and {van der Marel}, R.~P. and {Guhathakurta}, P. and {Gilbert}, K.~M. and {Kalirai}, J. and {Kirby}, E.~N.},
        title = "{Local Group Dwarf Elliptical Galaxies. II. Stellar Kinematics to Large Radii in NGC 147 and NGC 185}",
      journal = {\apj},
         year = 2010,
        month = mar,
       volume = {711},
       number = {1},
        pages = {361-373},
          doi = {10.1088/0004-637X/711/1/361},
archivePrefix = {arXiv},
       eprint = {0911.3654},
 primaryClass = {astro-ph.CO},
       adsurl = {https://ui.adsabs.harvard.edu/abs/2010ApJ...711..361G}
}

@ARTICLE{Walter2002,
       author = {{Walter}, Fabian and {Weiss}, Axel and {Martin}, Crystal and {Scoville}, Nick},
        title = "{The Interacting Dwarf Galaxy NGC 3077: The Interplay of Atomic and Molecular Gas with Violent Star Formation}",
      journal = {\aj},
         year = 2002,
        month = jan,
       volume = {123},
       number = {1},
        pages = {225-237},
          doi = {10.1086/324633},
archivePrefix = {arXiv},
       eprint = {astro-ph/0110588},
 primaryClass = {astro-ph},
       adsurl = {https://ui.adsabs.harvard.edu/abs/2002AJ....123..225W}
}

@ARTICLE{Okamoto2023,
       author = {{Okamoto}, Sakurako and {Arimoto}, Nobuo and {Ferguson}, Annette M.~N. and {Irwin}, Mike J. and {{\v{Z}}emaitis}, Rokas},
        title = "{The Progenitor of the Peculiar Galaxy NGC 3077}",
      journal = {\apj},
         year = 2023,
        month = jul,
       volume = {952},
       number = {1},
          eid = {77},
        pages = {77},
          doi = {10.3847/1538-4357/acdad1},
archivePrefix = {arXiv},
       eprint = {2306.04102},
 primaryClass = {astro-ph.GA},
       adsurl = {https://ui.adsabs.harvard.edu/abs/2023ApJ...952...77O}
}

@ARTICLE{Walker2011,
       author = {{Walker}, Matthew G. and {Pe{\~n}arrubia}, Jorge},
        title = "{A Method for Measuring (Slopes of) the Mass Profiles of Dwarf Spheroidal Galaxies}",
      journal = {\apj},
         year = 2011,
        month = nov,
       volume = {742},
       number = {1},
          eid = {20},
        pages = {20},
          doi = {10.1088/0004-637X/742/1/20},
archivePrefix = {arXiv},
       eprint = {1108.2404},
 primaryClass = {astro-ph.CO},
       adsurl = {https://ui.adsabs.harvard.edu/abs/2011ApJ...742...20W}
}

@ARTICLE{yagi2016,
       author = {{Yagi}, Masafumi and {Koda}, Jin and {Komiyama}, Yutaka and {Yamanoi},
        Hitomo},
        title = "{Catalog of Ultra-diffuse Galaxies in the Coma Clusters from Subaru
        Imaging Data}",
      journal = {The Astrophysical Journal Supplement Series},
         year = 2016,
        month = Jul,
       volume = {225},
          eid = {11},
        pages = {11},
          doi = {10.3847/0067-0049/225/1/11},
       adsurl = {https://ui.adsabs.harvard.edu/#abs/2016ApJS..225...11Y}
}

@ARTICLE{merritt2016,
       author = {{Merritt}, Allison and {van Dokkum}, Pieter and {Danieli}, Shany and
        {Abraham}, Roberto and {Zhang}, Jielai and {Karachentsev}, I.~D.
        and {Makarova}, L.~N.},
        title = "{The Dragonfly Nearby Galaxies Survey. II. Ultra-Diffuse Galaxies near
        the Elliptical Galaxy NGC 5485}",
      journal = {\apj},
         year = 2016,
        month = Dec,
       volume = {833},
          eid = {168},
        pages = {168},
          doi = {10.3847/1538-4357/833/2/168},
       adsurl = {https://ui.adsabs.harvard.edu/#abs/2016ApJ...833..168M}
}

@ARTICLE{chan2018,
       author = {{Chan}, T.~K. and {Kere{\v{s}}}, D. and {Wetzel}, A. and {Hopkins},
        P.~F. and {Faucher-Gigu{\`e}re}, C. -A. and {El-Badry}, K. and
        {Garrison-Kimmel}, S. and {Boylan-Kolchin}, M.},
        title = "{The origin of ultra diffuse galaxies: stellar feedback and quenching}",
      journal = {\mnras},
         year = 2018,
        month = Jul,
       volume = {478},
        pages = {906-925},
          doi = {10.1093/mnras/sty1153},
       adsurl = {https://ui.adsabs.harvard.edu/#abs/2018MNRAS.478..906C}
}

@ARTICLE{toloba2018,
       author = {{Toloba}, Elisa and {Lim}, Sungsoon and {Peng}, Eric and {Sales}, Laura
        V. and {Guhathakurta}, Puragra and {Mihos}, J. Christopher and
        {C{\^o}t{\'e}}, Patrick and {Boselli}, Alessandro and
        {Cuillandre}, Jean-Charles and {Ferrarese}, Laura and {Gwyn},
        Stephen and {Lan{\c{c}}on}, Ariane and {Mu{\~n}oz}, Roberto and
        {Puzia}, Thomas},
        title = "{Dark Matter in Ultra-diffuse Galaxies in the Virgo Cluster from Their
        Globular Cluster Populations}",
      journal = {\apj},
         year = 2018,
        month = Apr,
       volume = {856},
          eid = {L31},
        pages = {L31},
          doi = {10.3847/2041-8213/aab603},
       adsurl = {https://ui.adsabs.harvard.edu/#abs/2018ApJ...856L..31T}
}

@ARTICLE{vandokkum2018,
       author = {{van Dokkum}, Pieter and {Danieli}, Shany and {Cohen}, Yotam and
        {Merritt}, Allison and {Romanowsky}, Aaron J. and {Abraham},
        Roberto and {Brodie}, Jean and {Conroy}, Charlie and {Lokhorst},
        Deborah and {Mowla}, Lamiya and {O'Sullivan}, Ewan and {Zhang},
        Jielai},
        title = "{A galaxy lacking dark matter}",
      journal = {Nature},
         year = 2018,
        month = Mar,
       volume = {555},
        pages = {629-632},
          doi = {10.1038/nature25767},
       adsurl = {https://ui.adsabs.harvard.edu/#abs/2018Natur.555..629V}
}

@ARTICLE{dalcanton1997,
       author = {{Dalcanton}, Julianne J. and {Spergel}, David N. and {Gunn}, James E.
        and {Schmidt}, Maarten and {Schneider}, Donald P.},
        title = "{The Number Density of Low-Surface Brightness Galaxies with 23 {\&}lt;
        mu{\_}0 {\&}lt; 25 V Mag/arcsec^2.}",
      journal = {\aj},
         year = 1997,
        month = Aug,
       volume = {114},
        pages = {635-654},
          doi = {10.1086/118499},
archivePrefix = {arXiv},
       eprint = {astro-ph/9705088},
       adsurl = {https://ui.adsabs.harvard.edu/#abs/1997AJ....114..635D}
}

@ARTICLE{koda2015,
       author = {{Koda}, Jin and {Yagi}, Masafumi and {Yamanoi}, Hitomi and {Komiyama},
        Yutaka},
        title = "{Approximately a Thousand Ultra-diffuse Galaxies in the Coma Cluster}",
      journal = {\apj},
         year = 2015,
        month = Jul,
       volume = {807},
          eid = {L2},
        pages = {L2},
          doi = {10.1088/2041-8205/807/1/L2},
       adsurl = {https://ui.adsabs.harvard.edu/#abs/2015ApJ...807L...2K}
}

@ARTICLE{mihos2015,
       author = {{Mihos}, J. Christopher and {Durrell}, Patrick R. and {Ferrarese}, Laura
        and {Feldmeier}, John J. and {C{\^o}t{\'e}}, Patrick and {Peng},
        Eric W. and {Harding}, Paul and {Liu}, Chengze and {Gwyn},
        Stephen and {Cuillandre}, Jean-Charles},
        title = "{Galaxies at the Extremes: Ultra-diffuse Galaxies in the Virgo Cluster}",
      journal = {\apj},
         year = 2015,
        month = Aug,
       volume = {809},
          eid = {L21},
        pages = {L21},
          doi = {10.1088/2041-8205/809/2/L21},
       adsurl = {https://ui.adsabs.harvard.edu/#abs/2015ApJ...809L..21M}
}

@ARTICLE{el-badry2016,
       author = {{El-Badry}, Kareem and {Wetzel}, Andrew and {Geha}, Marla and {Hopkins},
        Philip F. and {Kere{\v{s}}}, Dusan and {Chan}, T.~K. and
        {Faucher-Gigu{\`e}re}, Claude-Andr{\'e}},
        title = "{Breathing FIRE: How Stellar Feedback Drives Radial Migration, Rapid Size
        Fluctuations, and Population Gradients in Low-mass Galaxies}",
      journal = {\apj},
         year = 2016,
        month = Apr,
       volume = {820},
          eid = {131},
        pages = {131},
          doi = {10.3847/0004-637X/820/2/131},
       adsurl = {https://ui.adsabs.harvard.edu/#abs/2016ApJ...820..131E}
}

@ARTICLE{amorisco&loeb2016,
       author = {{Amorisco}, N.~C. and {Loeb}, A.},
        title = "{Ultradiffuse galaxies: the high-spin tail of the abundant dwarf galaxy
        population}",
      journal = {\mnras},
         year = 2016,
        month = Jun,
       volume = {459},
        pages = {L51-L55},
          doi = {10.1093/mnrasl/slw055},
       adsurl = {https://ui.adsabs.harvard.edu/#abs/2016MNRAS.459L..51A}
}

@ARTICLE{peng&lim2016,
       author = {{Peng}, Eric W. and {Lim}, Sungsoon},
        title = "{A Rich Globular Cluster System in Dragonfly 17: Are Ultra-diffuse
        Galaxies Pure Stellar Halos?}",
      journal = {\apj},
         year = 2016,
        month = May,
       volume = {822},
          eid = {L31},
        pages = {L31},
          doi = {10.3847/2041-8205/822/2/L31},
       adsurl = {https://ui.adsabs.harvard.edu/#abs/2016ApJ...822L..31P}
}

@ARTICLE{papastergis2017,
       author = {{Papastergis}, E. and {Adams}, E.~A.~K. and {Romanowsky}, A.~J.},
        title = "{The HI content of isolated ultra-diffuse galaxies: A sign of multiple
        formation mechanisms?}",
      journal = {\aap},
         year = 2017,
        month = May,
       volume = {601},
          eid = {L10},
        pages = {L10},
          doi = {10.1051/0004-6361/201730795},
       adsurl = {https://ui.adsabs.harvard.edu/#abs/2017A&A...601L..10P}
}

@ARTICLE{safarzadeh2017,
       author = {{Safarzadeh}, Mohammadtaher and {Scannapieco}, Evan},
        title = "{The Fate of Gas-rich Satellites in Clusters}",
      journal = {\apj},
         year = 2017,
        month = Nov,
       volume = {850},
          eid = {99},
        pages = {99},
          doi = {10.3847/1538-4357/aa94c8},
       adsurl = {https://ui.adsabs.harvard.edu/#abs/2017ApJ...850...99S}
}

@ARTICLE{vandokkum2015b,
       author = {{van Dokkum}, Pieter G. and {Romanowsky}, Aaron J. and {Abraham},
        Roberto and {Brodie}, Jean P. and {Conroy}, Charlie and {Geha},
        Marla and {Merritt}, Allison and {Villaume}, Alexa and {Zhang},
        Jielai},
        title = "{Spectroscopic Confirmation of the Existence of Large, Diffuse Galaxies
        in the Coma Cluster}",
      journal = {\apj},
         year = 2015,
        month = May,
       volume = {804},
          eid = {L26},
        pages = {L26},
          doi = {10.1088/2041-8205/804/1/L26},
       adsurl = {https://ui.adsabs.harvard.edu/#abs/2015ApJ...804L..26V}
}

@ARTICLE{dicintio2017,
       author = {{Di Cintio}, Arianna and {Brook}, Chris B. and {Dutton}, Aaron A. and
        {Macci{\`o}}, Andrea V. and {Obreja}, Aura and {Dekel}, Avishai},
        title = "{NIHAO - XI. Formation of ultra-diffuse galaxies by outflows}",
      journal = {\mnras},
         year = 2017,
        month = Mar,
       volume = {466},
        pages = {L1-L6},
          doi = {10.1093/mnrasl/slw210},
       adsurl = {https://ui.adsabs.harvard.edu/#abs/2017MNRAS.466L...1D}
}

@ARTICLE{burkert2017,
       author = {{Burkert}, A.},
        title = "{The Geometry and Origin of Ultra-diffuse Ghost Galaxies}",
      journal = {\apj},
         year = 2017,
        month = Apr,
       volume = {838},
          eid = {93},
        pages = {93},
          doi = {10.3847/1538-4357/aa671c},
       adsurl = {https://ui.adsabs.harvard.edu/#abs/2017ApJ...838...93B}
}

@ARTICLE{mcconnachie2008,
       author = {{McConnachie}, Alan W. and {Huxor}, Avon and {Martin}, Nicolas F. and
        {Irwin}, Mike J. and {Chapman}, Scott C. and {Fahlman}, Gregory
        and {Ferguson}, Annette M.~N. and {Ibata}, Rodrigo A. and
        {Lewis}, Geraint F. and {Richer}, Harvey and {Tanvir}, Nial R.},
        title = "{A Trio of New Local Group Galaxies with Extreme Properties}",
      journal = {\apj},
         year = 2008,
        month = Dec,
       volume = {688},
        pages = {1009-1020},
          doi = {10.1086/591313},
archivePrefix = {arXiv},
       eprint = {0806.3988},
       adsurl = {https://ui.adsabs.harvard.edu/#abs/2008ApJ...688.1009M}
}

@ARTICLE{yozin2015,
       author = {{Yozin}, C. and {Bekki}, K.},
        title = "{The quenching and survival of ultra diffuse galaxies in the Coma cluster}",
      journal = {\mnras},
         year = 2015,
        month = Sep,
       volume = {452},
        pages = {937-943},
          doi = {10.1093/mnras/stv1073},
       adsurl = {https://ui.adsabs.harvard.edu/#abs/2015MNRAS.452..937Y}
}

@ARTICLE{weisz2014,
       author = {{Weisz}, Daniel R. and {Dolphin}, Andrew E. and {Skillman}, Evan D. and
        {Holtzman}, Jon and {Gilbert}, Karoline M. and {Dalcanton},
        Julianne J. and {Williams}, Benjamin F.},
        title = "{The Star Formation Histories of Local Group Dwarf Galaxies. I. Hubble
        Space Telescope/Wide Field Planetary Camera 2 Observations}",
      journal = {\apj},
         year = 2014,
        month = Jul,
       volume = {789},
          eid = {147},
        pages = {147},
          doi = {10.1088/0004-637X/789/2/147},
archivePrefix = {arXiv},
       eprint = {1404.7144},
       adsurl = {https://ui.adsabs.harvard.edu/#abs/2014ApJ...789..147W}
}

@ARTICLE{ferre-mateu2018,
       author = {{Ferr{\'e}-Mateu}, Anna and {Alabi}, Adebusola and {Forbes}, Duncan A.
        and {Romanowsky}, Aaron J. and {Brodie}, Jean and {Pandya},
        Viraj and {Mart{\'\i}n-Navarro}, Ignacio and {Bellstedt}, Sabine
        and {Wasserman}, Asher and {Stone}, Maria B. and {Okabe},
        Nobuhiro},
        title = "{Origins of ultradiffuse galaxies in the Coma cluster - II. Constraints
        from their stellar populations}",
      journal = {\mnras},
         year = 2018,
        month = Oct,
       volume = {479},
        pages = {4891-4906},
          doi = {10.1093/mnras/sty1597},
       adsurl = {https://ui.adsabs.harvard.edu/#abs/2018MNRAS.479.4891F}
}

@ARTICLE{deblok&mcgaugh1997,
       author = {{de Blok}, W.~J.~G. and {McGaugh}, S.~S.},
        title = "{The dark and visible matter content of low surface brightness disc
        galaxies}",
      journal = {\mnras},
         year = 1997,
        month = Sep,
       volume = {290},
        pages = {533-552},
          doi = {10.1093/mnras/290.3.533},
archivePrefix = {arXiv},
       eprint = {astro-ph/9704274},
 primaryClass = {Astrophysics},
       adsurl = {https://ui.adsabs.harvard.edu/#abs/1997MNRAS.290..533D}
}

@ARTICLE{ibata1995,
       author = {{Ibata}, Rodrigo A. and {Gilmore}, Gerard and {Irwin}, Michael J.},
        title = "{Sagittarius: the nearest dwarf galaxy}",
      journal = {\mnras},
         year = 1995,
        month = Dec,
       volume = {277},
        pages = {781-800},
          doi = {10.1093/mnras/277.3.781},
archivePrefix = {arXiv},
       eprint = {astro-ph/9506071},
 primaryClass = {Astrophysics},
       adsurl = {https://ui.adsabs.harvard.edu/#abs/1995MNRAS.277..781I}
}

@ARTICLE{weisz2011,
       author = {{Weisz}, Daniel R. and {Dalcanton}, Julianne J. and {Williams}, Benjamin
        F. and {Gilbert}, Karoline M. and {Skillman}, Evan D. and
        {Seth}, Anil C. and {Dolphin}, Andrew E. and {McQuinn}, Kristen
        B.~W. and {Gogarten}, Stephanie M. and {Holtzman}, Jon and
        {Rosema}, Keith and {Cole}, Andrew and {Karachentsev}, Igor D.
        and {Zaritsky}, Dennis},
        title = "{The ACS Nearby Galaxy Survey Treasury. VIII. The Global Star Formation
        Histories of 60 Dwarf Galaxies in the Local Volume}",
      journal = {\apj},
         year = 2011,
        month = Sep,
       volume = {739},
          eid = {5},
        pages = {5},
          doi = {10.1088/0004-637X/739/1/5},
archivePrefix = {arXiv},
       eprint = {1101.1093},
 primaryClass = {Astrophysics - Cosmology and Nongalactic Astrophysics},
       adsurl = {https://ui.adsabs.harvard.edu/#abs/2011ApJ...739....5W}
}

@ARTICLE{dolphin2002,
       author = {{Dolphin}, A.~E.},
        title = "{Numerical methods of star formation history measurement and applications
        to seven dwarf spheroidals}",
      journal = {\mnras},
         year = 2002,
        month = May,
       volume = {332},
        pages = {91-108},
          doi = {10.1046/j.1365-8711.2002.05271.x},
archivePrefix = {arXiv},
       eprint = {astro-ph/0112331},
 primaryClass = {Astrophysics},
       adsurl = {https://ui.adsabs.harvard.edu/#abs/2002MNRAS.332...91D}
}

@ARTICLE{toloba2016b,
       author = {{Toloba}, Elisa and {Sand}, David J. and {Spekkens}, Kristine and
        {Crnojevi{\'c}}, Denija and {Simon}, Joshua D. and
        {Guhathakurta}, Puragra and {Strader}, Jay and {Caldwell},
        Nelson and {McLeod}, Brian and {Seth}, Anil C.},
        title = "{A Tidally Disrupting Dwarf Galaxy in the Halo of NGC 253}",
      journal = {\apj},
         year = 2016,
        month = Jan,
       volume = {816},
          eid = {L5},
        pages = {L5},
          doi = {10.3847/2041-8205/816/1/L5},
 primaryClass = {Astrophysics - Astrophysics of Galaxies},
       adsurl = {https://ui.adsabs.harvard.edu/#abs/2016ApJ...816L...5T}
}

@ARTICLE{harmsen2017,
       author = {{Harmsen}, Benjamin and {Monachesi}, Antonela and {Bell}, Eric F. and
        {de Jong}, Roelof S. and {Bailin}, Jeremy and {Radburn-Smith},
        David J. and {Holwerda}, Benne W.},
        title = "{Diverse stellar haloes in nearby Milky Way mass disc galaxies}",
      journal = {\mnras},
         year = 2017,
        month = Apr,
       volume = {466},
        pages = {1491-1512},
          doi = {10.1093/mnras/stw2992},
 primaryClass = {Astrophysics - Astrophysics of Galaxies},
       adsurl = {https://ui.adsabs.harvard.edu/#abs/2017MNRAS.466.1491H}
}

@ARTICLE{torrealba2019,
       author = {{Torrealba}, G. and {Belokurov}, V. and {Koposov}, S.~E. and
         {Li}, T.~S. and {Walker}, M.~G. and {Sanders}, J.~L. and
         {Geringer-Sameth}, A. and {Zucker}, D.~B. and {Kuehn}, K. and
         {Evans}, N.~W. and {Dehnen}, W.},
        title = "{The hidden giant: discovery of an enormous Galactic dwarf satellite in Gaia DR2}",
      journal = {\mnras},
         year = 2019,
        month = sep,
       volume = {488},
       number = {2},
        pages = {2743-2766},
          doi = {10.1093/mnras/stz1624},
archivePrefix = {arXiv},
       eprint = {1811.04082},
 primaryClass = {astro-ph.GA},
       adsurl = {https://ui.adsabs.harvard.edu/abs/2019MNRAS.488.2743T}
}

@ARTICLE{liao2019,
       author = {{Liao}, Shihong and {Gao}, Liang and {Frenk}, Carlos S. and {Grand
        }, Robert J.~J. and {Guo}, Qi and {G{\'o}mez}, Facundo A. and
         {Marinacci}, Federico and {Pakmor}, R{\"u}diger and {Shao}, Shi and
         {Springel}, Volker},
        title = "{Ultra-diffuse galaxies in the Auriga simulations}",
      journal = {\mnras},
         year = 2019,
        month = dec,
       volume = {490},
       number = {4},
        pages = {5182-5195},
          doi = {10.1093/mnras/stz2969},
archivePrefix = {arXiv},
       eprint = {1904.06356},
 primaryClass = {astro-ph.GA},
       adsurl = {https://ui.adsabs.harvard.edu/abs/2019MNRAS.490.5182L}
}

@ARTICLE{barbosa2020,
       author = {{Barbosa}, C.~E. and {Zaritsky}, D. and {Donnerstein}, R. and
         {Zhang}, H. and {Dey}, A. and {Mendes de Oliveira}, C. and
         {Sampedro}, L. and {Molino}, A. and {Costa-Duarte}, M.~V. and
         {Coelho}, P. and {Cortesi}, A. and {Herpich}, F.~R. and {Hernand
        ez-Jimenez}, J.~A. and {Santos-Silva}, T. and {Pereira}, E. and
         {Werle}, A. and {Overzier}, R.~A. and {Cid Fernandes}, R. and
         {Smith Castelli}, A.~V. and {Ribeiro}, T. and {Schoenell}, W. and
         {Kanaan}, A.},
        title = "{One hundred SMUDGes in S-PLUS: ultra-diffuse galaxies flourish in the field}",
      journal = {arXiv e-prints},
         year = 2020,
        month = feb,
          eid = {arXiv:2002.05171},
        pages = {arXiv:2002.05171},
archivePrefix = {arXiv},
       eprint = {2002.05171},
 primaryClass = {astro-ph.GA},
       adsurl = {https://ui.adsabs.harvard.edu/abs/2020arXiv200205171B}
}

@ARTICLE{vandokkum2019,
       author = {{van Dokkum}, Pieter and {Danieli}, Shany and {Abraham}, Roberto and
         {Conroy}, Charlie and {Romanowsky}, Aaron J.},
        title = "{A Second Galaxy Missing Dark Matter in the NGC 1052 Group}",
      journal = {\apjl},
         year = 2019,
        month = mar,
       volume = {874},
       number = {1},
          eid = {L5},
        pages = {L5},
          doi = {10.3847/2041-8213/ab0d92},
archivePrefix = {arXiv},
       eprint = {1901.05973},
 primaryClass = {astro-ph.GA},
       adsurl = {https://ui.adsabs.harvard.edu/abs/2019ApJ...874L...5V}
}

@book{numpy-guide,
 title={A guide to NumPy},
 author={Oliphant, Travis E},
 volume={1},
 year={2006},
 publisher={Trelgol Publishing USA}
}

@article{numpy,
  title={The NumPy array: a structure for efficient numerical computation},
  author={Van Der Walt, Stefan and Colbert, S Chris and Varoquaux, Gael},
  journal={Computing in Science \& Engineering},
  volume={13},
  number={2},
  pages={22},
  year={2011},
  publisher={IEEE Computer Society}
}

@ARTICLE{matplotlib,
       author = {{Hunter}, John D.},
        title = "{Matplotlib: A 2D Graphics Environment}",
      journal = {Computing in Science and Engineering},
         year = 2007,
        month = may,
       volume = {9},
       number = {3},
        pages = {90-95},
          doi = {10.1109/MCSE.2007.55},
       adsurl = {https://ui.adsabs.harvard.edu/abs/2007CSE.....9...90H}
}

@ARTICLE{scipy,
       author = {{Virtanen}, Pauli and {Gommers}, Ralf and {Oliphant}, Travis E. and
         {Haberland}, Matt and {Reddy}, Tyler and {Cournapeau}, David and
         {Burovski}, Evgeni and {Peterson}, Pearu and {Weckesser}, Warren and
         {Bright}, Jonathan and {van der Walt}, St{\'e}fan J. and
         {Brett}, Matthew and {Wilson}, Joshua and {Millman}, K. Jarrod and
         {Mayorov}, Nikolay and {Nelson}, Andrew R.~J. and {Jones}, Eric and
         {Kern}, Robert and {Larson}, Eric and {Carey}, C.~J. and
         {Polat}, {\.I}lhan and {Feng}, Yu and {Moore}, Eric W. and {Vand
        erPlas}, Jake and {Laxalde}, Denis and {Perktold}, Josef and
         {Cimrman}, Robert and {Henriksen}, Ian and {Quintero}, E.~A. and
         {Harris}, Charles R. and {Archibald}, Anne M. and
         {Ribeiro}, Ant{\^o}nio H. and {Pedregosa}, Fabian and
         {van Mulbregt}, Paul and {SciPy 1. 0 Contributors}},
        title = "{SciPy 1.0: fundamental algorithms for scientific computing in Python}",
      journal = {Nature Methods},
         year = 2020,
        month = feb,
       volume = {17},
        pages = {261-272},
          doi = {10.1038/s41592-019-0686-2},
archivePrefix = {arXiv},
       eprint = {1907.10121},
 primaryClass = {cs.MS},
       adsurl = {https://ui.adsabs.harvard.edu/abs/2020NatMe..17..261V}
}

@MISC{ds9,
       author = {{Smithsonian Astrophysical Observatory}},
        title = "{SAOImage DS9: A utility for displaying astronomical images in the X11 window environment}",
         year = 2000,
        month = mar,
          eid = {ascl:0003.002},
        pages = {ascl:0003.002},
archivePrefix = {ascl},
       eprint = {0003.002},
       adsurl = {https://ui.adsabs.harvard.edu/abs/2000ascl.soft03002S}
}

@ARTICLE{astropy,
       author = {{Astropy Collaboration} and {Price-Whelan}, A.~M. and
         {Sip{\H{o}}cz}, B.~M. and {G{\"u}nther}, H.~M. and {Lim}, P.~L. and
         {Crawford}, S.~M. and {Conseil}, S. and {Shupe}, D.~L. and
         {Craig}, M.~W. and {Dencheva}, N. and {Ginsburg}, A. and {Vand
        erPlas}, J.~T. and {Bradley}, L.~D. and {P{\'e}rez-Su{\'a}rez}, D. and
         {de Val-Borro}, M. and {Aldcroft}, T.~L. and {Cruz}, K.~L. and
         {Robitaille}, T.~P. and {Tollerud}, E.~J. and {Ardelean}, C. and
         {Babej}, T. and {Bach}, Y.~P. and {Bachetti}, M. and {Bakanov}, A.~V. and
         {Bamford}, S.~P. and {Barentsen}, G. and {Barmby}, P. and
         {Baumbach}, A. and {Berry}, K.~L. and {Biscani}, F. and {Boquien}, M. and
         {Bostroem}, K.~A. and {Bouma}, L.~G. and {Brammer}, G.~B. and
         {Bray}, E.~M. and {Breytenbach}, H. and {Buddelmeijer}, H. and
         {Burke}, D.~J. and {Calderone}, G. and {Cano Rodr{\'\i}guez}, J.~L. and
         {Cara}, M. and {Cardoso}, J.~V.~M. and {Cheedella}, S. and {Copin}, Y. and
         {Corrales}, L. and {Crichton}, D. and {D'Avella}, D. and {Deil}, C. and
         {Depagne}, {\'E}. and {Dietrich}, J.~P. and {Donath}, A. and
         {Droettboom}, M. and {Earl}, N. and {Erben}, T. and {Fabbro}, S. and
         {Ferreira}, L.~A. and {Finethy}, T. and {Fox}, R.~T. and
         {Garrison}, L.~H. and {Gibbons}, S.~L.~J. and {Goldstein}, D.~A. and
         {Gommers}, R. and {Greco}, J.~P. and {Greenfield}, P. and
         {Groener}, A.~M. and {Grollier}, F. and {Hagen}, A. and {Hirst}, P. and
         {Homeier}, D. and {Horton}, A.~J. and {Hosseinzadeh}, G. and {Hu}, L. and
         {Hunkeler}, J.~S. and {Ivezi{\'c}}, {\v{Z}}. and {Jain}, A. and
         {Jenness}, T. and {Kanarek}, G. and {Kendrew}, S. and {Kern}, N.~S. and
         {Kerzendorf}, W.~E. and {Khvalko}, A. and {King}, J. and {Kirkby}, D. and
         {Kulkarni}, A.~M. and {Kumar}, A. and {Lee}, A. and {Lenz}, D. and
         {Littlefair}, S.~P. and {Ma}, Z. and {Macleod}, D.~M. and
         {Mastropietro}, M. and {McCully}, C. and {Montagnac}, S. and
         {Morris}, B.~M. and {Mueller}, M. and {Mumford}, S.~J. and {Muna}, D. and
         {Murphy}, N.~A. and {Nelson}, S. and {Nguyen}, G.~H. and
         {Ninan}, J.~P. and {N{\"o}the}, M. and {Ogaz}, S. and {Oh}, S. and
         {Parejko}, J.~K. and {Parley}, N. and {Pascual}, S. and {Patil}, R. and
         {Patil}, A.~A. and {Plunkett}, A.~L. and {Prochaska}, J.~X. and
         {Rastogi}, T. and {Reddy Janga}, V. and {Sabater}, J. and
         {Sakurikar}, P. and {Seifert}, M. and {Sherbert}, L.~E. and
         {Sherwood-Taylor}, H. and {Shih}, A.~Y. and {Sick}, J. and
         {Silbiger}, M.~T. and {Singanamalla}, S. and {Singer}, L.~P. and
         {Sladen}, P.~H. and {Sooley}, K.~A. and {Sornarajah}, S. and
         {Streicher}, O. and {Teuben}, P. and {Thomas}, S.~W. and
         {Tremblay}, G.~R. and {Turner}, J.~E.~H. and {Terr{\'o}n}, V. and
         {van Kerkwijk}, M.~H. and {de la Vega}, A. and {Watkins}, L.~L. and
         {Weaver}, B.~A. and {Whitmore}, J.~B. and {Woillez}, J. and
         {Zabalza}, V. and {Astropy Contributors}},
        title = "{The Astropy Project: Building an Open-science Project and Status of the v2.0 Core Package}",
      journal = {\aj},
         year = 2018,
        month = sep,
       volume = {156},
       number = {3},
          eid = {123},
        pages = {123},
          doi = {10.3847/1538-3881/aabc4f},
archivePrefix = {arXiv},
       eprint = {1801.02634},
 primaryClass = {astro-ph.IM},
       adsurl = {https://ui.adsabs.harvard.edu/abs/2018AJ....156..123A}
}

@ARTICLE{alves&sarajedini1999,
       author = {{Alves}, David R. and {Sarajedini}, Ata},
        title = "{The Age-dependent Luminosities of the Red Giant Branch Bump, Asymptotic Giant Branch Bump, and Horizontal Branch Red Clump}",
      journal = {\apj},
         year = 1999,
        month = jan,
       volume = {511},
       number = {1},
        pages = {225-234},
          doi = {10.1086/306655},
archivePrefix = {arXiv},
       eprint = {astro-ph/9808253},
 primaryClass = {astro-ph},
       adsurl = {https://ui.adsabs.harvard.edu/abs/1999ApJ...511..225A}
}

@ARTICLE{girardi2016,
       author = {{Girardi}, L{\'e}o},
        title = "{Red Clump Stars}",
      journal = {\araa},
         year = 2016,
        month = sep,
       volume = {54},
        pages = {95-133},
          doi = {10.1146/annurev-astro-081915-023354},
       adsurl = {https://ui.adsabs.harvard.edu/abs/2016ARA&A..54...95G}
}

@ARTICLE{dolphin2000,
       author = {{Dolphin}, Andrew E.},
        title = "{WFPC2 Stellar Photometry with HSTPHOT}",
      journal = {\pasp},
         year = 2000,
        month = oct,
       volume = {112},
       number = {776},
        pages = {1383-1396},
          doi = {10.1086/316630},
archivePrefix = {arXiv},
       eprint = {astro-ph/0006217},
 primaryClass = {astro-ph},
       adsurl = {https://ui.adsabs.harvard.edu/abs/2000PASP..112.1383D}
}

@ARTICLE{williams2014,
       author = {{Williams}, Benjamin F. and {Lang}, Dustin and {Dalcanton}, Julianne J. and {Dolphin}, Andrew E. and {Weisz}, Daniel R. and {Bell}, Eric F. and {Bianchi}, Luciana and {Byler}, Nell and {Gilbert}, Karoline M. and {Girardi}, L{\'e}o and {Gordon}, Karl and {Gregersen}, Dylan and {Johnson}, L.~C. and {Kalirai}, Jason and {Lauer}, Tod R. and {Monachesi}, Antonela and {Rosenfield}, Philip and {Seth}, Anil and {Skillman}, Eva},
        title = "{The Panchromatic Hubble Andromeda Treasury. X. Ultraviolet to Infrared Photometry of 117 Million Equidistant Stars}",
      journal = {\apjs},
         year = 2014,
        month = nov,
       volume = {215},
       number = {1},
          eid = {9},
        pages = {9},
          doi = {10.1088/0067-0049/215/1/9},
archivePrefix = {arXiv},
       eprint = {1409.0899},
 primaryClass = {astro-ph.GA},
       adsurl = {https://ui.adsabs.harvard.edu/abs/2014ApJS..215....9W}
}

@ARTICLE{gordon2016,
       author = {{Gordon}, Karl D. and {Fouesneau}, Morgan and {Arab}, Heddy and {Tchernyshyov}, Kirill and {Weisz}, Daniel R. and {Dalcanton}, Julianne J. and {Williams}, Benjamin F. and {Bell}, Eric F. and {Bianchi}, Luciana and {Boyer}, Martha and {Choi}, Yumi and {Dolphin}, Andrew and {Girardi}, L{\'e}o and {Hogg}, David W. and {Kalirai}, Jason S. and {Kapala}, Maria and {Lewis}, Alexia R. and {Rix}, Hans-Walter and {Sandstrom}, Karin and {Skillman}, Evan D.},
        title = "{The Panchromatic Hubble Andromeda Treasury. XV. The BEAST: Bayesian Extinction and Stellar Tool}",
      journal = {\apj},
         year = 2016,
        month = aug,
       volume = {826},
       number = {2},
          eid = {104},
        pages = {104},
          doi = {10.3847/0004-637X/826/2/104},
archivePrefix = {arXiv},
       eprint = {1606.06182},
 primaryClass = {astro-ph.GA},
       adsurl = {https://ui.adsabs.harvard.edu/abs/2016ApJ...826..104G}
}

@ARTICLE{danieli2019,
       author = {{Danieli}, Shany and {van Dokkum}, Pieter and {Conroy}, Charlie and {Abraham}, Roberto and {Romanowsky}, Aaron J.},
        title = "{Still Missing Dark Matter: KCWI High-resolution Stellar Kinematics of NGC1052-DF2}",
      journal = {\apjl},
         year = 2019,
        month = apr,
       volume = {874},
       number = {2},
          eid = {L12},
        pages = {L12},
          doi = {10.3847/2041-8213/ab0e8c},
archivePrefix = {arXiv},
       eprint = {1901.03711},
 primaryClass = {astro-ph.GA},
       adsurl = {https://ui.adsabs.harvard.edu/abs/2019ApJ...874L..12D}
}

@ARTICLE{weisz2012,
       author = {{Weisz}, Daniel R. and {Zucker}, Daniel B. and {Dolphin}, Andrew E. and {Martin}, Nicolas F. and {de Jong}, Jelte T.~A. and {Holtzman}, Jon A. and {Dalcanton}, Julianne J. and {Gilbert}, Karoline M. and {Williams}, Benjamin F. and {Bell}, Eric F. and {Belokurov}, Vasily and {Evans}, N. Wyn},
        title = "{The Star Formation History of Leo T from Hubble Space Telescope Imaging}",
      journal = {\apj},
         year = 2012,
        month = apr,
       volume = {748},
       number = {2},
          eid = {88},
        pages = {88},
          doi = {10.1088/0004-637X/748/2/88},
archivePrefix = {arXiv},
       eprint = {1201.4859},
 primaryClass = {astro-ph.CO},
       adsurl = {https://ui.adsabs.harvard.edu/abs/2012ApJ...748...88W}
}

@ARTICLE{weisz2008,
       author = {{Weisz}, Daniel R. and {Skillman}, Evan D. and {Cannon}, John M. and {Dolphin}, Andrew E. and {Kennicutt}, Robert C., Jr. and {Lee}, Janice and {Walter}, Fabian},
        title = "{The Recent Star Formation Histories of M81 Group Dwarf Irregular Galaxies}",
      journal = {\apj},
         year = 2008,
        month = dec,
       volume = {689},
       number = {1},
        pages = {160-183},
          doi = {10.1086/592323},
archivePrefix = {arXiv},
       eprint = {0809.5059},
 primaryClass = {astro-ph},
       adsurl = {https://ui.adsabs.harvard.edu/abs/2008ApJ...689..160W}
}

@ARTICLE{albers2019,
       author = {{Albers}, Saundra M. and {Weisz}, Daniel R. and {Cole}, Andrew A. and {Dolphin}, Andrew E. and {Skillman}, Evan D. and {Williams}, Benjamin F. and {Boylan-Kolchin}, Michael and {Bullock}, James S. and {Dalcanton}, Julianne J. and {Hopkins}, Philip F. and {Leaman}, Ryan and {McConnachie}, Alan W. and {Vogelsberger}, Mark and {Wetzel}, Andrew},
        title = "{Star formation at the edge of the Local Group: a rising star formation history in the isolated galaxy WLM}",
      journal = {\mnras},
         year = 2019,
        month = dec,
       volume = {490},
       number = {4},
        pages = {5538-5550},
          doi = {10.1093/mnras/stz2903},
archivePrefix = {arXiv},
       eprint = {1909.04040},
 primaryClass = {astro-ph.GA},
       adsurl = {https://ui.adsabs.harvard.edu/abs/2019MNRAS.490.5538A}
}

@ARTICLE{rusakov2021,
       author = {{Rusakov}, V. and {Monelli}, M. and {Gallart}, C. and {Fritz}, T.~K. and {Ruiz-Lara}, T. and {Bernard}, E.~J. and {Cassisi}, S.},
        title = "{The bursty star formation history of the Fornax dwarf spheroidal galaxy revealed with the HST}",
      journal = {\mnras},
         year = 2021,
        month = mar,
       volume = {502},
       number = {1},
        pages = {642-661},
          doi = {10.1093/mnras/stab006},
archivePrefix = {arXiv},
       eprint = {2002.09714},
 primaryClass = {astro-ph.GA},
       adsurl = {https://ui.adsabs.harvard.edu/abs/2021MNRAS.502..642R}
}

@ARTICLE{gallart2021,
       author = {{Gallart}, C. and {Monelli}, M. and {Ruiz-Lara}, T. and {Calamida}, A. and {Cassisi}, S. and {Cignoni}, M. and {Anderson}, J. and {Battaglia}, G. and {Bermejo-Climent}, J.~R. and {Bernard}, E.~J. and {Mart{\'\i}nez-V{\'a}zquez}, C.~E. and {Mayer}, L. and {Salvadori}, S. and {Monachesi}, A. and {Navarro}, J.~F. and {Shen}, S. and {Surot}, F. and {Tosi}, M. and {Bajaj}, V. and {Strinfellow}, G.~S.},
        title = "{The Star Formation History of Eridanus II: On the Role of Supernova Feedback in the Quenching of Ultrafaint Dwarf Galaxies}",
      journal = {\apj},
         year = 2021,
        month = mar,
       volume = {909},
       number = {2},
          eid = {192},
        pages = {192},
          doi = {10.3847/1538-4357/abddbe},
archivePrefix = {arXiv},
       eprint = {2101.04464},
 primaryClass = {astro-ph.GA},
       adsurl = {https://ui.adsabs.harvard.edu/abs/2021ApJ...909..192G}
}

@ARTICLE{collins2022,
       author = {{Collins}, Michelle L.~M. and {Williams}, Benjamin F. and {Tollerud}, Erik J. and {Balbinot}, Eduardo and {Gilbert}, Karoline M. and {Dolphin}, Andrew},
        title = "{A detailed star formation history for the extremely diffuse Andromeda XIX dwarf galaxy}",
      journal = {\mnras},
         year = 2022,
        month = dec,
       volume = {517},
       number = {3},
        pages = {4382-4388},
          doi = {10.1093/mnras/stac2794},
archivePrefix = {arXiv},
       eprint = {2209.12912},
 primaryClass = {astro-ph.GA},
       adsurl = {https://ui.adsabs.harvard.edu/abs/2022MNRAS.517.4382C}
}

@ARTICLE{mcquinn2023,
       author = {{McQuinn}, Kristen B.~W. and {Mao}, Yao-Yuan and {Buckley}, Matthew R. and {Shih}, David and {Cohen}, Roger E. and {Dolphin}, Andrew E.},
        title = "{Pegasus W: An Ultrafaint Dwarf Galaxy Outside the Halo of M31 Not Quenched by Reionization}",
      journal = {\apj},
         year = 2023,
        month = feb,
       volume = {944},
       number = {1},
          eid = {14},
        pages = {14},
          doi = {10.3847/1538-4357/acaec9},
archivePrefix = {arXiv},
       eprint = {2301.04157},
 primaryClass = {astro-ph.GA},
       adsurl = {https://ui.adsabs.harvard.edu/abs/2023ApJ...944...14M}
}

@ARTICLE{savino2023,
       author = {{Savino}, A. and {Weisz}, D.~R. and {Skillman}, E.~D. and {Dolphin}, A. and {Cole}, A.~A. and {Kallivayalil}, N. and {Wetzel}, A. and {Anderson}, J. and {Besla}, G. and {Boylan-Kolchin}, M. and {Brown}, T.~M. and {Bullock}, J.~S. and {Collins}, M.~L.~M. and {Cooper}, M.~C. and {Deason}, A.~J. and {Dotter}, A.~L. and {Fardal}, M. and {Ferguson}, A.~M.~N. and {Fritz}, T.~K. and {Geha}, M.~C. and {Gilbert}, K.~M. and {Guhathakurta}, P. and {Ibata}, R. and {Irwin}, M.~J. and {Jeon}, M. and {Kirby}, E.~N. and {Lewis}, G.~F. and {Mackey}, D. and {Majewski}, S.~R. and {Martin}, N. and {McConnachie}, A. and {Patel}, E. and {Rich}, R.~M. and {Simon}, J.~D. and {Sohn}, S.~T. and {Tollerud}, E.~J. and {van der Marel}, R.~P.},
        title = "{The Hubble Space Telescope Survey of M31 Satellite Galaxies II. The Star Formation Histories of Ultra-Faint Dwarf Galaxies}",
      journal = {arXiv e-prints},
         year = 2023,
        month = may,
          eid = {arXiv:2305.13360},
        pages = {arXiv:2305.13360},
          doi = {10.48550/arXiv.2305.13360},
archivePrefix = {arXiv},
       eprint = {2305.13360},
 primaryClass = {astro-ph.GA},
       adsurl = {https://ui.adsabs.harvard.edu/abs/2023arXiv230513360S}
}

@ARTICLE{mcquinn2015,
       author = {{McQuinn}, Kristen B.~W. and {Skillman}, Evan D. and {Dolphin}, Andrew and {Cannon}, John M. and {Salzer}, John J. and {Rhode}, Katherine L. and {Adams}, Elizabeth A.~K. and {Berg}, Danielle and {Giovanelli}, Riccardo and {Girardi}, L{\'e}o and {Haynes}, Martha P.},
        title = "{Leo P: An Unquenched Very Low-mass Galaxy}",
      journal = {\apj},
         year = 2015,
        month = oct,
       volume = {812},
       number = {2},
          eid = {158},
        pages = {158},
          doi = {10.1088/0004-637X/812/2/158},
archivePrefix = {arXiv},
       eprint = {1506.05495},
 primaryClass = {astro-ph.GA},
       adsurl = {https://ui.adsabs.harvard.edu/abs/2015ApJ...812..158M}
}

@ARTICLE{cioni2006,
       author = {{Cioni}, M. -R.~L. and {Girardi}, L. and {Marigo}, P. and {Habing}, H.~J.},
        title = "{AGB stars in the Magellanic Clouds. II. The rate of star formation across the LMC}",
      journal = {\aap},
         year = 2006,
        month = mar,
       volume = {448},
       number = {1},
        pages = {77-91},
          doi = {10.1051/0004-6361:20053933},
archivePrefix = {arXiv},
       eprint = {astro-ph/0509881},
 primaryClass = {astro-ph},
       adsurl = {https://ui.adsabs.harvard.edu/abs/2006A&A...448...77C}
}

@ARTICLE{girardi2000,
       author = {{Girardi}, L. and {Bressan}, A. and {Bertelli}, G. and {Chiosi}, C.},
        title = "{Evolutionary tracks and isochrones for low- and intermediate-mass stars: From 0.15 to 7 M$_{sun}$, and from Z=0.0004 to 0.03}",
      journal = {\aaps},
         year = 2000,
        month = feb,
       volume = {141},
        pages = {371-383},
          doi = {10.1051/aas:2000126},
archivePrefix = {arXiv},
       eprint = {astro-ph/9910164},
 primaryClass = {astro-ph},
       adsurl = {https://ui.adsabs.harvard.edu/abs/2000A&AS..141..371G}
}

@ARTICLE{marigo2008,
       author = {{Marigo}, P. and {Girardi}, L. and {Bressan}, A. and {Groenewegen}, M.~A.~T. and {Silva}, L. and {Granato}, G.~L.},
        title = "{Evolution of asymptotic giant branch stars. II. Optical to far-infrared isochrones with improved TP-AGB models}",
      journal = {\aap},
         year = 2008,
        month = may,
       volume = {482},
       number = {3},
        pages = {883-905},
          doi = {10.1051/0004-6361:20078467},
archivePrefix = {arXiv},
       eprint = {0711.4922},
 primaryClass = {astro-ph},
       adsurl = {https://ui.adsabs.harvard.edu/abs/2008A&A...482..883M}
}

@ARTICLE{girardi2010,
       author = {{Girardi}, L{\'e}o and {Williams}, Benjamin F. and {Gilbert}, Karoline M. and {Rosenfield}, Philip and {Dalcanton}, Julianne J. and {Marigo}, Paola and {Boyer}, Martha L. and {Dolphin}, Andrew and {Weisz}, Daniel R. and {Melbourne}, Jason and {Olsen}, Knut A.~G. and {Seth}, Anil C. and {Skillman}, Evan},
        title = "{The ACS Nearby Galaxy Survey Treasury. IX. Constraining Asymptotic Giant Branch Evolution with Old Metal-poor Galaxies}",
      journal = {\apj},
         year = 2010,
        month = dec,
       volume = {724},
       number = {2},
        pages = {1030-1043},
          doi = {10.1088/0004-637X/724/2/1030},
archivePrefix = {arXiv},
       eprint = {1009.4618},
 primaryClass = {astro-ph.SR},
       adsurl = {https://ui.adsabs.harvard.edu/abs/2010ApJ...724.1030G}
}

@ARTICLE{harmsen2023,
       author = {{Harmsen}, Benjamin and {Bell}, Eric F. and {D'Souza}, Richard and {Monachesi}, Antonela and {de Jong}, Roelof S. and {Smercina}, Adam and {Jang}, In Sung and {Holwerda}, Benne W.},
        title = "{Constraining the assembly time of the stellar haloes of nearby Milky Way-mass galaxies through AGB populations}",
      journal = {\mnras},
         year = 2023,
        month = nov,
       volume = {525},
       number = {3},
        pages = {4497-4514},
          doi = {10.1093/mnras/stad2480},
archivePrefix = {arXiv},
       eprint = {2308.11499},
 primaryClass = {astro-ph.GA},
       adsurl = {https://ui.adsabs.harvard.edu/abs/2023MNRAS.525.4497H}
}

@ARTICLE{jones2023,
       author = {{Jones}, Michael G. and {Karunakaran}, Ananthan and {Bennet}, Paul and {Sand}, David J. and {Spekkens}, Kristine and {Mutlu-Pakdil}, Bur{\c{c}}in and {Crnojevi{\'c}}, Denija and {Janowiecki}, Steven and {Leisman}, Lukas and {Fielder}, Catherine E.},
        title = "{Gas-rich, Field Ultra-diffuse Galaxies Host Few Globular Clusters}",
      journal = {\apjl},
         year = 2023,
        month = jan,
       volume = {942},
       number = {1},
          eid = {L5},
        pages = {L5},
          doi = {10.3847/2041-8213/acaaab},
archivePrefix = {arXiv},
       eprint = {2211.00651},
 primaryClass = {astro-ph.GA},
       adsurl = {https://ui.adsabs.harvard.edu/abs/2023ApJ...942L...5J}
}

@ARTICLE{mancera-pina2019,
       author = {{Mancera Pi{\~n}a}, Pavel E. and {Fraternali}, Filippo and {Adams}, Elizabeth A.~K. and {Marasco}, Antonino and {Oosterloo}, Tom and {Oman}, Kyle A. and {Leisman}, Lukas and {di Teodoro}, Enrico M. and {Posti}, Lorenzo and {Battipaglia}, Michael and {Cannon}, John M. and {Gault}, Lexi and {Haynes}, Martha P. and {Janowiecki}, Steven and {McAllan}, Elizabeth and {Pagel}, Hannah J. and {Reiter}, Kameron and {Rhode}, Katherine L. and {Salzer}, John J. and {Smith}, Nicholas J.},
        title = "{Off the Baryonic Tully-Fisher Relation: A Population of Baryon-dominated Ultra-diffuse Galaxies}",
      journal = {\apjl},
         year = 2019,
        month = oct,
       volume = {883},
       number = {2},
          eid = {L33},
        pages = {L33},
          doi = {10.3847/2041-8213/ab40c7},
archivePrefix = {arXiv},
       eprint = {1909.01363},
 primaryClass = {astro-ph.GA},
       adsurl = {https://ui.adsabs.harvard.edu/abs/2019ApJ...883L..33M}
}

@ARTICLE{ferre-mateu2023,
       author = {{Ferr{\'e}-Mateu}, Anna and {Gannon}, Jonah S. and {Forbes}, Duncan A. and {Buzzo}, Maria Luisa and {Romanowsky}, Aaron J. and {Brodie}, Jean P.},
        title = "{The star formation histories of quiescent ultra-diffuse galaxies and their dependence on environment and globular cluster richness}",
      journal = {\mnras},
         year = 2023,
        month = dec,
       volume = {526},
       number = {3},
        pages = {4735-4754},
          doi = {10.1093/mnras/stad3102},
archivePrefix = {arXiv},
       eprint = {2309.15148},
 primaryClass = {astro-ph.GA},
       adsurl = {https://ui.adsabs.harvard.edu/abs/2023MNRAS.526.4735F}
}

@ARTICLE{vannest2022,
       author = {{Van Nest}, Jordan D. and {Munshi}, F. and {Wright}, A.~C. and {Tremmel}, M. and {Brooks}, A.~M. and {Nagai}, D. and {Quinn}, T.},
        title = "{What's in a Name? Quantifying the Interplay between the Definition, Orientation, and Shape of Ultra-diffuse Galaxies Using the Romulus Simulations}",
      journal = {\apj},
         year = 2022,
        month = feb,
       volume = {926},
       number = {1},
          eid = {92},
        pages = {92},
          doi = {10.3847/1538-4357/ac43b7},
archivePrefix = {arXiv},
       eprint = {2108.12985},
 primaryClass = {astro-ph.GA},
       adsurl = {https://ui.adsabs.harvard.edu/abs/2022ApJ...926...92V}
}

@ARTICLE{mendez-abreu2017,
       author = {{M{\'e}ndez-Abreu}, J. and {Ruiz-Lara}, T. and {S{\'a}nchez-Menguiano}, L. and {de Lorenzo-C{\'a}ceres}, A. and {Costantin}, L. and {Catal{\'a}n-Torrecilla}, C. and {Florido}, E. and {Aguerri}, J.~A.~L. and {Bland-Hawthorn}, J. and {Corsini}, E.~M. and {Dettmar}, R.~J. and {Galbany}, L. and {Garc{\'\i}a-Benito}, R. and {Marino}, R.~A. and {M{\'a}rquez}, I. and {Ortega-Minakata}, R.~A. and {Papaderos}, P. and {S{\'a}nchez}, S.~F. and {S{\'a}nchez-Blazquez}, P. and {Spekkens}, K. and {van de Ven}, G. and {Wild}, V. and {Ziegler}, B.},
        title = "{Two-dimensional multi-component photometric decomposition of CALIFA galaxies}",
      journal = {\aap},
         year = 2017,
        month = feb,
       volume = {598},
          eid = {A32},
        pages = {A32},
          doi = {10.1051/0004-6361/201629525},
archivePrefix = {arXiv},
       eprint = {1610.05324},
 primaryClass = {astro-ph.GA},
       adsurl = {https://ui.adsabs.harvard.edu/abs/2017A&A...598A..32M}
}

@ARTICLE{marleau2021,
       author = {{Marleau}, Francine R. and {Habas}, Rebecca and {Poulain}, M{\'e}lina and {Duc}, Pierre-Alain and {M{\"u}ller}, Oliver and {Lim}, Sungsoon and {Durrell}, Patrick R. and {S{\'a}nchez-Janssen}, Rub{\'e}n and {Paudel}, Sanjaya and {Ahad}, Syeda Lammim and {Chougule}, Abhishek and {B{\'\i}lek}, Michal and {Fensch}, J{\'e}r{\'e}my},
        title = "{Ultra diffuse galaxies in the MATLAS low-to-moderate density fields}",
      journal = {\aap},
         year = 2021,
        month = oct,
       volume = {654},
          eid = {A105},
        pages = {A105},
          doi = {10.1051/0004-6361/202141432},
archivePrefix = {arXiv},
       eprint = {2109.13173},
 primaryClass = {astro-ph.GA},
       adsurl = {https://ui.adsabs.harvard.edu/abs/2021A&A...654A.105M}
}

@ARTICLE{mcconnachie2012,
       author = {{McConnachie}, Alan W.},
        title = "{The Observed Properties of Dwarf Galaxies in and around the Local Group}",
      journal = {\aj},
         year = 2012,
        month = jul,
       volume = {144},
       number = {1},
          eid = {4},
        pages = {4},
          doi = {10.1088/0004-6256/144/1/4},
archivePrefix = {arXiv},
       eprint = {1204.1562},
 primaryClass = {astro-ph.CO},
       adsurl = {https://ui.adsabs.harvard.edu/abs/2012AJ....144....4M}
}

@ARTICLE{zemaitis2023,
       author = {{{\v{Z}}emaitis}, Rokas and {Ferguson}, Annette M.~N. and {Okamoto}, Sakurako and {Cuillandre}, Jean-Charles and {Stone}, Connor J. and {Arimoto}, Nobuo and {Irwin}, Mike J.},
        title = "{A tale of a tail: a tidally disrupting ultra-diffuse galaxy in the M81 group}",
      journal = {\mnras},
         year = 2023,
        month = jan,
       volume = {518},
       number = {2},
        pages = {2497-2510},
          doi = {10.1093/mnras/stac3133},
archivePrefix = {arXiv},
       eprint = {2209.09713},
 primaryClass = {astro-ph.GA},
       adsurl = {https://ui.adsabs.harvard.edu/abs/2023MNRAS.518.2497Z}
}

@ARTICLE{smercina2020,
       author = {{Smercina}, Adam and {Bell}, Eric F. and {Price}, Paul A. and {Slater}, Colin T. and {D'Souza}, Richard and {Bailin}, Jeremy and {de Jong}, Roelof S. and {Jang}, In Sung and {Monachesi}, Antonela and {Nidever}, David},
        title = "{The Saga of M81: Global View of a Massive Stellar Halo in Formation}",
      journal = {\apj},
         year = 2020,
        month = dec,
       volume = {905},
       number = {1},
          eid = {60},
        pages = {60},
          doi = {10.3847/1538-4357/abc485},
archivePrefix = {arXiv},
       eprint = {1910.14672},
 primaryClass = {astro-ph.GA},
       adsurl = {https://ui.adsabs.harvard.edu/abs/2020ApJ...905...60S}
}

@ARTICLE{williams2010,
       author = {{Williams}, Benjamin F. and {Dalcanton}, Julianne J. and {Stilp}, Adrienne and {Gilbert}, Karoline M. and {Ro{\v{s}}kar}, Rok and {Seth}, Anil C. and {Weisz}, Daniel and {Dolphin}, Andrew and {Gogarten}, Stephanie M. and {Skillman}, Evan and {Holtzman}, Jon},
        title = "{The Advanced Camera for Surveys Nearby Galaxy Survey Treasury. IV. The Star Formation History of NGC 2976}",
      journal = {\apj},
         year = 2010,
        month = jan,
       volume = {709},
       number = {1},
        pages = {135-148},
          doi = {10.1088/0004-637X/709/1/135},
archivePrefix = {arXiv},
       eprint = {0911.4121},
 primaryClass = {astro-ph.GA},
       adsurl = {https://ui.adsabs.harvard.edu/abs/2010ApJ...709..135W}
}

@ARTICLE{yun1994,
       author = {{Yun}, M.~S. and {Ho}, P.~T.~P. and {Lo}, K.~Y.},
        title = "{A high-resolution image of atomic hydrogen in the M81 group of galaxies}",
      journal = {\nat},
         year = 1994,
        month = dec,
       volume = {372},
       number = {6506},
        pages = {530-532},
          doi = {10.1038/372530a0},
       adsurl = {https://ui.adsabs.harvard.edu/abs/1994Natur.372..530Y}
}

@ARTICLE{okamoto2015,
       author = {{Okamoto}, Sakurako and {Arimoto}, Nobuo and {Ferguson}, Annette M.~N. and {Bernard}, Edouard J. and {Irwin}, Mike J. and {Yamada}, Yoshihiko and {Utsumi}, Yousuke},
        title = "{A Hyper Suprime-Cam View of the Interacting Galaxies of the M81 Group}",
      journal = {\apjl},
         year = 2015,
        month = aug,
       volume = {809},
       number = {1},
          eid = {L1},
        pages = {L1},
          doi = {10.1088/2041-8205/809/1/L1},
archivePrefix = {arXiv},
       eprint = {1507.04889},
 primaryClass = {astro-ph.GA},
       adsurl = {https://ui.adsabs.harvard.edu/abs/2015ApJ...809L...1O}
}

@ARTICLE{dolphin2012,
       author = {{Dolphin}, Andrew E.},
        title = "{On the Estimation of Systematic Uncertainties of Star Formation Histories}",
      journal = {\apj},
         year = 2012,
        month = may,
       volume = {751},
       number = {1},
          eid = {60},
        pages = {60},
          doi = {10.1088/0004-637X/751/1/60},
archivePrefix = {arXiv},
       eprint = {1203.4634},
 primaryClass = {astro-ph.IM},
       adsurl = {https://ui.adsabs.harvard.edu/abs/2012ApJ...751...60D}
}

@ARTICLE{dolphin2013,
       author = {{Dolphin}, Andrew E.},
        title = "{On the Estimation of Random Uncertainties of Star Formation Histories}",
      journal = {\apj},
         year = 2013,
        month = sep,
       volume = {775},
       number = {1},
          eid = {76},
        pages = {76},
          doi = {10.1088/0004-637X/775/1/76},
archivePrefix = {arXiv},
       eprint = {1308.1740},
 primaryClass = {astro-ph.IM},
       adsurl = {https://ui.adsabs.harvard.edu/abs/2013ApJ...775...76D}
}

@ARTICLE{duane1987,
       author = {{Duane}, Simon and {Kennedy}, A.~D. and {Pendleton}, Brian J. and {Roweth}, Duncan},
        title = "{Hybrid Monte Carlo}",
      journal = {Physics Letters B},
         year = 1987,
        month = sep,
       volume = {195},
       number = {2},
        pages = {216-222},
          doi = {10.1016/0370-2693(87)91197-X},
       adsurl = {https://ui.adsabs.harvard.edu/abs/1987PhLB..195..216D}
}

@ARTICLE{williams2017,
       author = {{Williams}, Benjamin F. and {Dolphin}, Andrew E. and {Dalcanton}, Julianne J. and {Weisz}, Daniel R. and {Bell}, Eric F. and {Lewis}, Alexia R. and {Rosenfield}, Philip and {Choi}, Yumi and {Skillman}, Evan and {Monachesi}, Antonela},
        title = "{PHAT. XIX. The Ancient Star Formation History of the M31 Disk}",
      journal = {\apj},
         year = 2017,
        month = sep,
       volume = {846},
       number = {2},
          eid = {145},
        pages = {145},
          doi = {10.3847/1538-4357/aa862a},
       adsurl = {https://ui.adsabs.harvard.edu/abs/2017ApJ...846..145W}
}

@ARTICLE{lazzarini2022,
       author = {{Lazzarini}, Margaret and {Williams}, Benjamin F. and {Durbin}, Meredith J. and {Dalcanton}, Julianne J. and {Smercina}, Adam and {Bell}, Eric F. and {Choi}, Yumi and {Dolphin}, Andrew and {Gilbert}, Karoline and {Guhathakurta}, Puragra and {Rosolowsky}, Erik and {Skillman}, Evan and {Telford}, O. Grace and {Weisz}, Daniel},
        title = "{The Panchromatic Hubble Andromeda Treasury: Triangulum Extended Region (PHATTER). II. The Spatially Resolved Recent Star Formation History of M33}",
      journal = {\apj},
         year = 2022,
        month = jul,
       volume = {934},
       number = {1},
          eid = {76},
        pages = {76},
          doi = {10.3847/1538-4357/ac7568},
archivePrefix = {arXiv},
       eprint = {2206.11393},
 primaryClass = {astro-ph.GA},
       adsurl = {https://ui.adsabs.harvard.edu/abs/2022ApJ...934...76L}
}

@ARTICLE{mcquinn2024b,
       author = {{McQuinn}, Kristen. B.~W. and {B. Newman}, Max J. and {Savino}, Alessandro and {Dolphin}, Andrew E. and {Weisz}, Daniel R. and {Williams}, Benjamin F. and {Boyer}, Martha L. and {Cohen}, Roger E. and {Correnti}, Matteo and {Cole}, Andrew A. and {Geha}, Marla C. and {Gennaro}, Mario and {Kallivayalil}, Nitya and {Sandstrom}, Karin M. and {Skillman}, Evan D. and {Anderson}, Jay and {Bolatto}, Alberto and {Boylan-Kolchin}, Michael and {Garling}, Christopher T. and {Gilbert}, Karoline M. and {Girardi}, L{\'e}o and {Kalirai}, Jason S. and {Mazzi}, Alessandro and {Pastorelli}, Giada and {Richstein}, Hannah and {Warfield}, Jack T.},
        title = "{The JWST Resolved Stellar Populations Early Release Science Program. IV. The Star Formation History of the Local Group Galaxy WLM}",
      journal = {\apj},
         year = 2024,
        month = jan,
       volume = {961},
       number = {1},
          eid = {16},
        pages = {16},
          doi = {10.3847/1538-4357/ad1105},
archivePrefix = {arXiv},
       eprint = {2312.03060},
 primaryClass = {astro-ph.GA},
       adsurl = {https://ui.adsabs.harvard.edu/abs/2024ApJ...961...16M}
}

@ARTICLE{mcgaugh2005,
       author = {{McGaugh}, Stacy S.},
        title = "{The Baryonic Tully-Fisher Relation of Galaxies with Extended Rotation Curves and the Stellar Mass of Rotating Galaxies}",
      journal = {\apj},
         year = 2005,
        month = oct,
       volume = {632},
       number = {2},
        pages = {859-871},
          doi = {10.1086/432968},
archivePrefix = {arXiv},
       eprint = {astro-ph/0506750},
 primaryClass = {astro-ph},
       adsurl = {https://ui.adsabs.harvard.edu/abs/2005ApJ...632..859M}
}

@ARTICLE{kroupa2001,
       author = {{Kroupa}, Pavel},
        title = "{On the variation of the initial mass function}",
      journal = {\mnras},
         year = 2001,
        month = apr,
       volume = {322},
       number = {2},
        pages = {231-246},
          doi = {10.1046/j.1365-8711.2001.04022.x},
archivePrefix = {arXiv},
       eprint = {astro-ph/0009005},
 primaryClass = {astro-ph},
       adsurl = {https://ui.adsabs.harvard.edu/abs/2001MNRAS.322..231K}
}

@ARTICLE{hogg2010,
       author = {{Hogg}, David W. and {Bovy}, Jo and {Lang}, Dustin},
        title = "{Data analysis recipes: Fitting a model to data}",
      journal = {arXiv e-prints},
         year = 2010,
        month = aug,
          eid = {arXiv:1008.4686},
        pages = {arXiv:1008.4686},
          doi = {10.48550/arXiv.1008.4686},
archivePrefix = {arXiv},
       eprint = {1008.4686},
 primaryClass = {astro-ph.IM},
       adsurl = {https://ui.adsabs.harvard.edu/abs/2010arXiv1008.4686H}
}

@ARTICLE{weisz2015,
       author = {{Weisz}, Daniel R. and {Dolphin}, Andrew E. and {Skillman}, Evan D. and {Holtzman}, Jon and {Gilbert}, Karoline M. and {Dalcanton}, Julianne J. and {Williams}, Benjamin F.},
        title = "{The Star Formation Histories of Local Group Dwarf Galaxies. III. Characterizing Quenching in Low-mass Galaxies}",
      journal = {\apj},
         year = 2015,
        month = may,
       volume = {804},
       number = {2},
          eid = {136},
        pages = {136},
          doi = {10.1088/0004-637X/804/2/136},
archivePrefix = {arXiv},
       eprint = {1503.05195},
 primaryClass = {astro-ph.GA},
       adsurl = {https://ui.adsabs.harvard.edu/abs/2015ApJ...804..136W}
}

@ARTICLE{skillman2017,
       author = {{Skillman}, Evan D. and {Monelli}, Matteo and {Weisz}, Daniel R. and {Hidalgo}, Sebastian L. and {Aparicio}, Antonio and {Bernard}, Edouard J. and {Boylan-Kolchin}, Michael and {Cassisi}, Santi and {Cole}, Andrew A. and {Dolphin}, Andrew E. and {Ferguson}, Henry C. and {Gallart}, Carme and {Irwin}, Mike J. and {Martin}, Nicolas F. and {Mart{\'\i}nez-V{\'a}zquez}, Clara E. and {Mayer}, Lucio and {McConnachie}, Alan W. and {McQuinn}, Kristen B.~W. and {Navarro}, Julio F. and {Stetson}, Peter B.},
        title = "{The ISLAndS Project. II. The Lifetime Star Formation Histories of Six Andomeda dSphS}",
      journal = {\apj},
         year = 2017,
        month = mar,
       volume = {837},
       number = {2},
          eid = {102},
        pages = {102},
          doi = {10.3847/1538-4357/aa60c5},
archivePrefix = {arXiv},
       eprint = {1606.01207},
 primaryClass = {astro-ph.GA},
       adsurl = {https://ui.adsabs.harvard.edu/abs/2017ApJ...837..102S}
}

@ARTICLE{radburn-smith2011,
       author = {{Radburn-Smith}, D.~J. and {de Jong}, R.~S. and {Seth}, A.~C. and {Bailin}, J. and {Bell}, E.~F. and {Brown}, T.~M. and {Bullock}, J.~S. and {Courteau}, S. and {Dalcanton}, J.~J. and {Ferguson}, H.~C. and {Goudfrooij}, P. and {Holfeltz}, S. and {Holwerda}, B.~W. and {Purcell}, C. and {Sick}, J. and {Streich}, D. and {Vlajic}, M. and {Zucker}, D.~B.},
        title = "{The GHOSTS Survey. I. Hubble Space Telescope Advanced Camera for Surveys Data}",
      journal = {\apjs},
         year = 2011,
        month = aug,
       volume = {195},
       number = {2},
          eid = {18},
        pages = {18},
          doi = {10.1088/0067-0049/195/2/18},
       adsurl = {https://ui.adsabs.harvard.edu/abs/2011ApJS..195...18R}
}

@ARTICLE{battinelli2007,
       author = {{Battinelli}, P. and {Demers}, S. and {Artigau}, {\'E}.},
        title = "{The size and structure of the spheroid of IC 1613}",
      journal = {\aap},
         year = 2007,
        month = may,
       volume = {466},
       number = {3},
        pages = {875-881},
          doi = {10.1051/0004-6361:20066578},
archivePrefix = {arXiv},
       eprint = {astro-ph/0702489},
 primaryClass = {astro-ph},
       adsurl = {https://ui.adsabs.harvard.edu/abs/2007A&A...466..875B}
}

@BOOK{sersic1968,
       author = {{Sersic}, Jose Luis},
        title = "{Atlas de Galaxias Australes}",
         year = 1968,
       adsurl = {https://ui.adsabs.harvard.edu/abs/1968adga.book.....S}
}

@ARTICLE{forbes2024,
       author = {{Forbes}, Duncan A. and {Lyon}, Daniel and {Gannon}, Jonah and {Romanowsky}, Aaron J. and {Brodie}, Jean P.},
        title = "{Keck/KCWI spectroscopy of globular clusters in local volume dwarf galaxies}",
      journal = {\pasa},
         year = 2024,
        month = may,
       volume = {41},
          eid = {e044},
        pages = {e044},
          doi = {10.1017/pasa.2024.41},
archivePrefix = {arXiv},
       eprint = {2405.11749},
 primaryClass = {astro-ph.GA},
       adsurl = {https://ui.adsabs.harvard.edu/abs/2024PASA...41...44F}
}

@BOOK{deVaucouleurs1991,
       author = {{de Vaucouleurs}, Gerard and {de Vaucouleurs}, Antoinette and {Corwin}, Jr., Herold G. and {Buta}, Ronald J. and {Paturel}, Georges and {Fouque}, Pascal},
        title = "{Third Reference Catalogue of Bright Galaxies}",
         year = 1991,
       adsurl = {https://ui.adsabs.harvard.edu/abs/1991rc3..book.....D}
}

@ARTICLE{Besla2012,
       author = {{Besla}, Gurtina and {Kallivayalil}, Nitya and {Hernquist}, Lars and {van der Marel}, Roeland P. and {Cox}, T.~J. and {Kere{\v{s}}}, Du{\v{s}}an},
        title = "{The role of dwarf galaxy interactions in shaping the Magellanic System and implications for Magellanic Irregulars}",
      journal = {\mnras},
         year = 2012,
        month = apr,
       volume = {421},
       number = {3},
        pages = {2109-2138},
          doi = {10.1111/j.1365-2966.2012.20466.x},
archivePrefix = {arXiv},
       eprint = {1201.1299},
 primaryClass = {astro-ph.GA},
       adsurl = {https://ui.adsabs.harvard.edu/abs/2012MNRAS.421.2109B}
}

@ARTICLE{Keim2025,
       author = {{Keim}, Michael A. and {van Dokkum}, Pieter and {Shen}, Zili and {Souchereau}, Harrison and {Pasha}, Imad and {Danieli}, Shany and {Abraham}, Roberto and {Romanowsky}, Aaron J. and {Tang}, Yimeng},
        title = "{Kinematic Confirmation of a Remarkable Linear Trail of Galaxies in the NGC 1052 Field, Consistent with Formation in a High-Speed Bullet Dwarf Collision}",
      journal = {arXiv e-prints},
         year = 2025,
        month = jun,
          eid = {arXiv:2506.10220},
        pages = {arXiv:2506.10220},
          doi = {10.48550/arXiv.2506.10220},
archivePrefix = {arXiv},
       eprint = {2506.10220},
 primaryClass = {astro-ph.GA},
       adsurl = {https://ui.adsabs.harvard.edu/abs/2025arXiv250610220K}
}

@ARTICLE{vanDokkum2022,
       author = {{van Dokkum}, Pieter and {Shen}, Zili and {Keim}, Michael A. and {Trujillo-Gomez}, Sebastian and {Danieli}, Shany and {Dutta Chowdhury}, Dhruba and {Abraham}, Roberto and {Conroy}, Charlie and {Kruijssen}, J.~M. Diederik and {Nagai}, Daisuke and {Romanowsky}, Aaron},
        title = "{A trail of dark-matter-free galaxies from a bullet-dwarf collision}",
      journal = {\nat},
         year = 2022,
        month = may,
       volume = {605},
       number = {7910},
        pages = {435-439},
          doi = {10.1038/s41586-022-04665-6},
archivePrefix = {arXiv},
       eprint = {2205.08552},
 primaryClass = {astro-ph.GA},
       adsurl = {https://ui.adsabs.harvard.edu/abs/2022Natur.605..435V}
}

@ARTICLE{Watts2024,
       author = {{Watts}, Chandan and {Barway}, Sudhanshu and {Bait}, Omkar and {Wadadekar}, Yogesh},
        title = "{A Tale of NGC 3785: The formation of an ultra-diffuse galaxy at the end of the longest tidal tail}",
      journal = {\aap},
         year = 2024,
        month = nov,
       volume = {691},
          eid = {L13},
        pages = {L13},
          doi = {10.1051/0004-6361/202452002},
archivePrefix = {arXiv},
       eprint = {2410.18811},
 primaryClass = {astro-ph.GA},
       adsurl = {https://ui.adsabs.harvard.edu/abs/2024A&A...691L..13W}
}

@ARTICLE{deBlok2000,
       author = {{de Blok}, W.~J.~G. and {Walter}, F.},
        title = "{Evidence for Tidal Interaction and a Supergiant H I Shell in the Local Group Dwarf Galaxy NGC 6822}",
      journal = {\apjl},
         year = 2000,
        month = jul,
       volume = {537},
       number = {2},
        pages = {L95-L98},
          doi = {10.1086/312777},
archivePrefix = {arXiv},
       eprint = {astro-ph/0005473},
 primaryClass = {astro-ph},
       adsurl = {https://ui.adsabs.harvard.edu/abs/2000ApJ...537L..95D}
}

@ARTICLE{Silich2006,
       author = {{Silich}, S. and {Lozinskaya}, T. and {Moiseev}, A. and {Podorvanuk}, N. and {Rosado}, M. and {Borissova}, J. and {Valdez-Gutierrez}, M.},
        title = "{On the neutral gas distribution and kinematics in the dwarf irregular galaxy IC 1613}",
      journal = {\aap},
         year = 2006,
        month = mar,
       volume = {448},
       number = {1},
        pages = {123-131},
          doi = {10.1051/0004-6361:20053326},
archivePrefix = {arXiv},
       eprint = {astro-ph/0510812},
 primaryClass = {astro-ph},
       adsurl = {https://ui.adsabs.harvard.edu/abs/2006A&A...448..123S}
}

@ARTICLE{Drzazga2016,
       author = {{Drzazga}, R.~T. and {Chy{\.z}y}, K.~T. and {Heald}, G.~H. and {Elstner}, D. and {Gallagher}, J.~S.},
        title = "{Seeking large-scale magnetic fields in a pure-disk dwarf galaxy NGC 2976}",
      journal = {\aap},
         year = 2016,
        month = may,
       volume = {589},
          eid = {A12},
        pages = {A12},
          doi = {10.1051/0004-6361/201527236},
archivePrefix = {arXiv},
       eprint = {1603.00482},
 primaryClass = {astro-ph.GA},
       adsurl = {https://ui.adsabs.harvard.edu/abs/2016A&A...589A..12D}
}

@software{Dolphin2016,
       author = {{Dolphin}, Andrew},
        title = "{DOLPHOT: Stellar photometry}",
 howpublished = {Astrophysics Source Code Library, record ascl:1608.013},
         year = 2016,
        month = aug,
          eid = {ascl:1608.013},
       adsurl = {https://ui.adsabs.harvard.edu/abs/2016ascl.soft08013D}
}

@ARTICLE{bressan2012,
       author = {{Bressan}, Alessandro and {Marigo}, Paola and {Girardi}, L{\'e}o. and {Salasnich}, Bernardo and {Dal Cero}, Claudia and {Rubele}, Stefano and {Nanni}, Ambra},
        title = "{PARSEC: stellar tracks and isochrones with the PAdova and TRieste Stellar Evolution Code}",
      journal = {\mnras},
         year = 2012,
        month = nov,
       volume = {427},
       number = {1},
        pages = {127-145},
          doi = {10.1111/j.1365-2966.2012.21948.x},
archivePrefix = {arXiv},
       eprint = {1208.4498},
 primaryClass = {astro-ph.SR},
       adsurl = {https://ui.adsabs.harvard.edu/abs/2012MNRAS.427..127B}
}

@ARTICLE{Lim2020,
       author = {{Lim}, Sungsoon and {C{\^o}t{\'e}}, Patrick and {Peng}, Eric W. and {Ferrarese}, Laura and {Roediger}, Joel C. and {Durrell}, Patrick R. and {Mihos}, J. Christopher and {Wang}, Kaixiang and {Gwyn}, S.~D.~J. and {Cuillandre}, Jean-Charles and {Liu}, Chengze and {S{\'a}nchez-Janssen}, Rub{\'e}n and {Toloba}, Elisa and {Sales}, Laura V. and {Guhathakurta}, Puragra and {Lan{\c{c}}on}, Ariane and {Puzia}, Thomas H.},
        title = "{The Next Generation Virgo Cluster Survey (NGVS). XXX. Ultra-diffuse Galaxies and Their Globular Cluster Systems}",
      journal = {\apj},
         year = 2020,
        month = aug,
       volume = {899},
       number = {1},
          eid = {69},
        pages = {69},
          doi = {10.3847/1538-4357/aba433},
archivePrefix = {arXiv},
       eprint = {2007.10565},
 primaryClass = {astro-ph.GA},
       adsurl = {https://ui.adsabs.harvard.edu/abs/2020ApJ...899...69L}
}

@ARTICLE{Mei2007,
       author = {{Mei}, Simona and {Blakeslee}, John P. and {C{\^o}t{\'e}}, Patrick and {Tonry}, John L. and {West}, Michael J. and {Ferrarese}, Laura and {Jord{\'a}n}, Andr{\'e}s and {Peng}, Eric W. and {Anthony}, Andr{\'e} and {Merritt}, David},
        title = "{The ACS Virgo Cluster Survey. XIII. SBF Distance Catalog and the Three-dimensional Structure of the Virgo Cluster}",
      journal = {\apj},
         year = 2007,
        month = jan,
       volume = {655},
       number = {1},
        pages = {144-162},
          doi = {10.1086/509598},
archivePrefix = {arXiv},
       eprint = {astro-ph/0702510},
 primaryClass = {astro-ph},
       adsurl = {https://ui.adsabs.harvard.edu/abs/2007ApJ...655..144M}
}

@ARTICLE{Blakeslee2009,
       author = {{Blakeslee}, John P. and {Jord{\'a}n}, Andr{\'e}s and {Mei}, Simona and {C{\^o}t{\'e}}, Patrick and {Ferrarese}, Laura and {Infante}, Leopoldo and {Peng}, Eric W. and {Tonry}, John L. and {West}, Michael J.},
        title = "{The ACS Fornax Cluster Survey. V. Measurement and Recalibration of Surface Brightness Fluctuations and a Precise Value of the Fornax-Virgo Relative Distance}",
      journal = {\apj},
         year = 2009,
        month = mar,
       volume = {694},
       number = {1},
        pages = {556-572},
          doi = {10.1088/0004-637X/694/1/556},
archivePrefix = {arXiv},
       eprint = {0901.1138},
 primaryClass = {astro-ph.CO},
       adsurl = {https://ui.adsabs.harvard.edu/abs/2009ApJ...694..556B}
}

@ARTICLE{Crnojevic2016,
       author = {{Crnojevi{\'c}}, D. and {Sand}, D.~J. and {Spekkens}, K. and {Caldwell}, N. and {Guhathakurta}, P. and {McLeod}, B. and {Seth}, A. and {Simon}, J.~D. and {Strader}, J. and {Toloba}, E.},
        title = "{The Extended Halo of Centaurus A: Uncovering Satellites, Streams, and Substructures}",
      journal = {\apj},
         year = 2016,
        month = may,
       volume = {823},
       number = {1},
          eid = {19},
        pages = {19},
          doi = {10.3847/0004-637X/823/1/19},
archivePrefix = {arXiv},
       eprint = {1512.05366},
 primaryClass = {astro-ph.GA},
       adsurl = {https://ui.adsabs.harvard.edu/abs/2016ApJ...823...19C}
}

@ARTICLE{McQuinn2009,
       author = {{McQuinn}, Kristen B.~W. and {Skillman}, Evan D. and {Cannon}, John M. and {Dalcanton}, Julianne J. and {Dolphin}, Andrew and {Stark}, David and {Weisz}, Daniel},
        title = "{The True Durations of Starbursts: Hubble Space Telescope Observations of Three Nearby Dwarf Starburst Galaxies}",
      journal = {\apj},
         year = 2009,
        month = apr,
       volume = {695},
       number = {1},
        pages = {561-573},
          doi = {10.1088/0004-637X/695/1/561},
archivePrefix = {arXiv},
       eprint = {0901.2361},
 primaryClass = {astro-ph.GA},
       adsurl = {https://ui.adsabs.harvard.edu/abs/2009ApJ...695..561M}
}

@ARTICLE{McQuinn2010,
       author = {{McQuinn}, Kristen B.~W. and {Skillman}, Evan D. and {Cannon}, John M. and {Dalcanton}, Julianne and {Dolphin}, Andrew and {Hidalgo-Rodr{\'\i}guez}, Sebastian and {Holtzman}, Jon and {Stark}, David and {Weisz}, Daniel and {Williams}, Benjamin},
        title = "{The Nature of Starbursts. I. The Star Formation Histories of Eighteen Nearby Starburst Dwarf Galaxies}",
      journal = {\apj},
         year = 2010,
        month = sep,
       volume = {721},
       number = {1},
        pages = {297-317},
          doi = {10.1088/0004-637X/721/1/297},
archivePrefix = {arXiv},
       eprint = {1008.1589},
 primaryClass = {astro-ph.CO},
       adsurl = {https://ui.adsabs.harvard.edu/abs/2010ApJ...721..297M}
}

@ARTICLE{Telford2020,
       author = {{Telford}, O. Grace and {Dalcanton}, Julianne J. and {Williams}, Benjamin F. and {Bell}, Eric F. and {Dolphin}, Andrew E. and {Durbin}, Meredith J. and {Choi}, Yumi},
        title = "{Mass-to-light Ratios of Spatially Resolved Stellar Populations in M31}",
      journal = {\apj},
         year = 2020,
        month = mar,
       volume = {891},
       number = {1},
          eid = {32},
        pages = {32},
          doi = {10.3847/1538-4357/ab701c},
archivePrefix = {arXiv},
       eprint = {2001.09157},
 primaryClass = {astro-ph.GA},
       adsurl = {https://ui.adsabs.harvard.edu/abs/2020ApJ...891...32T}
}

@ARTICLE{Karachentsev2014,
       author = {{Karachentsev}, Igor D. and {Kaisina}, Elena I. and {Makarov}, Dmitry I.},
        title = "{Suites of Dwarfs around nearby Giant Galaxies}",
      journal = {\aj},
         year = 2014,
        month = jan,
       volume = {147},
       number = {1},
          eid = {13},
        pages = {13},
          doi = {10.1088/0004-6256/147/1/13},
archivePrefix = {arXiv},
       eprint = {1310.6838},
 primaryClass = {astro-ph.CO},
       adsurl = {https://ui.adsabs.harvard.edu/abs/2014AJ....147...13K}
}

@ARTICLE{Amorisco2018,
       author = {{Amorisco}, N.~C. and {Monachesi}, A. and {Agnello}, A. and {White}, S.~D.~M.},
        title = "{The globular cluster systems of 54 Coma ultra-diffuse galaxies: statistical constraints from HST data}",
      journal = {\mnras},
         year = 2018,
        month = apr,
       volume = {475},
       number = {3},
        pages = {4235-4251},
          doi = {10.1093/mnras/sty116},
archivePrefix = {arXiv},
       eprint = {1610.01595},
 primaryClass = {astro-ph.GA},
       adsurl = {https://ui.adsabs.harvard.edu/abs/2018MNRAS.475.4235A}
}

@ARTICLE{Toloba2023,
       author = {{Toloba}, Elisa and {Sales}, Laura V. and {Lim}, Sungsoon and {Peng}, Eric W. and {Guhathakurta}, Puragra and {Roediger}, Joel and {Wang}, Kaixiang and {Mihos}, J. Christopher and {C{\^o}t{\'e}}, Patrick and {Durrell}, Patrick R. and {Ferrarese}, Laura},
        title = "{The Next Generation Virgo Cluster Survey (NGVS). XXXV. First Kinematical Clues of Overly Massive Dark Matter Halos in Several Ultradiffuse Galaxies in the Virgo Cluster}",
      journal = {\apj},
         year = 2023,
        month = jul,
       volume = {951},
       number = {1},
          eid = {77},
        pages = {77},
          doi = {10.3847/1538-4357/acd336},
archivePrefix = {arXiv},
       eprint = {2305.06369},
 primaryClass = {astro-ph.GA},
       adsurl = {https://ui.adsabs.harvard.edu/abs/2023ApJ...951...77T}
}

@ARTICLE{Benavides2024,
       author = {{Benavides}, Jos{\'e} A. and {Sales}, Laura V. and {Abadi}, Mario. G. and {Vogelsberger}, Mark and {Marinacci}, Federico and {Hernquist}, Lars},
        title = "{Large Dark Matter Content and Steep Metallicity Profile Predicted for Ultradiffuse Galaxies Formed in High-spin Halos}",
      journal = {\apj},
         year = 2024,
        month = dec,
       volume = {977},
       number = {2},
          eid = {169},
        pages = {169},
          doi = {10.3847/1538-4357/ad8de8},
archivePrefix = {arXiv},
       eprint = {2407.15938},
 primaryClass = {astro-ph.GA},
       adsurl = {https://ui.adsabs.harvard.edu/abs/2024ApJ...977..169B}
}

@ARTICLE{Chan2015,
       author = {{Chan}, T.~K. and {Kere{\v{s}}}, D. and {O{\~n}orbe}, J. and {Hopkins}, P.~F. and {Muratov}, A.~L. and {Faucher-Gigu{\`e}re}, C. -A. and {Quataert}, E.},
        title = "{The impact of baryonic physics on the structure of dark matter haloes: the view from the FIRE cosmological simulations}",
      journal = {\mnras},
         year = 2015,
        month = dec,
       volume = {454},
       number = {3},
        pages = {2981-3001},
          doi = {10.1093/mnras/stv2165},
archivePrefix = {arXiv},
       eprint = {1507.02282},
 primaryClass = {astro-ph.GA},
       adsurl = {https://ui.adsabs.harvard.edu/abs/2015MNRAS.454.2981C}
}

@ARTICLE{Lazar2020,
       author = {{Lazar}, Alexandres and {Bullock}, James S. and {Boylan-Kolchin}, Michael and {Chan}, T.~K. and {Hopkins}, Philip F. and {Graus}, Andrew S. and {Wetzel}, Andrew and {El-Badry}, Kareem and {Wheeler}, Coral and {Straight}, Maria C. and {Kere{\v{s}}}, Du{\v{s}}an and {Faucher-Gigu{\`e}re}, Claude-Andr{\'e} and {Fitts}, Alex and {Garrison-Kimmel}, Shea},
        title = "{A dark matter profile to model diverse feedback-induced core sizes of {\ensuremath{\Lambda}}CDM haloes}",
      journal = {\mnras},
         year = 2020,
        month = sep,
       volume = {497},
       number = {2},
        pages = {2393-2417},
          doi = {10.1093/mnras/staa2101},
archivePrefix = {arXiv},
       eprint = {2004.10817},
 primaryClass = {astro-ph.GA},
       adsurl = {https://ui.adsabs.harvard.edu/abs/2020MNRAS.497.2393L}
}

@ARTICLE{Mateo1998,
       author = {{Mateo}, Mario L.},
        title = "{Dwarf Galaxies of the Local Group}",
      journal = {\araa},
         year = 1998,
        month = jan,
       volume = {36},
        pages = {435-506},
          doi = {10.1146/annurev.astro.36.1.435},
archivePrefix = {arXiv},
       eprint = {astro-ph/9810070},
 primaryClass = {astro-ph},
       adsurl = {https://ui.adsabs.harvard.edu/abs/1998ARA&A..36..435M}
}

@ARTICLE{Anand2021,
       author = {{Anand}, Gagandeep S. and {Rizzi}, Luca and {Tully}, R. Brent and {Shaya}, Edward J. and {Karachentsev}, Igor D. and {Makarov}, Dmitry I. and {Makarova}, Lidia and {Wu}, Po-Feng and {Dolphin}, Andrew E. and {Kourkchi}, Ehsan},
        title = "{The Extragalactic Distance Database: The Color-Magnitude Diagrams/Tip of the Red Giant Branch Distance Catalog}",
      journal = {\aj},
         year = 2021,
        month = aug,
       volume = {162},
       number = {2},
          eid = {80},
        pages = {80},
          doi = {10.3847/1538-3881/ac0440},
archivePrefix = {arXiv},
       eprint = {2104.02649},
 primaryClass = {astro-ph.GA},
       adsurl = {https://ui.adsabs.harvard.edu/abs/2021AJ....162...80A}
}

@ARTICLE{Ferrarese2000,
       author = {{Ferrarese}, Laura and {Ford}, Holland C. and {Huchra}, John and {Kennicutt}, Jr., Robert C. and {Mould}, Jeremy R. and {Sakai}, Shoko and {Freedman}, Wendy L. and {Stetson}, Peter B. and {Madore}, Barry F. and {Gibson}, Brad K. and {Graham}, John A. and {Hughes}, Shaun M. and {Illingworth}, Garth D. and {Kelson}, Daniel D. and {Macri}, Lucas and {Sebo}, Kim and {Silbermann}, N.~A.},
        title = "{A Database of Cepheid Distance Moduli and Tip of the Red Giant Branch, Globular Cluster Luminosity Function, Planetary Nebula Luminosity Function, and Surface Brightness Fluctuation Data Useful for Distance Determinations}",
      journal = {\apjs},
         year = 2000,
        month = jun,
       volume = {128},
       number = {2},
        pages = {431-459},
          doi = {10.1086/313391},
archivePrefix = {arXiv},
       eprint = {astro-ph/9910501},
 primaryClass = {astro-ph},
       adsurl = {https://ui.adsabs.harvard.edu/abs/2000ApJS..128..431F}
}

@ARTICLE{Karachentsev2000,
       author = {{Karachentsev}, I.~D. and {Karachentseva}, V.~E. and {Dolphin}, A.~E. and {Geisler}, D. and {Grebel}, E.~K. and {Guhathakurta}, P. and {Hodge}, P.~W. and {Sarajedini}, A. and {Seitzer}, P. and {Sharina}, M.~E.},
        title = "{Dwarf spheroidal galaxies in the M81 group imaged with WFPC2}",
      journal = {\aap},
         year = 2000,
        month = nov,
       volume = {363},
        pages = {117-129},
          doi = {10.48550/arXiv.astro-ph/0010146},
archivePrefix = {arXiv},
       eprint = {astro-ph/0010146},
 primaryClass = {astro-ph},
       adsurl = {https://ui.adsabs.harvard.edu/abs/2000A&A...363..117K}
}

@ARTICLE{Dalcanton2009,
       author = {{Dalcanton}, Julianne J. and {Williams}, Benjamin F. and {Seth}, Anil C. and {Dolphin}, Andrew and {Holtzman}, Jon and {Rosema}, Keith and {Skillman}, Evan D. and {Cole}, Andrew and {Girardi}, L{\'e}o and {Gogarten}, Stephanie M. and {Karachentsev}, Igor D. and {Olsen}, Knut and {Weisz}, Daniel and {Christensen}, Charlotte and {Freeman}, Ken and {Gilbert}, Karoline and {Gallart}, Carme and {Harris}, Jason and {Hodge}, Paul and {de Jong}, Roelof S. and {Karachentseva}, Valentina and {Mateo}, Mario and {Stetson}, Peter B. and {Tavarez}, Maritza and {Zaritsky}, Dennis and {Governato}, Fabio and {Quinn}, Thomas},
        title = "{The ACS Nearby Galaxy Survey Treasury}",
      journal = {\apjs},
         year = 2009,
        month = jul,
       volume = {183},
       number = {1},
        pages = {67-108},
          doi = {10.1088/0067-0049/183/1/67},
archivePrefix = {arXiv},
       eprint = {0905.3737},
 primaryClass = {astro-ph.GA},
       adsurl = {https://ui.adsabs.harvard.edu/abs/2009ApJS..183...67D}
}

@ARTICLE{Choi2016,
       author = {{Choi}, Jieun and {Dotter}, Aaron and {Conroy}, Charlie and {Cantiello}, Matteo and {Paxton}, Bill and {Johnson}, Benjamin D.},
        title = "{Mesa Isochrones and Stellar Tracks (MIST). I. Solar-scaled Models}",
      journal = {\apj},
         year = 2016,
        month = jun,
       volume = {823},
       number = {2},
          eid = {102},
        pages = {102},
          doi = {10.3847/0004-637X/823/2/102},
archivePrefix = {arXiv},
       eprint = {1604.08592},
 primaryClass = {astro-ph.SR},
       adsurl = {https://ui.adsabs.harvard.edu/abs/2016ApJ...823..102C}
}

@ARTICLE{Dotter2016,
       author = {{Dotter}, Aaron},
        title = "{MESA Isochrones and Stellar Tracks (MIST) 0: Methods for the Construction of Stellar Isochrones}",
      journal = {\apjs},
         year = 2016,
        month = jan,
       volume = {222},
       number = {1},
          eid = {8},
        pages = {8},
          doi = {10.3847/0067-0049/222/1/8},
archivePrefix = {arXiv},
       eprint = {1601.05144},
 primaryClass = {astro-ph.SR},
       adsurl = {https://ui.adsabs.harvard.edu/abs/2016ApJS..222....8D}
}

\newpage
\appendix

\section{Red Clump Morphology and Age-Sensitivity}
\label{sec:rc-lfs}

\begin{figure*}[t]
    \centering
    \includegraphics[width=0.8\linewidth]{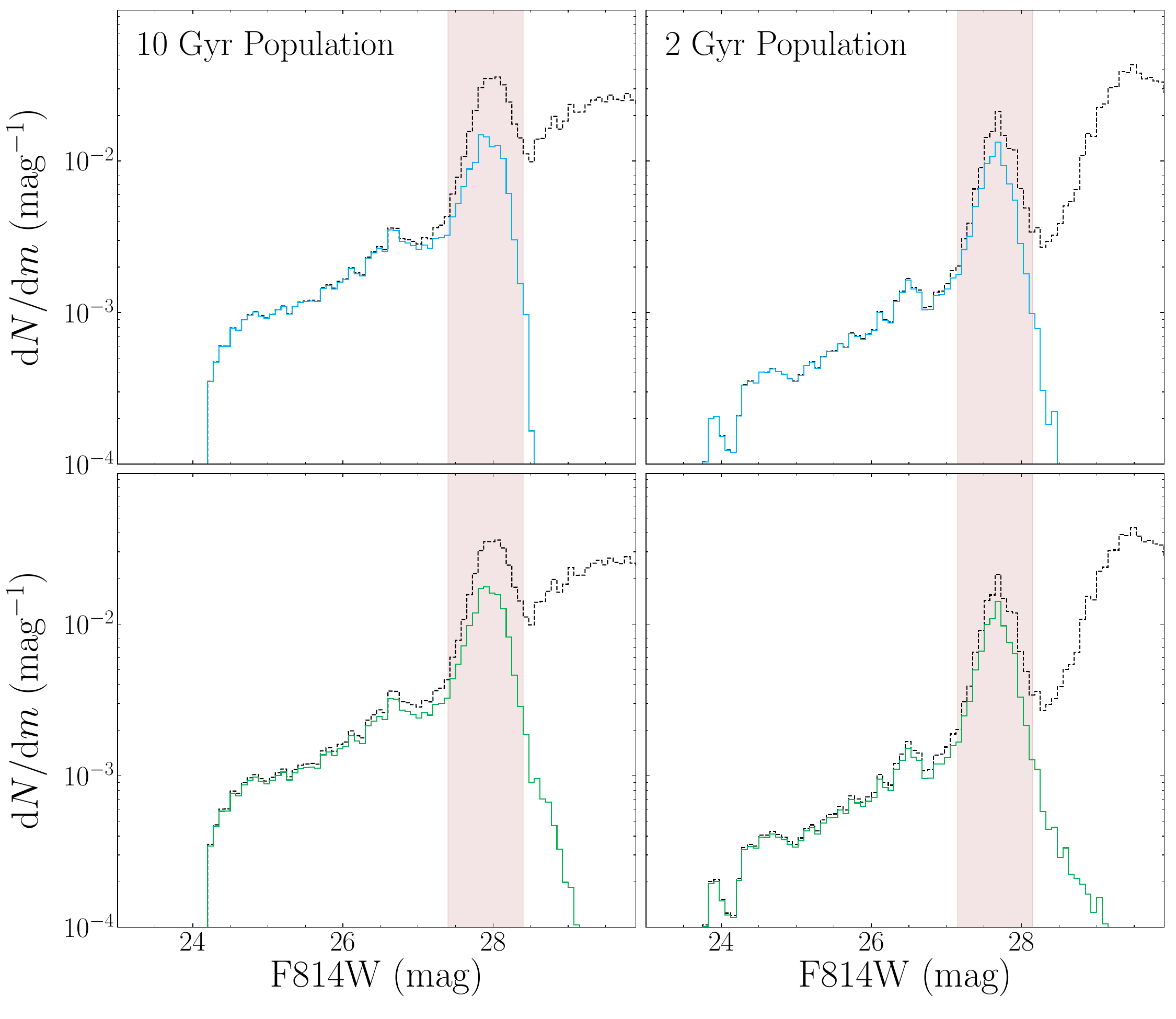}
    \caption{Model LFs in the F814W filter, constructed with \texttt{MATCH} using the Padova stellar libraries (as in \S\,\ref{sec:sfh}), and adjusted to the distance and reddening of our F8D1 observations. The left column shows LFs for a 10\,Gyr burst population, and the right column for a 2\,Gyr burst. Black dashed curves show the theoretical LFs, while the colored curves show the predicted observed LFs after applying the combined two-filter completeness curves for each camera: top row shows the results for WFC3 (blue), and the bottom row for ACS (green). The shaded red region shows the 1\,magnitude region dominated by the RC feature.}
    \label{fig:RC-LFs}
\end{figure*}

In this section, we demonstrate the ability of these observations to measure the Red Clump in F8D1, and illustrate its sensitivity to age. First, we show zoomed versions of the CMDs in Figure \ref{fig:RC-LFs}, with highly reduced color range, and overlaid with a series of isochrones from 1--6\,Gyr. The age dependence of the RC is clearly visible. In addition to the RC, one can see by-eye the contribution from the $\sim$1\,Gyr sub-giant branch to the observed CMDs. Second, we have constructed model luminosity functions (LFs) in F814W, of populations at two different ages, using \texttt{MATCH}: 10\,Gyr and 2\,Gyr. We constructed LFs for each camera, adjusted them to the distance modulus and reddening assumed for the observations, and applied the corresponding combined completeness curves for both filters (see Figure \ref{fig:completeness}). The resulting theoretical (black) and observed (colored) LFs are shown in Figure \ref{fig:RC-LFs}.

The difference in the location, shape, and prominence of the RC is easily visible between the two populations. The central position of the RC is $\sim$0.3\,magnitudes brighter, and is $\sim$50\% more prominent, relative to the RGB, for the 2\,Gyr population than the 10\,Gyr population. The peak of the RC is also noticeably sharper in the 2\,Gyr case. The RC as a feature is $\sim$1\,magnitude in width in both cases. At the depth of our observations, we predict that 39\% of 10\,Gyr RC stars and 47\% of 2\,Gyr RC stars pass our WFC3 GST selection criteria, while this rises to 62\% and 65\% for ACS. This reinforces that the depth of these observations is most appropriate to infer SFHs within the past $\sim$6\,Gyr, with reduced sensitivity to differentiate the most ancient populations. 

\begin{figure*}[!h]
    \centering
    \includegraphics[width=0.8\linewidth]{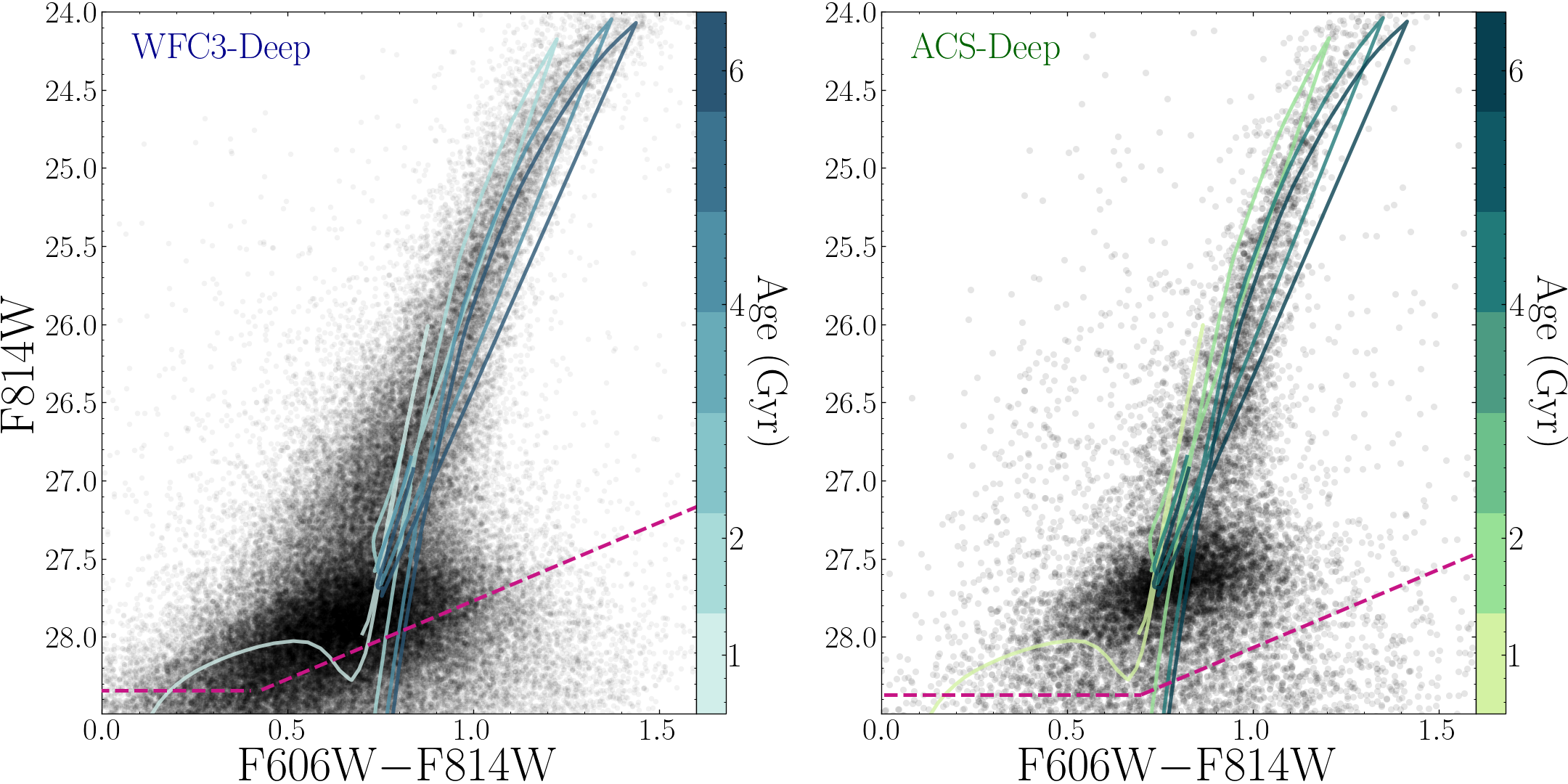}
    \caption{CMDs of the same stars as shown in Figure \ref{fig:deep-cmds}, but shown here as individual points with transparency. The WFC3 photometry is again shown on the left, and the ACS on the right. Completeness curves are shown in pink. Overlaid are several PARSEC isochrones \citep[e.g.,][]{bressan2012}, ranging from 1--6\,Gyr. Tracks are truncated at the core-Helium burning phase (label = 4 in PARSEC parlance).}
    \label{fig:RC-cmd}
\end{figure*}

\section{Distance--Reddening Degeneracy in SFH Fitting}
\label{sec:degen}

\begin{figure*}[t]
    \centering
    \includegraphics[width=\linewidth]{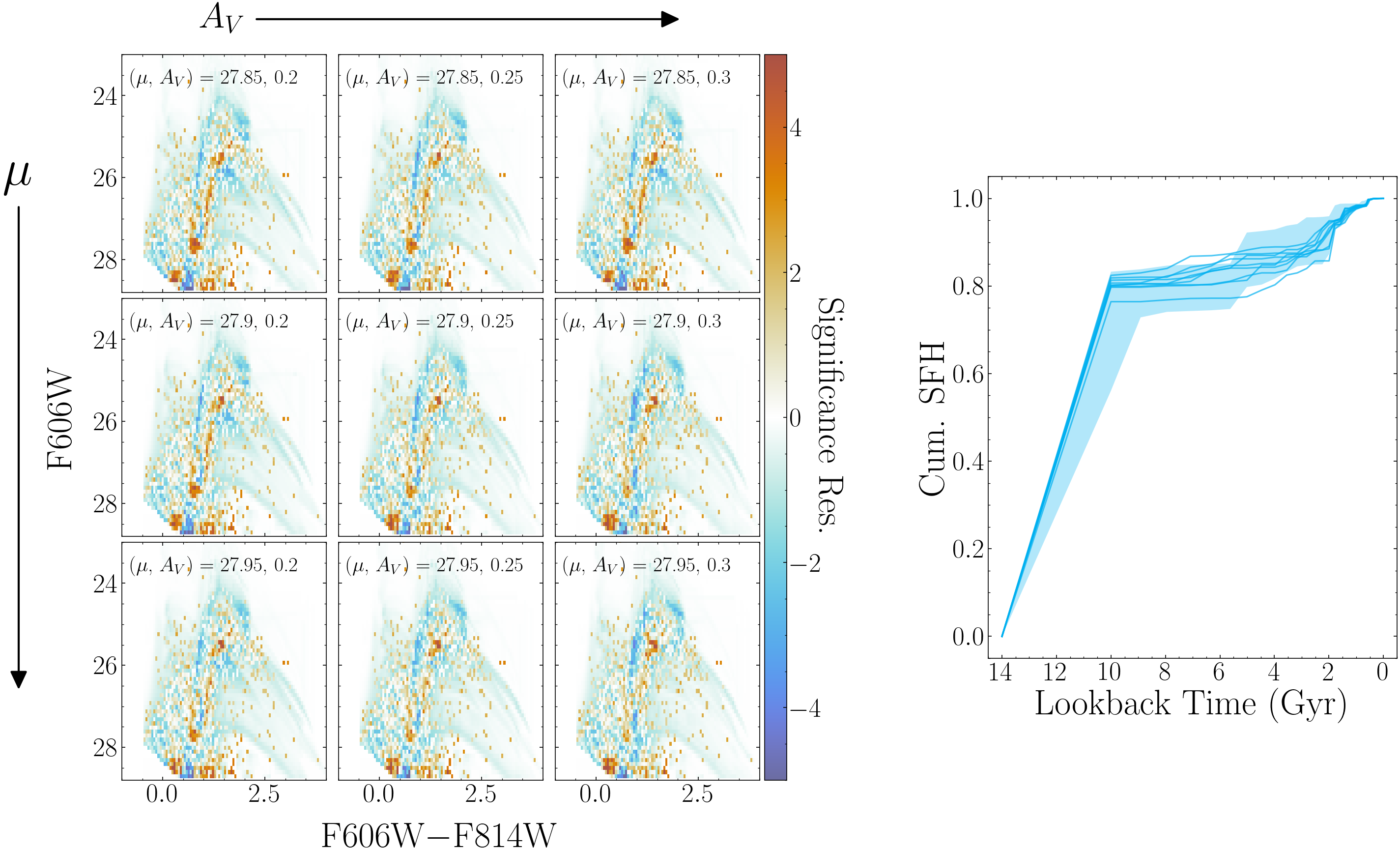}
    \caption{A demonstration of the impact of $A_V$\ and distance modulus variation on our SFH inferences. On the left, we show a 3$\times$3 grid of weighted ``significance'' residuals of fits to our central deep WFC3 CMD. $A_V$\ increases to the right and distance modulus ($\mu$) towards the bottom, each in steps of 0.05\,mag. Our best-fit model (\S\,\ref{sec:sfh}) is shown in the center panel. Variations in the fit residuals are visible on the scale of individual pixels. To the right we show the corresponding variation in cumulative SFH as solid blue curves. The systematic uncertainty envelope calculated in \S\,\ref{sec:uncertainties} and shown in Figure \ref{fig:sfhs} is again shown as a blue shaded region. The majority of the fits fall within this region.}
    \label{fig:av-dmod-grid}
\end{figure*}

Here we demonstrate how the degeneracy between distance and reddening, and the typical uncertainty on these parameters, impacts the inference of F8D1's SFH. We take our central deep WFC3 field as a test case. Using the same time bins, IMF, and color--magnitude limits as in our primary analysis (\S\,\ref{sec:sfh}), we fit the observed CMD with a 3$\times$3 grid of extinction ($A_V$) and distance modulus ($\mu$), in bins of 0.05\,mag and centered on our best-fit $A_V$\ of 0.25 and adopted distance modulus of 27.9, respectively. The 0.05\,mag uncertainty on distance is consistent with the spread in literature measurements of F8D1 (see \S\,\ref{sec:intro-f8d1}). Figure \ref{fig:av-dmod-grid} we show the results of the fits, including the variation in both the weighted (significance) residuals and cumulative SFH. As expected, we observe small variations in both. Shifts in the residual plots are particularly subtle, but are reflected in the $\chi^2$\ of the fits. Overall, the magnitude of variations in cumulative SFH is small and falls within the estimated systematic uncertainties on our primary fit (\S\,\ref{sec:uncertainties}). Most notably for this work, differences in SFH at ages $\lesssim$2\,Gyr are very small.

This is as expected, as uncertainty on both distance and reddening is implicitly captured in our systematic uncertainties, which mirror differences in the stellar atmospheres that occur when using different stellar models -- evidenced by how well the different SFHs track our estimated uncertainty envelope. Differences in models, distance, and reddening are not independent -- they are deeply covariant quantities. The goal of the systematic uncertainty estimates is to capture these degenerate sources of uncertainty in a non-parametric way. 

\section{Stellar Evolution Model Comparison}
\label{sec:mist}

\begin{figure*}[b]
    \centering
    \includegraphics[width=0.8\linewidth]{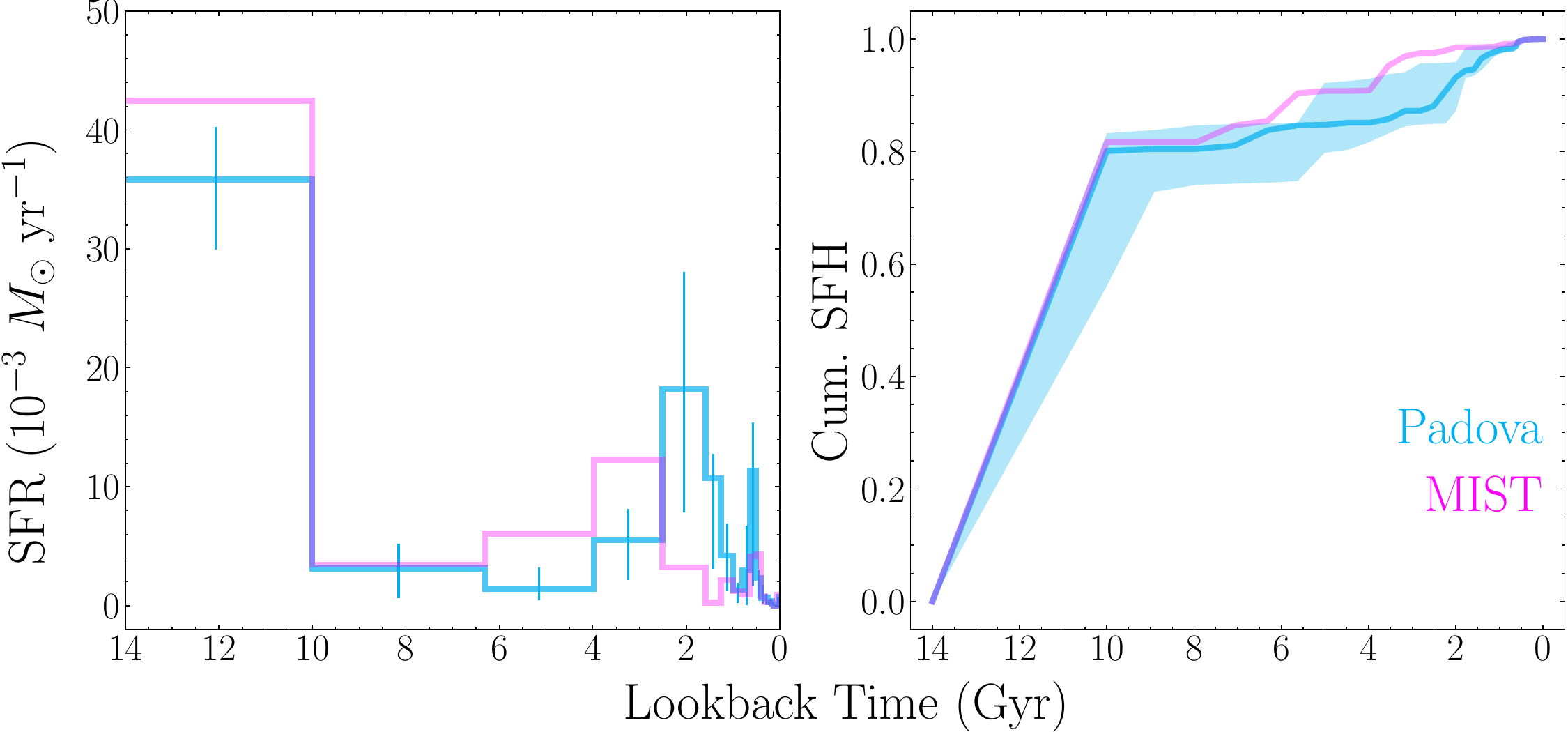}
    \caption{F8D1's SFH calculated in the deep central WFC3 field, using both the Padova (fiducial, blue) and MIST (pink) evolutionary models. The best-fit instantaneous SFHs are shown in the left column, while the cumulative SFHs are shown on the right. The total (statistical + systematic) uncertainties on the fiducial Padova fit are shown in both panels. Overall, the difference between the two fits is consistent with the total estimated uncertainty on the fiducial Padova model (see \S\,\ref{sec:uncertainties}), while Padova is an overall significantly better fit to the data. The most noticeable difference in the instantaneous SFH is a shift in the recent prominent burst from $\sim$2\,Gyr in the case of Padova to $\sim$3\,Gyr in the case of MIST.}
    \label{fig:mist}
\end{figure*}

As an independent test of the efficacy of the Gaussian perturbation approach to systematic uncertainty estimation (\S\,\ref{sec:uncertainties}), here we directly compare the SFH of F8D1's center, inferred using the Padova evolutionary models, to a fit using the MIST evolutionary models \citep{Dotter2016,Choi2016}. For the MIST fit, we follow an identical procedure and setup as in the main paper results (\S\,\ref{sec:sfh}). The best-fit SFH solution yields the same average extinction value of $A_V\,{=}\,0.25$\ as the Padova fit. Figure \ref{fig:mist} compares the instantaneous and cumulative SFHs between the two models directly. As expected, the difference between the MIST and Padova fits is consistent with the total estimated uncertainty envelope of the fiducial fit. The Padova model is an overall much better fit to the data, with a 2.5$\times$\ lower \texttt{MATCH} fit value. As the fit value reported by \texttt{MATCH} is related to a log-likelihood, this is equivalent to a many orders-of-magnitude difference in the quality of the fit. 

As discussed in \S\,\ref{sec:uncertainties}, the uncertainty estimated on the cumulative SFH can be naturally interpreted as an uncertainty on the timing of features in the SFH. This is born out in the explicit comparison between the MIST and Padova fit results: at intermediate ages the systematic uncertainty is larger (likely driven by uncertainty in modeling of the RC), and therefore there is a larger uncertainty on the timing of specific features in the SFH. In this specific case, this leads to a difference of $\sim$1\,Gyr in the timing of the dominant burst inferred from both Padova and MIST. The recent SFH is much more tightly constrained, however, and thus there is very little change to the timing of the 500\,Myr burst (though it is less pronounced in the MIST fit).

Overall, while this example demonstrates that the fidelity of timing specific features in the SFH at intermediate ages should not be over-interpreted at the depths of these data, it does not alter the fundamental inference that F8D1 experienced two distinct periods of enhancement over the past 6\,Gyr: one at intermediate ages ($\sim$2\,Gyr ago), and one much more recently, only $\sim$500\,Myr ago.

\end{document}